%% file: main.tex
\documentclass{article}
\usepackage{etoolbox}       %

\usepackage{graphicx}  %

\usepackage[parfill]{parskip}
\usepackage{amsmath,amsthm}
\usepackage{mathtools}  %
\usepackage[dvipsnames]{xcolor}         %
\usepackage{placeins} 

\newtoggle{arxiv}
\toggletrue{arxiv}

  \usepackage[parfill]{parskip}
  \usepackage[citestyle=authoryear-comp,sorting=nyt,maxbibnames=99,backend=biber]{biblatex}
  
  \newcommand{\citep}{\parencite}
  \newcommand{\citet}{\textcite}
  \usepackage{tcolorbox}
\tcbuselibrary{skins,breakable}

\definecolor{natureteal}{RGB}{0,115,150}   
\definecolor{natureblue}{RGB}{0,62,116}    

  \newlength{\defbaselineskip}
  \RequirePackage[T1]{fontenc}
  \RequirePackage[tt=false, type1=true]{libertine}
  \RequirePackage[varqu]{zi4}
  \RequirePackage[libertine]{newtxmath}

\usepackage{bm}
\usepackage{xurl}
\usepackage[inline]{enumitem}
\usepackage{booktabs}
\usepackage[dvipsnames]{xcolor}  %
\usepackage{subcaption}
\usepackage{multirow}
\usepackage{wrapfig}
\usepackage{algorithm,algorithmicx,algpseudocode}
\usepackage{nicematrix}

\usepackage[frozencache,cachedir=.]{minted} %
\usepackage[colorlinks=true,linkcolor=natureteal,citecolor=natureteal,urlcolor=natureteal]{hyperref}

\usepackage[capitalise,noabbrev]{cleveref}
\usepackage[strict]{changepage}

\usepackage{import}

\usepackage{pifont}%

\title{Directional coherence and effect magnitude in single-cell CRISPR perturbation responses}
\author{Prashant C. Raju\\{\footnotesize\texttt{rajuprashant@gmail.com}}}

\date{}

\begin{document}

\maketitle

\begin{abstract}
\noindent
\input{abstract}
\end{abstract}


\subsubsection*{\begin{center}
    \small Graphical Abstract
\end{center}}
\begin{figure}[H]
    \centering
        \includegraphics[width=\linewidth]{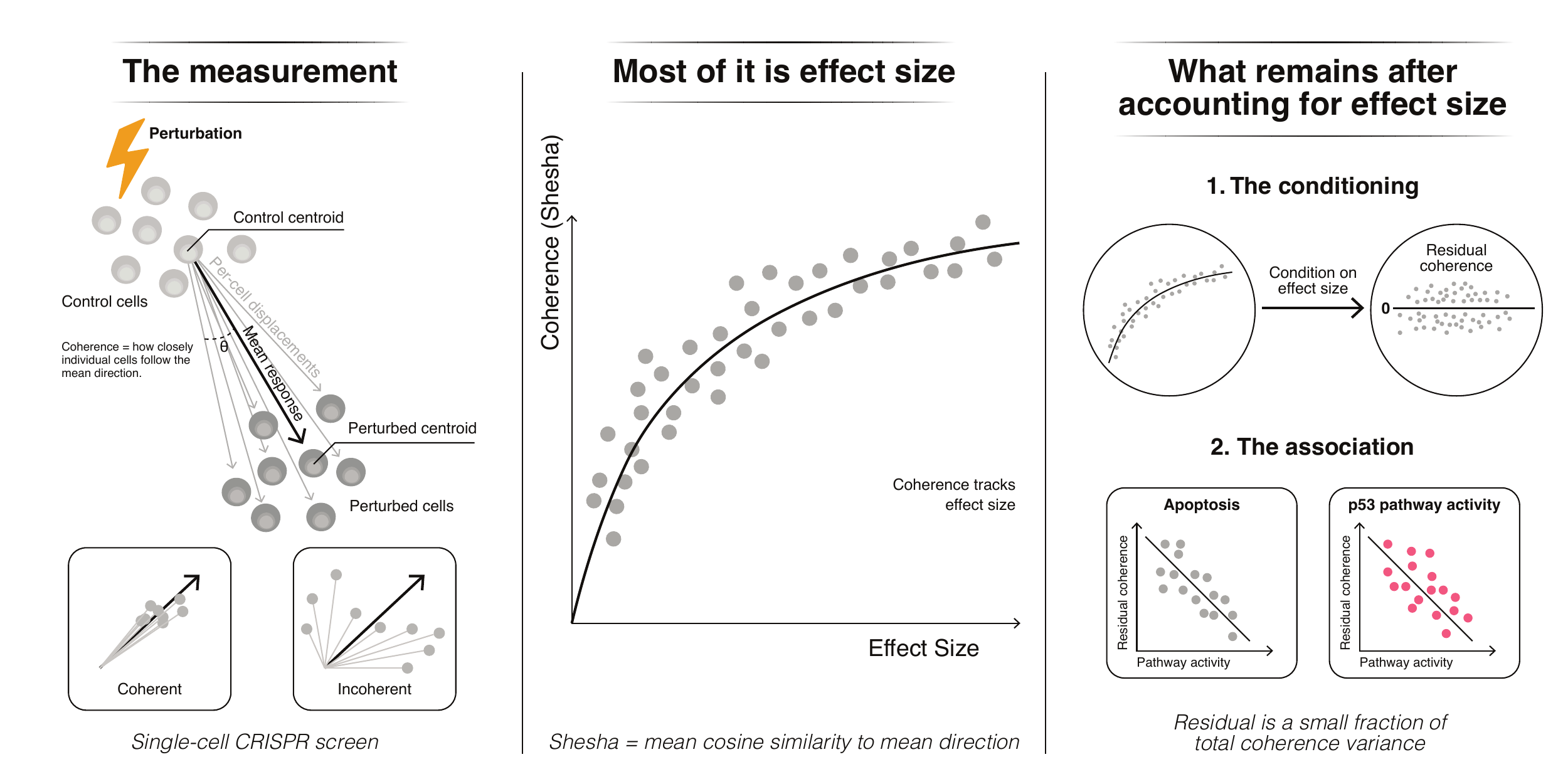}
\end{figure}

\maketitle
\setcounter{figure}{0}

\section{Introduction}
\label{sec:intro}
\input{intro-new}

\section{Materials and Methods}
\label{sec:methods}
\input{methods-new}

\section{Results}

\subsection{Coherence tracks effect magnitude across modalities and effect-size definitions}
\label{sec:pertubation}
\input{perturbation-new}


\subsection{Perturbation efficiency does not explain the coherence-magnitude relationship}
\label{sec:efficiency}
\input{effciency}

\subsection{Concordance depends on the representation}
\label{sec:scgpt}
\input{scgpt-new}


\subsection{Residual coherence is associated with apoptosis and p53 signaling}
\label{sec:stress}
\input{stress-new}


\subsection{Raw associations are dominated by effect size}
\label{sec:effect-size}
\input{effect-size}

\subsection{Limits of the measurement}
\label{sec:limits-of-measurement}
\input{limits-of-measurement}

\section{Discussion}
\label{sec:discussion}
\input{discussion-new}
\FloatBarrier

\section*{Conflicts of interest}
The authors declare that they have no competing interests.


\section*{Data availability}
\input{code}

\section*{Acknowledgments}
\label{sec:acknowledgements}
\input{acknowledgements-new}

\printbibliography

\newpage

\setcounter{figure}{0}

\renewcommand{\figurename}{Supplementary Figure}

\renewcommand{\thefigure}{S\arabic{figure}}

\setcounter{table}{0}

\renewcommand{\tablename}{Table}

\renewcommand{\thetable}{S\arabic{table}}

\appendix

\onecolumn

\input{si-new}

\input{si_figs_tables-new}

\end{document}

%% file: abstract.tex
Single-cell CRISPR screens summarize each perturbation by how far cells move from their unperturbed state, averaging over cell-to-cell variation. Whether the cells moved together is not captured: two perturbations with identical effect magnitude can differ qualitatively, one driving cells along a shared trajectory, the other scattering them around the same mean. We define Shesha\textsubscript{P} coherence ($S_p$), the mean cosine similarity between individual cell displacement vectors and their mean, and ask what it adds across six Perturb-seq datasets (2,285 perturbations; CRISPRa, CRISPRi, Cas9 knockout). Coherence and effect magnitude are closely coupled (Spearman $\rho =0.84$--$0.98$), and the coupling persists in a foundation-model embedding, under alternative effect-size definitions, and when analysis is restricted to responding cells. Associations reported without conditioning on effect size largely recover it, shown here for two comparisons that vanish under conditioning. Effect size accounts for 89--96\% of coherence variance; within the remainder, coherence is negatively associated with apoptosis and p53 signalling in the two largest screens. Two standard responder classifiers agree on only 68\% of cells. Directional coherence is largely a restatement of effect magnitude, and heterogeneity measures should report their redundancy with effect size as a matter of course. $S_p$ is implemented in the open-source \href{https://pypi.org/project/shesha-geometry/}{\texttt{shesha-geometry}} Python package.

%% file: intro-new.tex
Two cells carrying the same genetic perturbation do not respond the same way. In pooled CRISPR screens with single-cell readout, cells assigned an identical guide routinely diverge: some shift far from their unperturbed state and others barely move, some engage one downstream program and others another, and in developmental systems a shared perturbation can leave one cell differentiating and its neighbor stem-like~\citep{Replogle2022,Weinreb2020}. That variation reflects where each cell began, which programs were available to it, and how completely the perturbation took hold. Collapsing it into a mean is what single-cell readout was meant to avoid.

The analysis of these screens has not kept pace with that resolution. A perturbation's effect is still usually reported as the distance between mean perturbed and mean control expression profiles. Two perturbations with
the same summary can differ in kind: one may drive every cell along a shared trajectory, the other may scatter cells in many directions around the same mean. A hit that averages over subpopulations doing different things will not behave like a uniform response, whether the next step is mechanistic follow-up, a predictive model, or a therapeutic target.

Recent work has begun to recover what the mean discards. \cite{Nadig2024} decompose perturbation effects into shared and gene-specific components through transcriptome-wide differential expression, characterizing how broadly a perturbation acts. \cite{Song2025} estimate a per-cell perturbation-response score, identifying which cells within an assigned guide population responded strongly. Harmonized benchmarking resources~\citep{Peidli2022} have made comparison across screens possible. None of these asks whether the cells that responded moved in the same direction.

\begin{figure}[H]
    \centering
        \includegraphics[width=\linewidth]{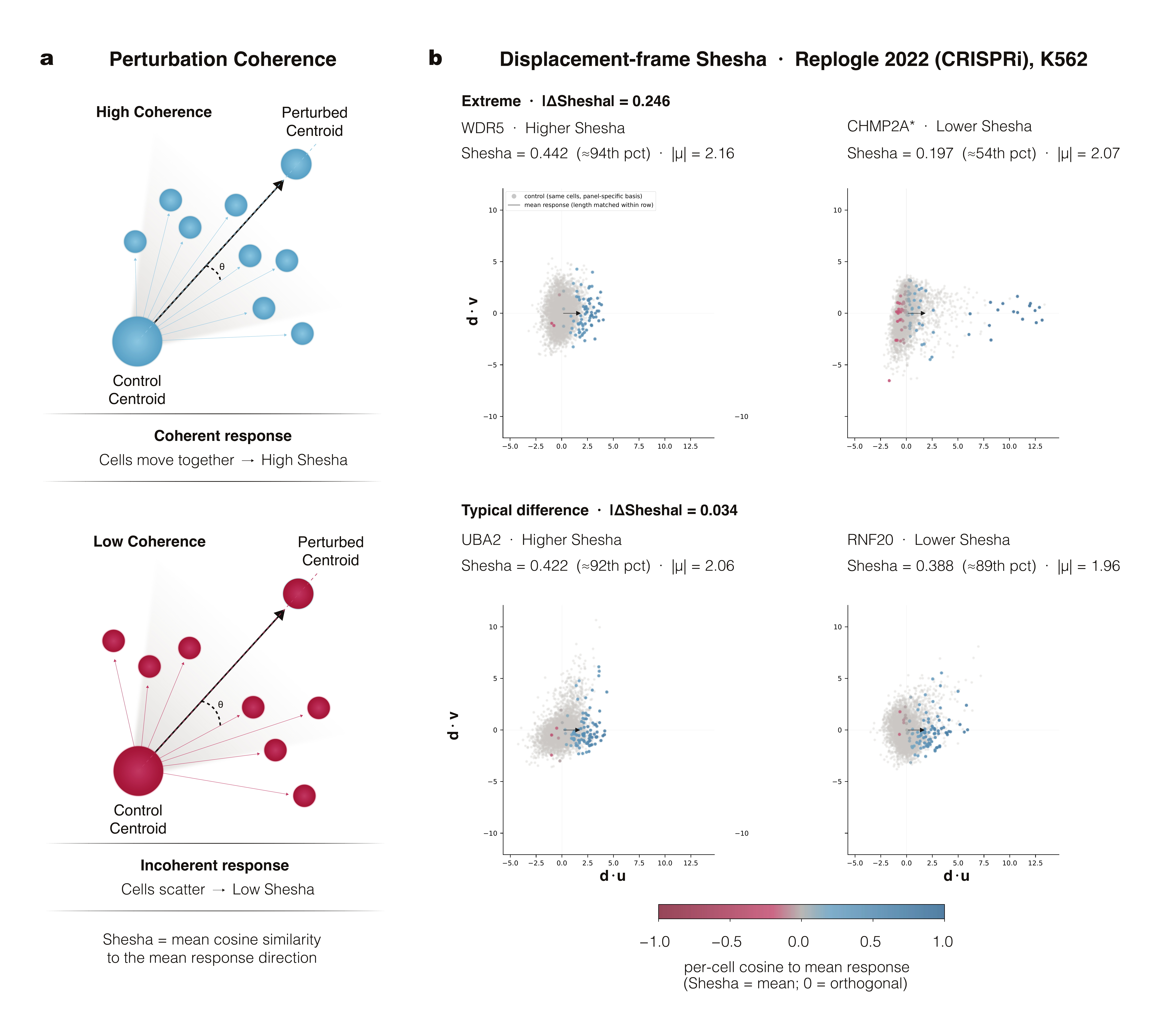}
\caption{\textbf{Directional coherence of single-cell perturbation responses.}
(a)~Shesha\textsubscript{P} ($S_p$) is the mean cosine similarity between each perturbed cell's displacement from the control centroid and the mean response direction. Both schematics share the same mean response vector; only the angular spread of individual displacements differs, so coherence and effect magnitude are in principle separable.
(b)~The same contrast in Replogle et al. 2022~\citep[CRISPRi, K562][]{Replogle2022}. Cells are plotted in each perturbation's own displacement frame, with the horizontal axis along the mean response direction $\mathbf{u}$ and the vertical axis the leading principal component of the residual displacements $\mathbf{v}$; perturbed cells are colored by their individual cosine to the mean response, the quantity $S_p$ averages. Control cells (grey) are the same cells in all four panels, projected into each panel's own basis. Black arrows mark the mean response, matched in length within each row. Upper row shows the pair with the largest coherence difference available at matched effect magnitude; lower row shows the median difference, where the two distributions are visually indistinguishable. Percentiles are within-dataset. Each panel has its own basis, so vertical positions are not comparable across panels, and the two-dimensional projection compresses 49 orthogonal dimensions into one, so visible angles understate the true angular spread. $^{*}$Partial penetrance on the $\mathbf{u}$ projection (a near-control cell mass alongside a responding tail); this is a per-perturbation observation and responder filtering does not remove the coherence--magnitude relationship in any dataset (\textbf{Fig.~\ref{fig:efficiency_filter}}).}
\label{fig:definition}
\end{figure}

That question is geometric. For each perturbed cell, compute its displacement from the control centroid; for the perturbation, compute the mean of those displacements; then take the average cosine similarity between the individual displacements and their mean. We call this Shesha\textsubscript{P} coherence ($S_p$), applying to perturbation biology a measure of geometric self-consistency developed for representational stability more generally~\citep{raju2026geometric,shesha2026,raju2026canary}. High values indicate cells moving together, low values cells scattering, and the two configurations are in principle separable from how far the cells moved (\textbf{Fig.~\ref{fig:definition}}). Coherence does bound magnitude from below, since aligned displacements produce a longer mean than scattered ones. But that constraint runs one way and says nothing about the joint distribution real perturbations occupy: cells moving a short distance in near-unison, and cells moving far in many directions, are both geometrically admissible and both biologically interpretable.

Separable in principle is not the same as separable in the data. A statistic can be well defined, cheap to compute, and still mostly recover a quantity already in routine use. We treat coherence as a quantity to characterize rather than a metric to promote: how much of it is effect magnitude, what remains after that is accounted for, and what the remainder is associated with. To our knowledge this has not been asked of a perturbation-heterogeneity measure. Answering it takes more than reporting one correlation. It requires conditioning against effect-size measures that do not share the metric's construction, testing whether the result depends on which cells count as responding, and checking whether rankings survive a change of representation. 

We address it in six single-cell CRISPR datasets~\citep{adamson2016multiplexed,dixit2016perturb,norman2019exploring,papalexi2021characterizing,Replogle2022} spanning activation, interference and Cas9 knockout in three cell types, $2{,}285$ perturbations under a single frozen configuration. Coherence and effect magnitude are closely coupled (Spearman $\rho = 0.843$ to $0.977$). The coupling survives recomputing both quantities in a transformer embedding~\citep[scGPT;][]{Cui2024}, redefining effect size as an energy distance or as transcriptome-wide impact from differential expression, and restricting analysis to cells classified as responding to their assigned guide. Associations reported without conditioning on effect size largely recover effect size, including two comparisons that are strongly positive until that step and vanish after it. What remains is a residual of four to eleven percent of coherence variance, within which coherence is negatively associated with apoptosis and p53 signalling in the two largest screens. Those associations are small, concentrated in the largest dataset, and conditional on which family of effect-size measure defines the residual.

%% file: methods-new.tex
\subsection{Datasets}

Six single-cell CRISPR perturbation datasets were analyzed (\textbf{SI Appendix, Table~\ref{tab:datasets}}): Norman et al.\ 2019 (CRISPRa, K562, $n = 236$ perturbations including 105 single-gene and 131 combinatorial)~\citep{norman2019exploring}; Adamson et al.\ 2016 UPR (CRISPRi, K562, $n = 87$) and Adamson et al.\ 2016 pilot (CRISPRi, K562, $n = 8$; secondary / underpowered for pathway claims)~\citep{adamson2016multiplexed}; Dixit et al.\ 2016 (CRISPR-KO, BMDCs, $n = 98$)~\citep{dixit2016perturb}; Papalexi et al.\ 2021 (CRISPR-KO, THP-1, $n = 24$)~\citep{papalexi2021characterizing}; and Replogle et al.\ 2022 (genome-scale CRISPRi, K562, $n = 1{,}832$ perturbations after filtering for $\geq 50$ cells per perturbation)~\citep{Replogle2022}. Together these comprise $n = 2{,}285$ perturbations in the frozen Sp table; pathway and stress partials use the five scoreable datasets ($n = 2{,}277$) after excluding the Adamson pilot. All datasets were accessed via pertpy (v1.0.4)~\citep{pertpy}; Papalexi was additionally rebuilt from GEO (GSE153056) as an independent check that scores are not pertpy-loader--dependent (\textbf{SI Appendix, Table~\ref{tab:papalexi_geo_ladder}}).

\subsection{Preprocessing}

Each dataset was preprocessed independently to prevent cross-dataset batch effects. All manuscript-facing Sp, pathway, stress, QC, and embedding tables use one frozen preprocessing and scoring pipeline (seed 320; version stamps and software pins in \textbf{SI Appendix, Tables~\ref{tab:pipeline_supersession} and~\ref{tab:repro_verification}}). The pipeline consisted of quality filtering (cells with fewer than 100
detected genes removed). Per-dataset matrix scale was pinned from inspection of \texttt{.X} (maximum and fraction near-integer): Norman and Dixit are already log-transformed (\texttt{matrix\_is\_log=True}; normalization and log1p skipped), whereas Adamson UPR/pilot, Papalexi, and Replogle are raw counts (\texttt{matrix\_is\_log=False}; \textbf{SI Appendix, Table~\ref{tab:matrix_is_log}}). Raw-count matrices were library-size normalized to the median library size (\texttt{target\_sum=None}) and log1p-transformed. The frozen 50-dimensional linear embedding used two recorded computational backends. Adamson UPR, Adamson pilot, and Papalexi used \texttt{scanpy.pp.highly\_variable\_genes} (2,000 genes) followed by \texttt{scanpy.pp.pca}. Norman, Dixit, and Replogle used a memory-safe sparse backend: genes were ranked by variance divided by mean on all retained cells for Dixit or on a seed-320 sample of 20,000 cells for Norman and Replogle, followed by \texttt{sklearn.decomposition.TruncatedSVD} (\texttt{random\_state=320}, \texttt{n\_iter=5}). Perturbations with fewer than 50 cells were excluded. Large perturbations and controls were downsampled in an order-invariant manner by hashing \texttt{seed|obs\_name} (at most 100 cells per perturbation and 5{,}000 controls), so cell selection does not depend on loader row order. All downstream coherence and magnitude computations were
performed on this frozen linear embedding. Robustness to embedding dimensionality (10, 20, 30, 50, 100 dimensions), refitting the same dataset-specific backend at each dimension, is confirmed in \textbf{SI Appendix, Fig.~\ref{fig:pca_ablation} and Table~\ref{tab:pca_ablation}}.

\subsection{Control group identification}

Control group assignment used a multi-stage matching protocol to accommodate heterogeneous labeling conventions across datasets. Labels were first matched case-insensitively against known control terms (``control'', ``ctrl'', ``non-targeting'', ``NT'', ``unperturbed''), then by delimiter-aware regex for short tokens, and finally by substring matching for embedded keywords. Dataset-specific handling: Replogle labels containing ``non-targeting'' or beginning with ``chr'' were assigned to control, and ``pos\_control'' labels were removed; Papalexi non-targeting guides (NTg1--NTg7) were pooled into a single control group (2,386 cells); Norman cells labeled ``control'' served as the control group. Full details are provided in \textbf{SI Appendix, Control Group Identification}.

\subsection{Shesha perturbation coherence}
\label{sec:shesha}

For each perturbation $p$ applied to $n_p$ cells, we computed the shift vector for each perturbed cell as the difference between its frozen linear-embedding coordinates and the control centroid. The mean perturbation direction is the average of these shift vectors. Shesha\textsubscript{P} coherence ($S_p$) is defined as the mean cosine similarity between individual cell shift vectors and this mean direction:

\begin{equation}
S_p = \frac{1}{n_p} \sum_{i=1}^{n_p}
\frac{\mathbf{d}_i \cdot \bar{\mathbf{d}}}
{\|\mathbf{d}_i\| \, \|\bar{\mathbf{d}}\|}
\label{eq:sp}
\end{equation}

where $\mathbf{d}_i = \mathbf{x}_i - \boldsymbol{\mu}_{\text{ctrl}}$ is the shift vector for cell $i$ and $\bar{\mathbf{d}} = \frac{1}{n_p} \sum_i \mathbf{d}_i$ is the mean perturbation direction. A perturbation with $S_p$ near 1 drives all cells in the same direction (high coherence); $S_p$ near 0 indicates scattering across expression space. Effect magnitude ($M_p$) is the Euclidean norm of the mean shift vector, $M_p = \|\bar{\mathbf{d}}\|$.

The metric adapts geometric self-consistency from the general Shesha framework~\citep{raju2026geometric,shesha2026} to perturbation biology. To avoid implying temporal or dynamical stability, we use ``coherence'' for the measured quantity; $S_p$ remains the mathematical symbol. The authoritative frozen manuscript scores are computed by \texttt{pipeline\_core.calculate\_sp}, the local implementation of Eq.~\eqref{eq:sp}. Every manuscript table and figure reporting the primary frozen $S_p$ uses those scores. The pinned \texttt{shesha.bio} package is used only by auxiliary callers and does not replace the frozen score table; versions are recorded in \textbf{SI Appendix, Reproducibility}. Robustness to distance metric choice (Euclidean, Mahalanobis, $k$-NN) is evaluated across all six datasets in \textbf{SI Appendix,
Fig.~\ref{fig:distance_robustness} and Table~\ref{tab:distance_robustness}}. The uniform frozen filter excludes perturbations represented by fewer than 50 cells. The Adamson pilot ($n=8$ perturbations) is retained in the six-dataset $S_p$ table but excluded from pathway and stress partials.

\subsection{E-distance competitor}
\label{sec:edistance}

To test whether the $S_p$--magnitude relationship is an artifact of centroid shift, we computed the scPerturb / pertpy energy distance~\citep{Peidli2022} from each perturbation to the control distribution in the same frozen 50-PC embedding and the same hash-selected cells used for $S_p$. E-distance is the Euclidean $U$-statistic $2\,\mathbb{E}\|X-Y\| - \mathbb{E}_{i\neq j}\|X_i-X_j\| - \mathbb{E}_{i\neq j}\|Y_i-Y_j\|$, using off-diagonal within-group means. The signed estimate is retained (it can be negative when the two distributions are close) and is never labelled magnitude; centroid magnitude $M_p$ remains the primary size
measure. The full frozen control set (at most 5{,}000 cells) is used for every perturbation. Per dataset we report Spearman $\rho(M_p,\text{E-distance})$, $\rho(S_p,M_p)$, $\rho(S_p,\text{E-distance})$, and the fraction of rank-$S_p$ variance remaining after rank-OLS residualization on each size measure. Pathway and stress associations were re-fit under three size models ($|$centroid, $|$E-distance, $|$centroid$+$E-distance) and three QC models ($|$centroid$+$QC, $|$E-distance$+$QC, joint). Primary survivors are the intersection of those three QC models, not their union. Per-dataset values are in \textbf{SI Appendix, Tables~\ref{tab:edistance_competitor} and~\ref{tab:trade_competitor}}.

\subsection{TRADE transcriptome-wide impact}
\label{sec:trade}

As a non-geometric size competitor we estimated transcriptome-wide impact (TWI) with TRADE~\citep{Nadig2024}. TWI is the mixture variance $\mathrm{Var}(\beta)$ of true gene-level $\log_2$ fold-changes from an ashr half-uniform mixture with a uniform prior~\citep{Stephens2016}. Inputs are Welch $\log_2$FC and standard errors versus the frozen control cells on the full log-normalized transcriptome (same hash-selected cells as $S_p$; no highly variable gene subset and no PCA). scanpy \texttt{log1p} is natural log and is converted to $\log_2$ by dividing by $\ln 2$. Genes with non-finite SE, $|\log_2\mathrm{FC}| > 10$, or matching the perturbed target are excluded. TWI values were computed via a Python implementation of the ashr half-uniform mixture variance, numerically consistent with R \texttt{ashr}~\citep{Stephens2016}. Per dataset we report $\rho(S_p,\mathrm{TWI})$, $\rho(M_p,\mathrm{TWI})$, and the geometric comparators $\rho(S_p,M_p)$ and $\rho(S_p,\text{E-distance})$. TRADE was added after the frozen pipeline lock and is a robustness column: it is not folded into the three-model QC intersection. TWI-conditioned pathway partials are interpreted only after a diagnostic Spearman of TWI against each Hallmark score (centroid magnitude as comparator), because TWI is estimated from gene-level differential expression and Hallmark scores are themselves expression-derived. The Adamson pilot ($n = 8$) is excluded from pathway claims. Details are in \textbf{SI Appendix, Tables~\ref{tab:trade_competitor} and~\ref{tab:twi_pathway}}.

\subsection{Perturbation-response score proxies}

Mean per-cell Euclidean and Mahalanobis distances from the control centroid were evaluated as geometric proxies for Song et al.'s perturbation-response score (PS)~\citep{Song2025}. Their weak correlations with the scMAGeCK-derived score (computed via a faithful Python reimplementation of the scMAGeCK constrained-optimisation algorithm; the official R package is also supported and yields numerically equivalent results; $\rho=0.097$ and $0.149$, respectively; \textbf{SI Appendix, Table~\ref{tab:ps_proxy_caution}}) show that neither distance is a substitute for PS. They are retained only as a proxy caution, not as alternate implementations of PS. The responder-filter sensitivity below uses Song PS as the primary caller and Mixscape as a secondary caller.

\subsection{Perturbation-efficiency sensitivity}
\label{sec:efficiency-filter}

To test whether the magnitude--coherence relationship is an artifact of unperturbed cells within assigned guide pools, we recomputed $S_p$ after retaining only putative responder cells identified by Song et al.'s
perturbation-response score (PS; primary arm; computed via the Python reimplementation of scMAGeCK described above) or Mixscape~\citep{papalexi2021characterizing} (secondary arm on Norman). Responder calling and Sp rescoring use the same frozen preprocess path as the manuscript Sp table; this sensitivity analysis does not replace the frozen Sp table. For PS, cells were labeled responders when within-perturbation rescaled PS $\geq 0.5$. Primary endpoints are the change in Sp--magnitude Spearman $\rho$ after filtering and the rank correlation between Sp on all cells and Sp on responders. Absolute $\Delta S_p$ after filtering is not interpreted as recovered coherence, because both callers select along the mean signature direction. The Adamson
pilot was excluded by the same thin-$n$ floor used elsewhere ($n < 30$); Papalexi ($n = 24$) is reported in Table~\ref{tab:efficiency_ps}. Full dataset-level results, PS--Mixscape per-cell agreement, and version stamps are reported in \textbf{SI Appendix, Tables~\ref{tab:efficiency_mixscape} and~\ref{tab:responder_agreement}}.

\subsection{Split-half directional-agreement assay}

We used split-half directional agreement as a descriptive check on sampling precision, not as experimental reproducibility. For each perturbation in Replogle ($n=1{,}832$) and Norman ($n=236$), cells were randomly divided into equal halves. We computed each half's mean shift relative to the same control centroid and recorded the cosine between the two half-shift vectors, repeating this procedure for 50 random splits. Because this endpoint and $S_p$ are derived from the same cell-level shifts and control reference, no predictive or biological-replication claim is made from their association.

The non-circular check used independent guides targeting the same gene. After conditioning on magnitude, gene-level $S_p$ did not predict between-guide directional agreement in Replogle ($n=104$ genes; $\rho=0.006$, 95\% CI $-0.220$, $0.225$) or Papalexi ($n=23$ genes; $\rho=0.183$, 95\% CI $-0.317$, $0.644$; \textbf{SI Appendix, Table~\ref{tab:between_guide}}). These null results are reported as the relevant test of the stronger reproducibility claim.

\subsection{Pathway-level stress scoring}

To assess the association between Shesha\textsubscript{P} coherence and cellular stress, we computed composite pathway scores for five MSigDB Hallmark gene sets~\citep{Liberzon2015}: Unfolded Protein Response (UPR), p53
Pathway, Apoptosis, Reactive Oxygen Species (ROS), and mTORC1 Signaling. Pathway scores were computed using \texttt{scanpy.tl.score\_genes} with curated gene lists (72--78 overlapping genes for UPR, 32--51 for p53, 53--73 for Apoptosis, 45--55 for ROS per dataset) on the full normalized transcriptome before highly variable gene selection, then correlated with perturbation coherence controlling for effect magnitude. Both raw Spearman correlations and partial correlations (controlling for magnitude) are reported, with Benjamini-Hochberg correction applied across the pathway--dataset combinations.

Per-perturbation cell-quality covariates are mean mitochondrial percentage, detected-gene count, and library depth \\(\texttt{percent\_mito}, \texttt{n\_genes}, \texttt{n\_counts}) over the same frozen cells. When $n < 40$, only mitochondrial percentage is used to avoid rank deficiency. Primary pathway and stress survivors
are the intersection of $|$centroid$+$QC, $|$E-distance$+$QC, and the joint model. TWI-conditioned partials are a separate robustness column and are not added to that intersection.

Magnitude-conditioned and QC-conditioned associations were adopted when bootstrap confidence intervals excluded 0, $|\rho| > 0.1$, and BH-FDR $< 0.05$. Benjamini--Hochberg FDR is computed once per dataset $\times$ covariate model across the nine scored features (five Hallmark pathway scores and four stress markers); the same family is used for $|$magnitude, $|$centroid+QC, $|$E-distance+QC, and the joint model. After observing knife-edge CI bounds and small-$n$ rank deficiency, three post hoc conservative rules were applied: (i)~demote otherwise-surviving rows whose nearer $|\mathrm{CI}|$ bound is strictly $< \varepsilon = 0.02$ to indeterminate; (ii)~use mito-only QC covariates when $n < 40$; (iii)~report point estimates only / exclude from the gate when $n < 30$. These post hoc rules and the adopted $\varepsilon=0.02$ threshold are reported explicitly in the SI.

At the individual marker level, four canonical stress response genes (DDIT3, ATF4, XBP1, HSPA5) were assessed using the same partial correlation framework.

\subsection{Unbiased discordance GSEA (Approach B)}

Approach A is the magnitude-conditioned Hallmark partial correlation above. Approach B asks whether unbiased Hallmark enrichment recovers apoptosis or p53 among terms associated with residual discordance. $S_p$ is rank-residualized on magnitude; perturbations are split into residual quartiles; each discordant (Q4) perturbation is caliper-matched to a concordant (Q1) neighbor on magnitude. The matching gate requires
$|\mathrm{SMD}| \le 0.25$ for magnitude, $n_{\mathrm{cells}}$, depth, gene detection, and mitochondrial fraction, and at least 15 pairs. Cells per perturbation are equalized, profiles are pseudobulked, and Mann--Whitney tests are run across perturbations (not cells). Pre-ranked GSEA uses Hallmark in the main report and KEGG/Reactome in the SI. Only Replogle and Norman are powered for this design. In the frozen run, magnitude balanced after matching, but cell-quality covariates failed the pre-specified gate in both datasets. Approach B was therefore dropped and Approach A carries the pathway result. This failure independently supports conditioning the primary partial models on cell-quality covariates; full SMDs are reported in \textbf{SI Appendix, Table~\ref{tab:approach_b_balance}}.

\subsection{scGPT validation}

To test whether the magnitude--coherence relationship is specific to the frozen linear embedding, we recomputed Shesha\textsubscript{P} coherence and magnitude in scGPT ``Whole Human'' pretrained embeddings~\citep{Cui2024} for four datasets (Norman, Adamson UPR, Dixit, Replogle). Raw counts (not log-normalized) were embedded using \texttt{embed\_data()} with deterministic settings. Coherence and magnitude were computed identically to the frozen linear-embedding pipeline. The within-scGPT $S_p$--magnitude correlations are a redundancy check on Figure~2 and are not conflated with cross-representation concordance
(\textbf{SI Appendix, Fig.~\ref{fig:scgpt_concordance}}; embedding comparison against frozen $S_p$ is \textbf{SI Appendix, Table~\ref{tab:embedding_concordance}}). Papalexi is excluded because scGPT ranks are not numerically reproducible at $n = 24$; the Adamson pilot is excluded ($n = 8$). Within-scGPT $S_p$--magnitude values are in \textbf{SI Appendix, Table~\ref{tab:embedding_magnitude}}.

\subsection{Diffusion maps and PHATE}

To test whether $S_p$ rankings are specific to the frozen linear embedding, we recomputed $S_p$ and magnitude on the same frozen cells in scanpy diffusion maps~\citep{Haghverdi2015, Haghverdi2016} (neighbors on 100 PCs, $k = 15$; trivial DC0 dropped so scoring uses the informative components) and in PHATE~\citep{Moon2019} ($k = 5$, internal PCA $= 100$, SGD-MDS, with SMACOF as the convergence control when SGD does not converge). Manuscript concordance is the Spearman correlation of each embedding's $S_p$ (and, separately, its magnitude) against frozen linear-embedding $S_p$ (respectively $M_p$). Within-embedding $S_p$--magnitude correlations
are reported separately and are not conflated with cross-representation concordance. Papalexi is excluded from scGPT concordance because Sp ranks are not numerically reproducible at $n = 24$. Per-dataset values and the PHATE convergence control are in \textbf{SI Appendix, Tables~\ref{tab:embedding_concordance} and \ref{tab:phate_convergence}}.

\subsection{Statistical analysis}

All confidence intervals were computed via bootstrap resampling (10,000 iterations, seed 320, percentile method). The magnitude--coherence scatter plots (\textbf{Fig.~\ref{fig:magnitude_coherence}} and \textbf{SI Appendix, Fig.~\ref{fig:si_mag_stab_loess}}) display a LOESS smooth (bandwidth fraction $= 0.4$) in place of a linear regression line to capture the nonlinear attenuation at low magnitudes visible in the data; reported Spearman $\rho$ values are point estimates with 95\% bootstrap CIs computed as described above. Cross-dataset generalization is the pooled Spearman after within-dataset $z$-scoring (\textbf{Fig.~\ref{fig:magnitude_coherence}g}). The SI reports a linear slope on those same $z$-scored axes (\textbf{SI Appendix, Table~\ref{tab:mixed_effects}}). That companion is not the random-intercept model of raw $S_p$ on globally $z$-scored magnitude, spread, and cell count. Partial correlations between coherence and stress markers were computed controlling for effect magnitude using rank-based partial Spearman throughout (not Spearman correlation of raw-scale residuals; \textbf{SI Appendix, Table~\ref{tab:partial_corr_arithmetic}}). All $p$-values are two-sided unless otherwise noted. Software pins, configuration version stamps, and digest checks are listed in \textbf{SI Appendix, Tables~\ref{tab:pipeline_supersession} and~\ref{tab:repro_verification}}.

\subsection*{{Generative AI Usage}}
Large language models (specifically the GPT, Claude, Gemini, and Grok families) were used to assist with factoring the initial code for reproducibility, code debugging, software dependency management, and text editing and polishing. All hypotheses, experimental designs, analyses, and interpretations were independently formulated and verified by the author, and the author assume full responsibility for all content and claims in this work.

%% file: perturbation-new.tex
\begin{figure}[H]
\centering
\includegraphics[width=\textwidth]{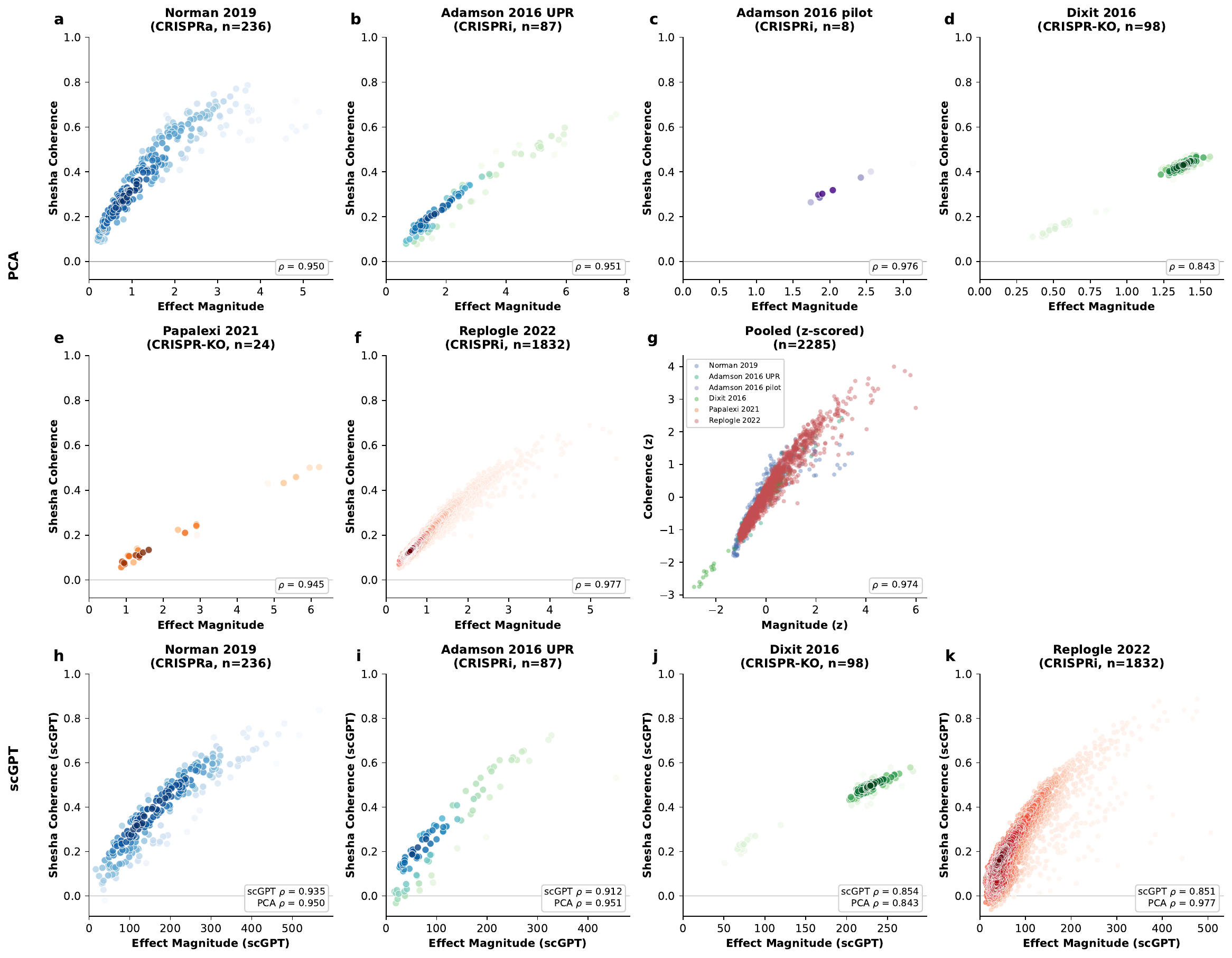}
\caption{\textbf{Coherence tracks effect magnitude in PCA and in a foundation-model embedding.}
Shesha\textsubscript{P} ($S_p$) versus effect magnitude, with Spearman $\rho$ annotated.
Rows~1--2, frozen PCA: (a)~Norman 2019 (CRISPRa), (b)~Adamson 2016 UPR (CRISPRi),
(c)~Adamson 2016 pilot (CRISPRi; $n=8$, uninformative), (d)~Dixit 2016 (CRISPR-KO),
(e)~Papalexi 2021 (CRISPR-KO), (f)~Replogle 2022 (CRISPRi), (g)~pooled after within-dataset
$z$-scoring. Row~3 recomputes both quantities in the scGPT embedding. The coupling survives
that change of representation: $\rho$ is $0.851$--$0.935$ in scGPT versus $0.843$--$0.977$
in PCA on the same four datasets (a different question from whether the two embeddings rank
perturbations the same way; \textbf{SI Appendix, Fig.~\ref{fig:scgpt_concordance}}). Papalexi and the Adamson pilot are
omitted from row~3. Axes start at the origin; point color encodes per-panel perturbation density.
Confidence intervals, sample sizes, and cell types are in Table~\ref{tab:six_datasets}.}
\label{fig:magnitude_coherence}
\end{figure}

Shesha\textsubscript{P} coherence tracks effect magnitude across all six datasets
(\textbf{Fig.~\ref{fig:magnitude_coherence}}; inventory in
\textbf{SI Appendix, Table~\ref{tab:datasets}}). Spearman correlations run from
$\rho = 0.843$ in Dixit (95\% CI $[0.754, 0.902]$) to $\rho = 0.977$ in
Replogle ($[0.974, 0.979]$); the two largest screens are $\rho = 0.977$
(Replogle, $n = 1{,}832$) and $\rho = 0.950$ (Norman, $[0.930, 0.961]$,
$n = 236$). The Adamson pilot ($n = 8$) is shown for completeness and is
uninformative at that sample size. The same pattern holds for CRISPRa,
CRISPRi and CRISPR-Cas9 knockout, in K562, BMDCs and THP-1, and in screens
that differ by two orders of magnitude in size. That spread of settings
argues against an idiosyncrasy of any one experiment: the coupling looks
like a property of perturbation geometry in single-cell expression space.

\begin{table}[H]
\centering
\caption{Directional coherence tracks effect magnitude across six CRISPR datasets.
Spearman $\rho$ between effect magnitude (Euclidean centroid displacement) and
Shesha\textsubscript{P} coherence ($S_p$, mean cosine similarity of per-cell displacement vectors to
the mean response direction) in PCA space. Bootstrap 95\% confidence intervals
(10,000 resamples, seed 320). Final row shows pooled ($z$-scored) results.
These rank correlations do not depend on a fitted curve; LOESS fits are in
\textbf{SI Appendix, Fig.~\ref{fig:si_mag_stab_loess}}. $^a$$n=8$ is below the $n \ge 15$ threshold applied in the pathway analyses; reported here for completeness and excluded elsewhere. The upper
confidence bound reaches 1.000 and the estimate should be treated as
uninformative.}
\label{tab:six_datasets}
\begin{tabular}{lllcccc}
\hline
Dataset & Modality & {Cell type} & $n$ & $\rho$ & 95\% CI & $p$\\
\hline
Norman      & CRISPRa & {K562} & 236   & 0.950  & [0.930, 0.961] & $6.8\times10^{-120}$ \\
{Adamson (UPR)}   & {CRISPRi}    & {K562}  & {87   } & {0.951} & {[0.909, 0.972]} & {$3.2\times10^{-45}$} \\
Adamson (pilot)$^a$    & CRISPRi &  K562  & 8     & 0.976 & [0.795, 1.000] &  $3.3\times10^{-5}$  \\
Dixit       & CRISPR-KO & BMDC & 98   & 0.843 & [0.754, 0.902] & $1.4\times10^{-27}$ \\
Papalexi    & CRISPR-KO  & {THP-1} &  24    & 0.945  & [0.836, 0.984] & $3.6\times10^{-12}$ \\
Replogle    & CRISPRi & {K562}  & 1,832 & 0.977 & [0.974, 0.979] & $< 10^{-300}$ \\
\hline
Pooled & ---   & {---} & 2,285 & 0.974 & [0.972, 0.976] & $< 10^{-300}$\\
\hline
\end{tabular}
\end{table}

Pooling all $2{,}285$ perturbations after within-dataset $z$-scoring gave
$\rho = 0.974$ ($[0.972, 0.976]$;
\textbf{Fig.~\ref{fig:magnitude_coherence}g}). A linear fit on those same
$z$-scored axes has slope $0.956$
(\textbf{SI Appendix, Table~\ref{tab:mixed_effects}}). LOESS on the same points
(\textbf{SI Appendix, Fig.~\ref{fig:si_mag_stab_loess}}) shows that this rank
correlation does not depend on a linear fit. Whitened (Mahalanobis) and
$k$-NN matched local controls remain high across all six datasets (whitened
$\rho = 0.963$--$0.988$; $k$-NN $\rho = 0.929$--$0.988$) and close the gap
where Euclidean is weakest, most clearly in Dixit ($0.843$ versus $0.965$ and
$0.949$; \textbf{SI Appendix, Fig.~\ref{fig:distance_robustness} and
Table~\ref{tab:distance_robustness}}). Embedding dimension from 10 to 100,
leave-one-out deletion, analysis seed, and restriction to PS-called responder
cells leave the same picture
(\textbf{SI Appendix, Figs.~\ref{fig:pca_ablation},~\ref{fig:seed_reproducibility},~\ref{fig:loo_influence}
and Tables~\ref{tab:pca_ablation},~\ref{tab:seed_reproducibility},~\ref{tab:loo_influence},~\ref{tab:efficiency_ps}}).

None of this is an artifact of how either quantity is computed.
Recomputing both coherence and magnitude inside the scGPT foundation-model
embedding still gives $\rho = 0.851$ to $0.935$ on the four datasets with
enough perturbations, against $0.843$ to $0.977$ in the frozen linear
embedding (\textbf{Fig.~\ref{fig:magnitude_coherence}h--k};
\textbf{SI Appendix, Table~\ref{tab:embedding_magnitude}}). That
within-scGPT redundancy is a different question from whether the two
embeddings rank the same perturbations the same way
(\textbf{SI Appendix, Fig.~\ref{fig:scgpt_concordance} and
Table~\ref{tab:embedding_concordance}}). Defining effect
size as energy distance---which additionally penalizes within-group
dispersion---ranks perturbations almost identically to centroid displacement
($\rho = 0.950$ to $1.000$) and accounts for slightly more coherence variance
in every dataset (\textbf{Table~\ref{tab:effect_size_competitors}};
\textbf{SI Appendix, Table~\ref{tab:edistance_competitor}}). The
stricter test is an effect-size measure that shares no geometry with
coherence: TRADE transcriptome-wide impact is estimated from gene-level
differential expression with a power correction, and its correlation with
coherence tracks its correlation with centroid displacement (median
difference $-0.01$; Replogle $0.912$ versus $0.919$;
\textbf{SI Appendix, Table~\ref{tab:trade_competitor}}). Where transcriptome-wide
impact diverges from displacement geometry, it diverges from coherence and
effect magnitude alike.

Coherence and effect magnitude are tightly coupled, and that coupling is a
property of the perturbation responses rather than of the representation or
the effect-size definition. A large mean shift requires that individual cells
move in a common direction, so perturbations that displace cells further from the control state necessarily do so more coherently. That constraint runs one
way, however: it bounds magnitude from below at a given coherence but does not
determine the joint distribution, and both a small coordinated response and a
large scattered one are geometrically admissible. What the data show is that Perturb-seq perturbations occupy mainly the diagonal. The rest of the paper asks what, if anything, coherence still carries once that coupling is accounted for.

\begin{figure}[H]
    \centering
        \includegraphics[width=\linewidth]{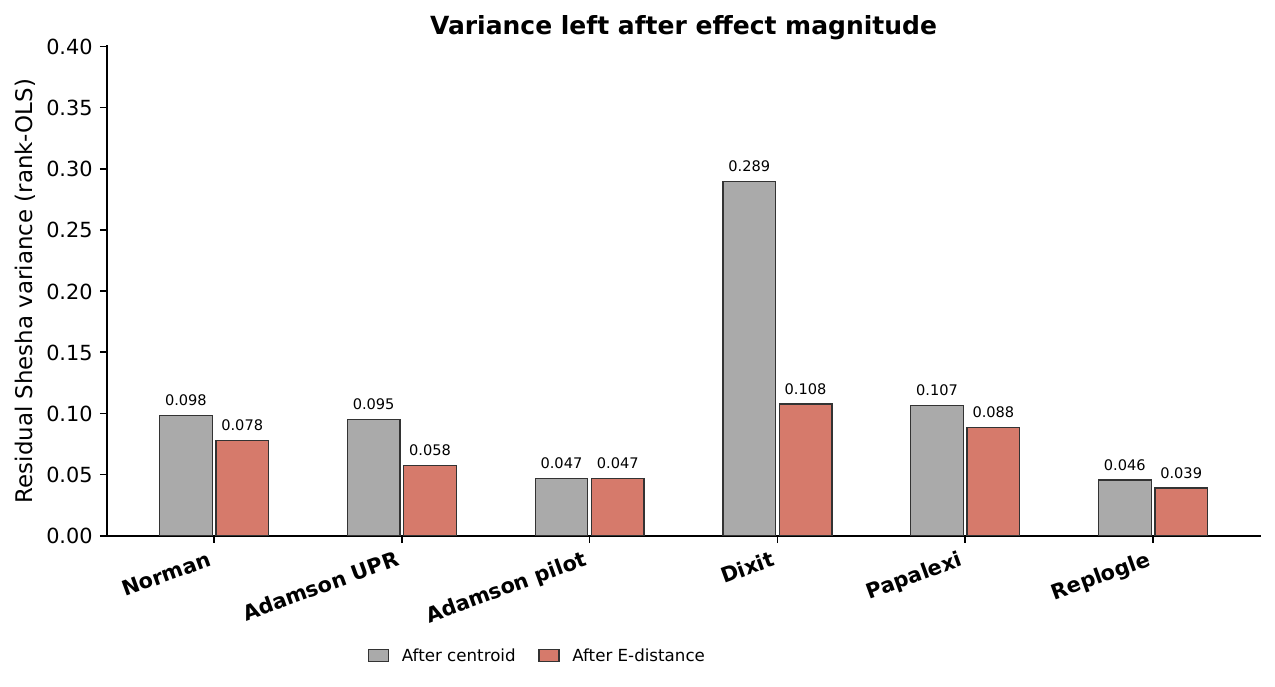}
\caption{\textbf{Residual coherence variance after conditioning on effect size.} Fraction of
Shesha variance unexplained by centroid displacement (grey) and by energy distance (red),
rank-OLS. Energy distance explains more in every dataset, most sharply in Dixit ($0.289$ to
$0.108$), whose apparent excess residual under centroid displacement is largely within-group
dispersion that energy distance captures. Conditional analyses therefore operate on $4$--$11\%$ of
coherence variance.}
\label{fig:edistance_competitor}
\end{figure}

\begin{table}[H]
\centering
\caption{Effect-size competitors on the frozen 50-PC embedding. $\rho$ is Spearman; residual
fractions are $1-R^{2}$ from rank-OLS of $\mathrm{rank}(S_p)$ on $\mathrm{rank}(\mathrm{size})$.
Centroid displacement and energy distance are both computed from the same displacement vectors as
Shesha\textsubscript{P}, so their high correlation with it is partly definitional. TRADE transcriptome-wide impact
(TWI) is estimated from gene-level differential expression with a power correction and shares no
geometry with coherence. The informative TRADE contrast is $\rho(S_p,\mathrm{TWI})$ against
$\rho(M_p,\mathrm{TWI})$ rather than against $\rho(S_p,M_p)$: the two track closely (median
difference $-0.01$; Replogle $0.912$ vs $0.919$), so where transcriptome-wide impact diverges from
displacement geometry it diverges from coherence and effect magnitude alike. The Adamson pilot
($n=8$) is uninformative.}
\label{tab:effect_size_competitors}
\smallskip
\begin{tabular}{lccccccc}
\hline
Dataset & $n$ & $\rho(S_p,M_p)$ & $\rho(S_p,\mathrm{Ed})$ & frac$|$cent. & frac$|$Ed & $\rho(S_p,\mathrm{TWI})$ & $\rho(M_p,\mathrm{TWI})$ \\
\hline
Norman & 236 & 0.950 & 0.960 & 0.098 & 0.078 & 0.688 & 0.812 \\
Adamson UPR & 87 & 0.951 & 0.971 & 0.095 & 0.058 & 0.954 & 0.968 \\
Adamson pilot & 8 & 0.976 & 0.976 & 0.047 & 0.047 & 0.857 & 0.833 \\
Dixit & 98 & 0.843 & 0.945 & 0.289 & 0.108 & 0.452 & 0.513 \\
Papalexi & 24 & 0.945 & 0.955 & 0.107 & 0.088 & 0.858 & 0.832 \\
Replogle & 1{,}832 & 0.977 & 0.980 & 0.046 & 0.039 & 0.912 & 0.919 \\
\hline
\end{tabular}
\end{table}

%% file: effciency.tex
\begin{figure}[h]
    \centering
        \includegraphics[width=\linewidth]{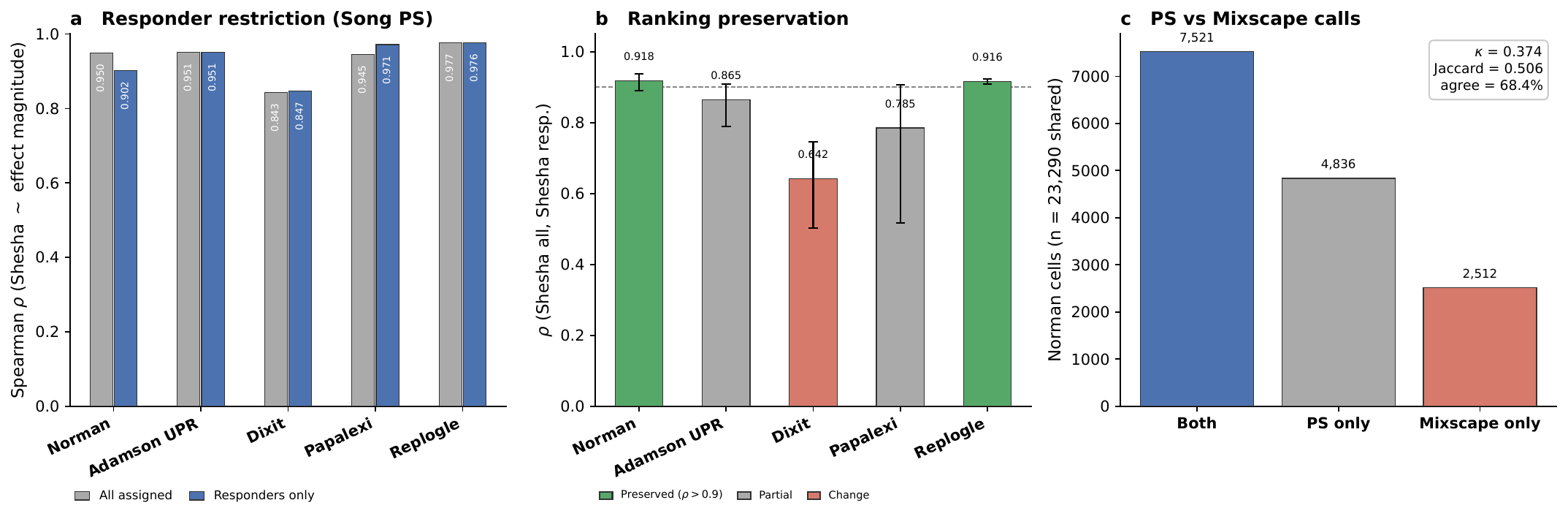}
\caption{\textbf{The coherence--magnitude relationship is not an artifact of variable perturbation
efficiency.}
(a)~Spearman $\rho$ between Shesha\textsubscript{P} and effect magnitude using all assigned cells (grey) and
restricting to cells classified as responders by perturbation scoring (blue). Change ranges
$-0.048$ to $+0.026$ across datasets, with no consistent direction.
(b)~Rank preservation between coherence computed on all cells and on responders only. Colors
indicate the pre-specified verdict: preserved ($\rho \ge 0.9$), partial, or changed ($\rho < 0.7$).
Dixit is the only dataset whose rankings change.
(c)~Per-cell responder calls in Norman disagree between the two standard classifiers
($\kappa = 0.374$, Jaccard $= 0.506$, $68.4\%$ agreement on $23{,}290$ shared cells), so
responder-only reanalysis is conditional on the classifier and is reported as a sensitivity
analysis rather than a correction. Error bars are bootstrap 95\% CIs.}
\label{fig:efficiency_filter}
\end{figure}

Pooled CRISPR screens deliver perturbations of variable efficiency: within a population assigned the same guide, some cells respond and others do not. Unperturbed cells contribute short, arbitrarily oriented displacement vectors, so a guide pool with many non-responders will look both small and incoherent. That mechanism could produce the coupling above even if magnitude and coherence were unrelated among cells that actually respond.

We recomputed both quantities on Song et al.\ perturbation-response score callers only (\textbf{Fig.~\ref{fig:efficiency_filter}a}; \textbf{SI Appendix, Table~\ref{tab:efficiency_ps}}). Median responder fractions ran from $0.36$ to $0.58$, so $42$--$64$ percent of assigned cells were called non-responding. Restricting to responders left the coherence--magnitude correlation in the same place in every dataset (\textbf{SI Appendix, Table~\ref{tab:efficiency_ps}}): Norman $0.950 \rightarrow 0.902$, Replogle $0.978 \rightarrow 0.976$, Adamson UPR $0.963 \rightarrow 0.951$, Dixit $0.841 \rightarrow 0.847$, Papalexi $0.945 \rightarrow 0.971$ (changes $-0.048$ to $+0.026$, no shared sign). These all-cell baselines are the efficiency-lane pair, not a substitute for the frozen Table~1 $\rho$. Rankings of perturbations were also largely preserved (\textbf{Fig.~\ref{fig:efficiency_filter}b}): Norman and Replogle sit above the pre-specified preservation threshold (rank $\rho = 0.918$ and $0.916$); Dixit is the only dataset whose rankings move (rank $\rho = 0.642$). The coupling is not an artifact of unperturbed cells in assigned guide pools.

The same analysis also limits how far any responder-only reanalysis can be pushed. Mixscape on Norman, as a second caller, moved the correlation by $+0.030$ --- opposite in sign to the perturbation-response score on that dataset, and on a $149/236$ survivor subset rather than the frozen Norman baseline (\textbf{SI Appendix, Table~\ref{tab:efficiency_mixscape}}). Cell-by-cell comparison shows why: on the $23{,}290$ Norman cells scored by both, the two callers agree on $68.4$ percent of labels (Cohen's $\kappa = 0.374$, Jaccard $= 0.506$; \textbf{Fig.~\ref{fig:efficiency_filter}c}; \textbf{SI Appendix, Table~\ref{tab:responder_agreement}}). Mixscape is the more conservative caller ($43.1$ versus $53.1$ percent responders) and is not nested inside the perturbation-response set: $2{,}512$ cells are Mixscape-only. Which cells count as responders is method-defined, and the sign of the change in the coherence--magnitude correlation sits inside caller-to-caller variation.

Two limits follow. Responder-only reanalysis is conditional on a classifier and is a sensitivity check, not a correction of the primary estimates; every frozen value elsewhere in this paper uses all assigned cells. Both callers also select cells whose displacement aligns with an estimated perturbation signature, which is close to what coherence measures, so coherence rises on the restricted set by construction and we do not treat that rise as a result. The endpoints here are the coherence--magnitude correlation and the preservation of rankings, neither of which is fixed by the selection rule.

%% file: scgpt-new.tex
\begin{figure*}[h]
\centering
\includegraphics[width=\textwidth]{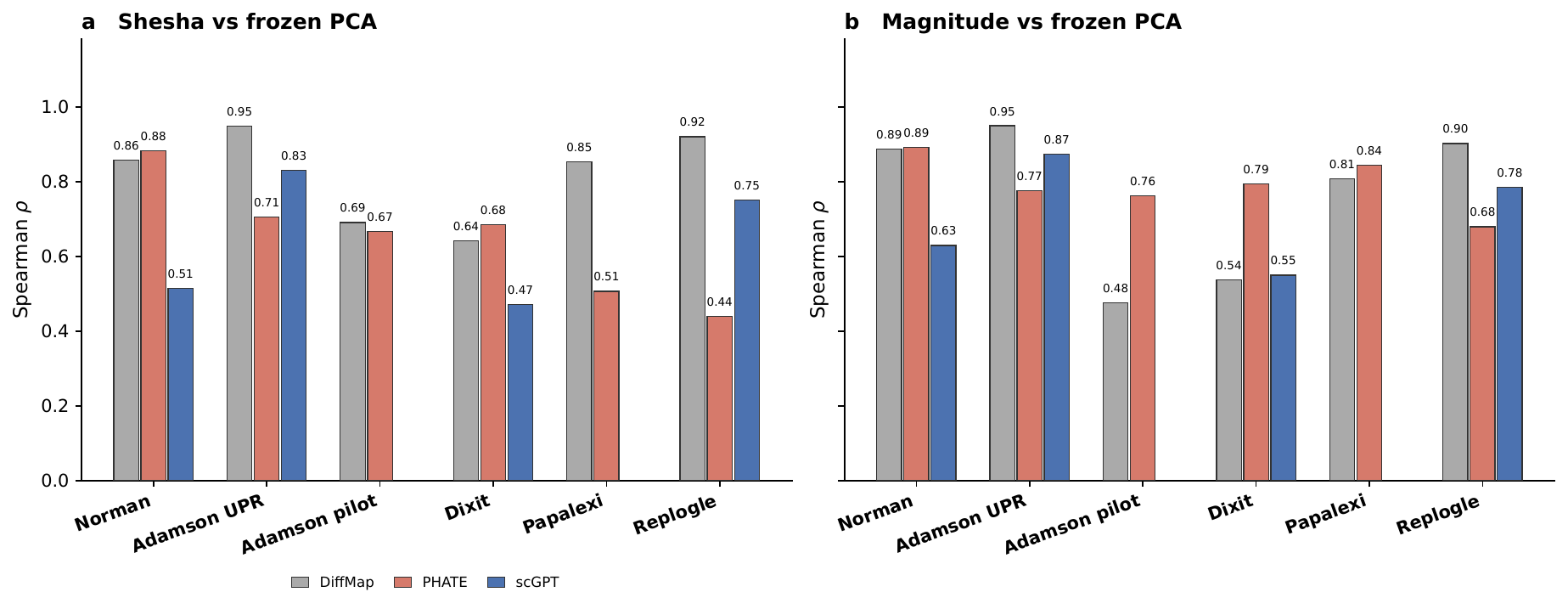}
\caption{\textbf{Coherence rankings transfer less reliably than effect-magnitude rankings when the representation changes.}
Spearman $\rho$ of each alternative embedding against the frozen linear embedding.
(a)~Shesha\textsubscript{P} concordance. (b)~Effect-magnitude concordance.
Arms: diffusion maps, PHATE, and scGPT. Diffusion maps are the only representation in which
coherence matches or exceeds magnitude in four of six datasets; under PHATE and scGPT,
magnitude transfers better in every dataset tested. scGPT is omitted for Papalexi ($n=24$)
and the Adamson pilot ($n=8$). Values are in
\textbf{SI Appendix, Table~\ref{tab:embedding_concordance}};
PHATE MDS convergence is in
\textbf{SI Appendix, Table~\ref{tab:phate_convergence}}.}
\label{fig:embedding_comparison}
\end{figure*}

Coherence is defined relative to a representation: the mean response direction, and each cell's angle to it, are properties of the space in which displacement vectors are computed. We therefore asked how far perturbation rankings are preserved when that space changes, using three alternatives to the frozen linear embedding: diffusion maps, PHATE, and the scGPT foundation-model embedding~\citep{Cui2024}. To separate a property of coherence from a property of perturbation geometry generally, we computed the same concordance for effect magnitude in each representation (\textbf{SI Appendix, Table~\ref{tab:embedding_concordance}}). Papalexi ($n=24$) and the Adamson pilot ($n=8$) are omitted from the scGPT arm: coherence ranks are not reproducible under GPU inference at those sample sizes (\textbf{SI Appendix, Fig.~\ref{fig:seed_reproducibility}}).

Concordance depends on the representation for both quantities, and coherence is the more sensitive of the two. Under diffusion maps, coherence transfers well (Spearman $\rho = 0.641$ to $0.948$ against frozen values) and matches or exceeds effect magnitude in four of six datasets. Under PHATE and scGPT, magnitude transfers better in every dataset tested. Across all sixteen dataset--representation comparisons, magnitude transfers at least as well as coherence in twelve.

The PHATE result is the most informative. Effect magnitude is largely preserved there ($\rho = 0.679$ to $0.891$) while coherence is not, falling to $0.439$ in Replogle and $0.506$ in Papalexi. PHATE applies a potential-distance transform followed by metric embedding, a construction designed to linearize continuous trajectories. An embedding that preserves how far cells have moved while reorganizing which way they moved is precisely the case in which a displacement-direction statistic should fail to transfer, and that is what is observed. The pattern is not attributable to optimizer failure: MDS convergence does not predict concordance, with the three datasets whose embeddings converged spanning $\rho = 0.51$ to $0.68$ and the three that did not spanning $0.44$ to $0.88$ (\textbf{SI Appendix, Table~\ref{tab:phate_convergence}}). Under scGPT the two quantities degrade together, with gaps of $0.03$ to $0.12$ (Norman $0.514$ versus $0.629$; Replogle $0.750$ versus $0.784$; \textbf{SI Appendix, Fig.~\ref{fig:scgpt_concordance}}), which points to the learned embedding not preserving displacement geometry for either statistic rather than to anything specific to coherence. That cross-representation question is separate from whether coherence and magnitude remain coupled \emph{inside} scGPT, which they do ($\rho = 0.851$ to $0.935$; \textbf{Fig.~\ref{fig:magnitude_coherence}h--k}; \textbf{SI Appendix, Table~\ref{tab:embedding_magnitude}}).

Two practical consequences follow. Coherence values should not be compared across representations, and in particular not across trajectory-linearizing embeddings, without establishing concordance first. And where a study reports coherence in a learned embedding, the effect-magnitude concordance should be reported alongside it, since without that comparator a moderate value cannot be attributed to the statistic rather than to the representation.

%% file: stress-new.tex
\begin{figure}[h]
    \centering
        \includegraphics[width=\linewidth]{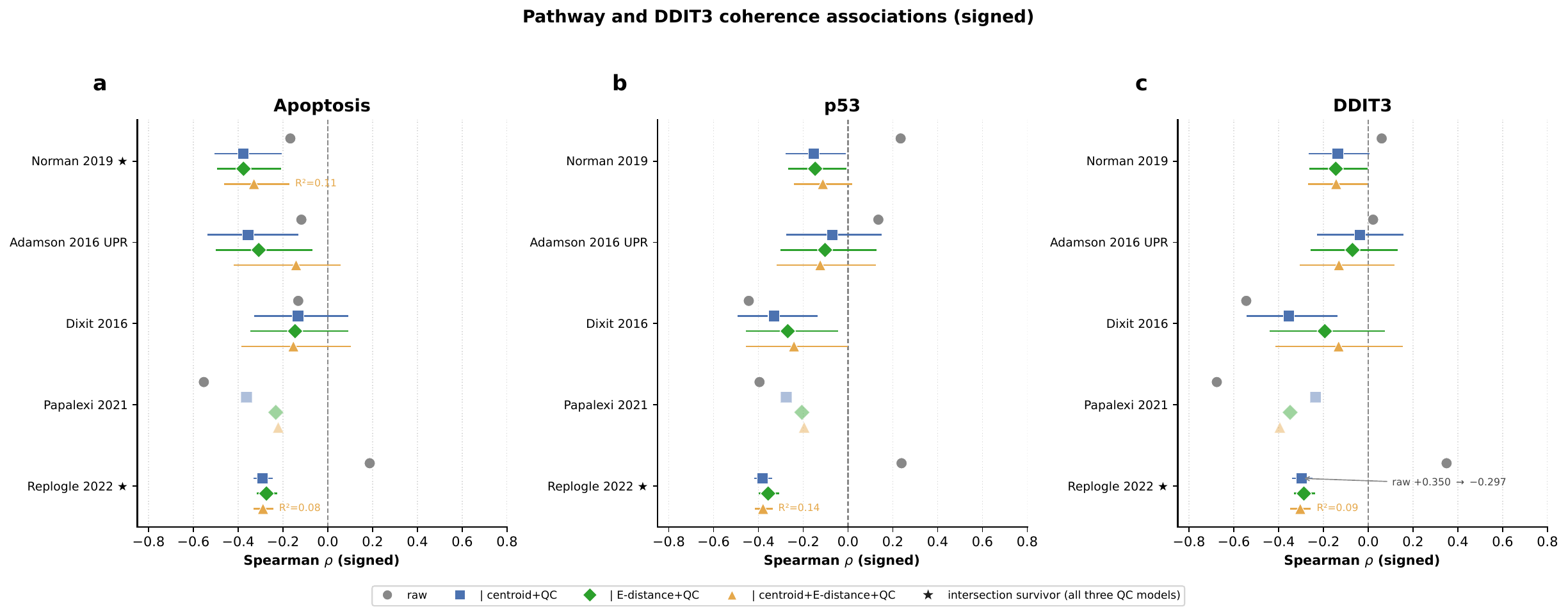}
\caption{\textbf{Associations surviving conditioning on effect size and cell quality.}
Signed Spearman $\rho$ between Shesha\textsubscript{P} and (a)~Hallmark Apoptosis, (b)~Hallmark p53 Pathway, and
(c)~DDIT3 expression. Grey, unconditioned. Blue, green and gold, rank partials conditioning on
centroid displacement, energy distance, or both, each with per-perturbation cell-quality
covariates. Stars mark associations surviving all three models; partial $R^2$ is annotated on
those rows under the joint model. Papalexi ($n=24$) is point-estimate only. Most surviving
associations reverse sign under conditioning, as expected when covariates explain $89$--$96\%$ of
coherence variance; Replogle DDIT3 is annotated as an example. Rows without stars include cases
significant in some models but not all, such as Norman p53 and Dixit DDIT3.
Values are in \textbf{SI Appendix, Tables~\ref{tab:pathway_stress} and
\ref{tab:stress_markers}}.}
\label{fig:pathway_forest}
\end{figure}

Coherence and effect magnitude are closely coupled, but the coupling is not complete: energy distance leaves four to eleven percent of coherence variance unexplained (\textbf{SI Appendix, Table~\ref{tab:edistance_competitor}}). We asked whether that residual carries biological signal, using composite scores for four MSigDB Hallmark gene sets (Apoptosis, p53 Pathway, Unfolded Protein Response, Reactive Oxygen Species) and four canonical stress markers.

Two considerations shape how this was tested. Residual coherence tracks cell quality: in a matched analysis, perturbations in the lowest and highest residual-coherence quartiles could be balanced on effect magnitude but not on detected genes or UMI counts (\textbf{Fig.~\ref{fig:approach_b_balance}}; \textbf{SI Appendix, Table~\ref{tab:approach_b_balance}}). All partial correlations reported here therefore condition on per-perturbation mitochondrial fraction, detected genes, and total counts alongside effect size. Because these covariates are themselves partly downstream of apoptotic biology, conditioning on them removes real signal along with technical artifact, and the resulting estimates are lower bounds. Coherence is also coupled to effect size under more than one definition, so each association was tested against centroid displacement, energy distance, and both jointly. We report the intersection across all three models, never the union (\textbf{Fig.~\ref{fig:pathway_forest}}; \textbf{SI Appendix, Tables~\ref{tab:stress_markers} and \ref{tab:pathway_stress}}).

After conditioning, Apoptosis and p53 are negative in all five scoreable datasets. Apoptosis survives all three models in Norman ($\rho = -0.329$ under the joint model, partial $R^2 = 0.11$) and Replogle ($-0.289$, $R^2 = 0.08$); p53 survives in Replogle ($-0.379$, $R^2 = 0.14$). Among individual markers, DDIT3 is negative in all fifteen dataset--model cells and survives in Replogle ($-0.303$, $R^2 = 0.09$). The two pathways are not redundant: their Hallmark gene sets share few members (Jaccard $= 0.097$) and their scores correlate weakly in every dataset contributing to the gate ($\rho = 0.145$ to $0.423$; \textbf{SI Appendix, Table~\ref{tab:apoptosis_p53_independence}}).

Three features constrain what these associations support. Most reverse sign under conditioning: in Replogle, the unconditioned correlations are $+0.187$ for Apoptosis, $+0.239$ for p53 and $+0.350$ for DDIT3, all becoming negative once effect size is controlled (\textbf{Fig.~\ref{fig:pathway_forest}}). This is expected when covariates explain $89$ to $96$ percent of coherence variance, but it means these are properties of the residual rather than of coherence as measured. Five of the six intersection survivors are in Replogle ($n = 1{,}832$); apoptosis replicates in Norman ($n = 236$) and no association survives the intersection in the three smaller datasets. This is therefore substantially a statement about what is detectable at that sample size, with the remaining datasets underpowered rather than negative. And the effects are small in absolute terms, operating on a residual of four to eleven percent of coherence variance.

The associations are also conditional on the family of effect-size measure. Conditioning instead on transcriptome-wide impact, estimated from differential expression rather than from displacement geometry, gives partial correlations of $+0.005$ for both Apoptosis ($[-0.040, 0.051]$) and p53 ($[-0.034, 0.045]$) in Replogle. The obvious explanation, that this measure is entangled with the pathway scores because both derive from expression, does not hold: its correlations with the two pathway scores ($0.203$ and $0.260$) are lower than those of centroid displacement ($0.254$ and $0.327$), and it is less correlated with coherence than centroid displacement is ($0.912$ versus $0.977$; \textbf{SI Appendix, Tables~\ref{tab:trade_competitor} and \ref{tab:twi_pathway}}). It should be the more conservative covariate and it produces nothing. The associations therefore survive conditioning on two geometric effect-size measures and on cell quality, and do not survive conditioning on a differential-expression-derived measure of transcriptome-wide impact.

Among the remaining markers, ATF4 illustrates why individual genes are unreliable here. It survives in Replogle at $\rho = -0.106$ to $-0.111$, marginal against the $|\rho| > 0.1$ threshold, while in Adamson UPR it survives with the opposite sign ($+0.340$ and $+0.315$ under two of three models). Both are CRISPRi screens in K562, so modality and cell type do not account for the reversal; screen composition is the more likely explanation, since Adamson UPR is a targeted 87-perturbation screen and Replogle is genome-scale. XBP1 survives in Replogle only, with positive point estimates in two other datasets. HSPA5 survives in no dataset under any QC-conditioned model (\textbf{SI Appendix, Table~\ref{tab:stress_markers}}). We do not treat any of them as a marker of residual coherence.

%% file: effect-size.tex
\begin{table}[h]
    \centering
    \caption{Raw associations with Shesha\textsubscript{P} and the same associations after conditioning on effect size. Combinatorial and between-guide rows are rank-partial Spearman given centroid magnitude;
    Apoptosis, p53, and DDIT3 are Replogle centroid+QC (magnitude plus mitochondrial fraction and
    gene-count covariates), not the joint model in Fig.~\ref{fig:pathway_forest}.
    In every case the unconditioned association is dominated by effect size: two vanish under conditioning and three reverse sign and persist on the residual.
    $^{a}$131 combinatorial and 105 single-gene perturbations. Combinatorial perturbations have
    substantially larger effects (magnitude SMD $=+0.924$, IQR overlap $0.139$), and coherence tracks
    magnitude at $\rho=0.950$ in this dataset, so the raw difference is what the effect-size gap alone
    predicts. Cell count is not the driver (SMD $=-0.37$; single-gene perturbations are uniformly
    $n=100$ cells). $^{b}$The multi-guide subset of the $1{,}832$ scored perturbations
    (\textbf{SI Appendix, Table~\ref{tab:between_guide}}).}
    \label{tab:raw_vs_conditioned}
    \smallskip
    \begin{tabular}{lccccc}
    \hline
    Comparison & Dataset & $n$ & raw $\rho$ & $\rho \mid$ effect size & 95\% CI \\
    \hline
    Combinatorial vs single-gene$^{a}$ & Norman & 131 / 105 & $+0.424$ & $-0.048$ & $[-0.176, 0.091]$ \\
    Between-guide agreement$^{b}$ & Replogle & 104 & $+0.332$ & $+0.006$ & $[-0.220, 0.225]$ \\
    Apoptosis & Replogle & 1{,}832 & $+0.187$ & $-0.291$ & $[-0.329, -0.248]$ \\
    p53 & Replogle & 1{,}832 & $+0.239$ & $-0.382$ & $[-0.415, -0.343]$ \\
    DDIT3 & Replogle & 1{,}832 & $+0.350$ & $-0.297$ & $[-0.337, -0.253]$ \\
    \hline
    \end{tabular}
\end{table}

The preceding sections each conditioned on effect size before reporting an association. Collecting those comparisons alongside two others makes the pattern visible: unconditioned correlations with coherence are substantially or entirely accounted for by how large the perturbation's effect was (\textbf{Table~\ref{tab:raw_vs_conditioned}}).

Two of the five vanish under conditioning. In Norman, combinatorial perturbations appear far more coherent than single-gene perturbations ($\rho = +0.424$; $n = 131$ versus $105$), a difference that does not survive controlling for effect magnitude ($-0.048$, $[-0.176, 0.091]$; adding cell-quality covariates and cell count gives $-0.030$, $[-0.162, 0.113]$). The mechanism is transparent: combinatorial perturbations have substantially larger effects (magnitude standardized mean difference $+0.924$, with interquartile ranges overlapping by only $0.139$), and coherence tracks magnitude at $\rho = 0.950$ in this dataset (\textbf{Table~\ref{tab:six_datasets}}), so the observed difference is what the effect-size gap alone predicts. Cell count does not account for it (standardized mean difference $-0.37$; single-gene perturbations are uniformly $n = 100$ cells). Norman also selected pairs expected to interact, which further limits generalization.

In Replogle the agreement between mean response directions estimated from independent guides targeting the same gene correlates with coherence at $\rho = +0.332$, and at $+0.006$ ($[-0.220, 0.225]$) once effect magnitude is controlled ($n = 104$ multi-guide genes of $1{,}832$ scored perturbations; \textbf{SI Appendix, Table~\ref{tab:between_guide}}). The same partial is null in Papalexi ($+0.183$, $[-0.317, 0.644]$; $n = 23$). Coherence carries no information about whether independent reagents produce agreeing responses beyond what effect size already provides.

The remaining three reverse sign and persist. Apoptosis, p53 and DDIT3 all correlate positively with coherence in Replogle before conditioning ($+0.187$, $+0.239$, $+0.350$) and negatively afterwards under the centroid-plus-QC model used in this table ($-0.291$, $-0.382$, $-0.297$; \textbf{Fig.~\ref{fig:pathway_forest}}; \textbf{SI Appendix, Tables~\ref{tab:stress_markers} and \ref{tab:pathway_stress}}). Those three values are the centroid-plus-QC partials, not the joint-model intersection quoted in the pathway section ($-0.289$, $-0.379$, $-0.303$); the sign and the conclusion are the same. Reversal of this kind is expected when the covariate explains most of the variance in the outcome, but it means these associations are properties of the residual rather than relationships a reader would find by correlating coherence with pathway activity directly.

Directional coherence in Perturb-seq data is largely a restatement of effect magnitude. Associations reported without conditioning on effect size will mostly recover that coupling rather than anything about response geometry. This holds whether the comparison is between perturbation classes, between independent reagents, or with pathway activity, and it is the reason every association in this paper is reported both raw and conditioned.

%% file: limits-of-measurement.tex
\begin{figure}[h]
    \centering
        \includegraphics[width=\linewidth]{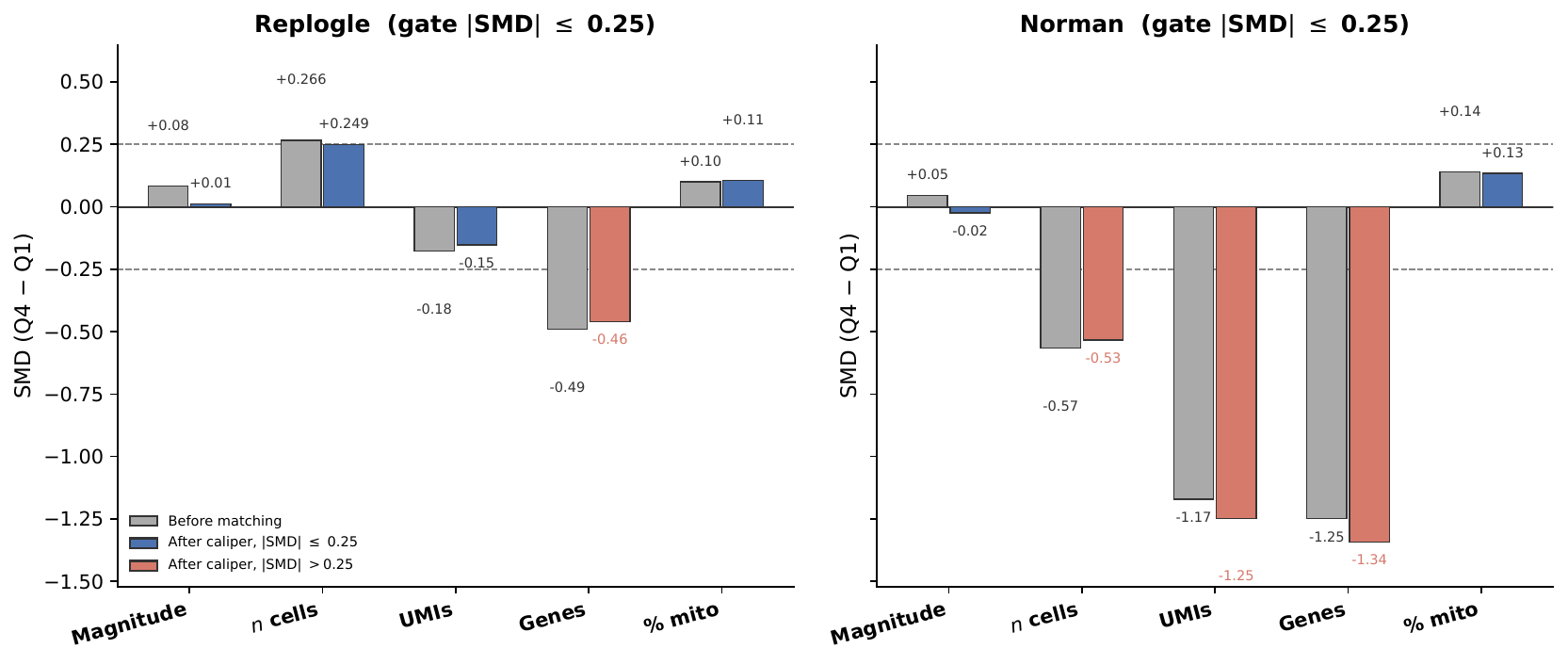}
\caption{\textbf{Covariate balance in the discordance-quartile comparison, before and after caliper
matching.} Standardized mean differences between the lowest and highest residual-coherence
quartiles. Matching balances effect magnitude ($|\mathrm{SMD}| \le 0.25$, blue) but not cell
quality: detected genes remain imbalanced in both datasets and UMI counts in Norman (red). The
pre-specified balance criterion was not met, so the enrichment analysis is not reported. The
residual association with cell quality is the reason primary partials condition on QC covariates.
Full SMDs are in \textbf{SI Appendix, Table~\ref{tab:approach_b_balance}}.}
\label{fig:approach_b_balance}
\end{figure}

Several analyses were run to completion and did not yield usable results. We report them here because they bound what directional coherence can resolve, and because two of them constrain how the primary estimates should be read.

\subsubsection{Unbiased enrichment could not be balanced on cell quality}

To identify biology associated with residual coherence without pre-specifying gene sets, we compared the lowest and highest residual-coherence quartiles by differential expression and enrichment, matching perturbations on effect magnitude with a caliper and requiring covariate balance before proceeding. Matching balanced effect magnitude in both datasets (standardized mean difference $+0.013$ in Replogle, $-0.024$ in Norman) but the pre-specified criterion failed on cell quality: detected genes remained imbalanced in both ($-0.46$ and $-1.34$) and UMI counts in Norman ($-1.25$; \textbf{Fig.~\ref{fig:approach_b_balance}}; \textbf{SI Appendix, Table~\ref{tab:approach_b_balance}}). The enrichment analysis is therefore not reported. Residual coherence remains associated with cell quality after effect magnitude is matched away, which is why the pathway partials reported above condition on quality covariates rather than on effect size alone.

\subsubsection{A controlled simulation could not be calibrated to the observed range}

We implemented a perturbation simulator to test whether coherence can distinguish responses that are heterogeneous because of variable perturbation efficiency from responses that are heterogeneous because cells engage different programs, a distinction the real data cannot separate. Across 120 simulated datasets and five seeds, the effect scale required to reproduce observed coherence proved dependent on which reference perturbation the pipeline selected, varying roughly fourfold at fixed settings when the reference changed. The generator also could not hold the magnitude of multi-program responses fixed while varying the number of programs, so the identifiability question it was built to answer was not testable as constructed. The arm is reported as attempted and not used (\textbf{SI Appendix, Table~\ref{tab:synthetic_calibration}}).

\subsubsection{The positive control does not clear conditioning} 

Perturbations targeting core unfolded protein response machinery in the Adamson UPR screen show lower coherence than the remainder (median $0.158$ versus $0.268$, $p = 0.013$), which is the expected direction for a perturbation class that should produce heterogeneous responses. Under the same survival criterion applied to the pathway analyses, the magnitude-conditioned partial is indeterminate ($\rho = -0.274$, $[-0.499, -0.008]$), with the nearer interval bound inside the knife-edge margin $\varepsilon = 0.02$. We were also unable to recover the gene set used in an earlier version of this analysis, so the previously reported value could not be reproduced. The control establishes a raw effect in the expected direction and does not establish one independent of effect size.

\subsubsection{Interval estimation fails at small sample size} 

In Papalexi ($n = 24$), the quality-covariate models are rank-deficient in bootstrap resamples and confidence intervals are not estimable (\textbf{SI Appendix, Tables~\ref{tab:stress_markers} and \ref{tab:pathway_stress}}). Point estimates for that dataset are reported descriptively throughout and are excluded from every survival gate. The Adamson pilot ($n = 8$) is below the same $n < 30$ floor and is shown only for completeness.

\subsubsection{Coherence rankings are not numerically reproducible at small $n$ under GPU inference} 

Coherence is a rank statistic over closely spaced cosine values, so small floating-point differences reorder perturbations. Repeated scGPT inference on identical inputs changes concordance in the fifth decimal at $n = 1{,}832$ and in the first decimal at $n = 24$, where three runs on a fixed input gave $0.575$, $0.633$ and $0.630$ (\textbf{SI Appendix, Fig.~\ref{fig:seed_reproducibility} and Tables~\ref{tab:repro_verification} and~\ref{tab:seed_reproducibility}}). Effect-magnitude concordance is exactly reproducible in both cases ($\rho = 0.753913$ on Papalexi). Deterministic algorithm settings reduced but did not eliminate the spread. Papalexi is therefore excluded from the scGPT concordance comparison, and coherence rank statistics computed on small perturbation sets under GPU inference should be treated as approximate.

%% file: discussion-new.tex
\subsection{Coherence and effect magnitude}

Directional coherence and effect magnitude are closely coupled in single-cell CRISPR screens. Across six Perturb-seq datasets spanning CRISPRa, CRISPRi and Cas9 knockout in three cell types, Spearman correlations run from $0.843$ to $0.977$ (\textbf{Fig.~\ref{fig:magnitude_coherence}}; \textbf{Table~\ref{tab:six_datasets}}). The coupling is not an artifact of how either quantity is computed. It remains when both are recomputed in a transformer embedding~\citep{Cui2024}, when effect size is an energy distance that also penalizes within-group dispersion~\citep{szekely_2013_energy,peidli_2024_scperturb} (\textbf{Fig.~\ref{fig:edistance_competitor}}), and when effect size comes from gene-level differential expression rather than from displacement geometry~\citep{Nadig2024}. Removing cells classified as non-responding does not remove it either~\citep{Song2025} (\textbf{Fig.~\ref{fig:efficiency_filter}}).

Unconditioned associations with coherence will therefore mostly recover effect size (\textbf{Table~\ref{tab:raw_vs_conditioned}}). Combinatorial perturbations look more coherent than single-gene perturbations until magnitude is controlled; so does agreement between independent guides targeting the same gene. Three associations with stress-pathway activity reverse sign under the same step. A claim that one class of perturbation is more coherent than another, without that control, is usually a claim that one class has larger effects.

What is left after conditioning is small. Effect size accounts for $89$ to $96$ percent of coherence variance in the datasets used for the pathway analyses, a residual of four to eleven percent (\textbf{Fig.~\ref{fig:edistance_competitor}}). Inside that residual, coherence is negatively associated with apoptosis and p53 signalling and with DDIT3, an effector of the integrated stress response~\citep{costamattioli_2020_isr} (\textbf{Fig.~\ref{fig:pathway_forest}}). Those associations survive two geometric effect-size measures and per-perturbation cell quality, but they are concentrated in the largest screen, they are small, and they disappear when the covariate is transcriptome-wide impact. The residual is not empty. It is also not a reason to treat coherence as a substitute for effect size, or as a useful complement to it, in routine screen analysis.

\subsection{Why the coupling is not a tautology}

A close relationship between coherence and magnitude can look guaranteed. Coherence is the mean cosine of individual displacement vectors to their own mean---a mean resultant length in the sense of directional statistics~\citep{mardia_2000_directional}---and the length of that mean \emph{is} the effect magnitude. Vectors that agree in direction produce a longer mean than vectors that cancel. On that reading the correlation is arithmetic.

It is not forced. For a fixed set of displacement lengths the mean shift is longest when the vectors align and shortest when they cancel, so high coherence bounds magnitude from below. That bound runs only one way, and it says nothing about the joint distribution real perturbations occupy. Cells that all move a short distance in nearly the same direction give high coherence and low magnitude. Cells that move far in scattered directions give low coherence and high magnitude. Both are geometrically admissible, and both would be easy to interpret: a small, uniformly executed response, or a strong perturbation that different cells resolve differently through non-genetic heterogeneity or variable penetrance~\citep{symmons_2016_luck}. The question is how often those configurations occur. In Perturb-seq they mostly do not; the data sit on the diagonal.

Two observations keep this empirical. The coupling is not equally tight everywhere: it is weakest in Dixit ($\rho = 0.843$) and strongest in Replogle ($0.977$), and the gap narrows when effect size is energy distance ($0.945$ and $0.980$), so some of the between-dataset variation is within-group dispersion rather than the geometry of the mean shift (\textbf{SI Appendix, Table~\ref{tab:edistance_competitor}}). And the residual is not noise. If coherence were a deterministic function of magnitude, the leftover four to eleven percent would be estimation error and would carry no structure. It carries associations with apoptosis and p53 signalling, in the same direction, in the two largest screens.

A quantity that could in principle vary independently of effect size largely does not, in this data. That is worth recording because the alternative looked plausible enough to motivate the measurement.

\subsection{Relationship to existing measures}

\cite{Nadig2024}'s TRADE framework estimates transcriptome-wide impact from gene-level differential expression, with a correction for error in log-fold-change estimates. It shares no construction with coherence, so it is the cleanest check we have on whether the coupling is definitional. It is not. Coherence tracks transcriptome-wide impact about as closely as centroid displacement does (median difference $-0.01$; Replogle $0.912$ versus $0.919$; Adamson UPR $0.954$ versus $0.968$; \textbf{SI Appendix, Table~\ref{tab:trade_competitor}}). Where transcriptome-wide impact diverges from displacement geometry, as in Dixit ($0.452$ and $0.513$), it diverges from both quantities together. The one dataset where coherence loses more than magnitude is Norman ($0.688$ versus $0.812$), the only CRISPRa screen. Breadth of effect across genes and agreement of direction across cells are different questions; in these data they are not independent of one another in the way that framing suggests.

\cite{Song2025}'s perturbation-response score asks how strongly each cell responded, and we use it to classify responders (\textbf{Fig.~\ref{fig:efficiency_filter}}). A narrower point is worth keeping: per-cell distance from the control centroid is not a lightweight stand-in for that score. In Replogle the Euclidean and whitened centroid distances correlate with the algorithm-derived score at $\rho = 0.097$ and $0.149$ (\textbf{SI Appendix, Table~\ref{tab:ps_proxy_caution}}). Any comparison of coherence to ``response strength'' that uses those proxies will recover the coherence--magnitude relationship, not a comparison to Song's score, and we do not report one.

The two responder classifiers we compared do not identify the same cells. On the $23{,}290$ Norman cells scored by both, they agree on $68.4$ percent of labels ($\kappa = 0.374$; \textbf{SI Appendix, Table~\ref{tab:responder_agreement}}). That disagreement has a counterpart in estimates of the unperturbed fraction in K562 Perturb-seq: Garbulowski et al.~\citep{Garbulowski2024} put it at roughly zero to $25$ percent from simulated and experimental $E$-test distributions, while our perturbation-score responder fractions imply $42$ to $64$ percent non-responders~\citep{papalexi2021characterizing}. The two ranges are not reconcilable. What counts as a responding cell is method-defined, and any analysis restricted to responders inherits that.

Between-reagent consistency has been used as a quality filter outside transcriptomic screens, reflecting the well-documented variability of CRISPR guide efficiency~\citep{doench_2016_optimized}. \cite{Seale2026} drop guides whose mutational-spectrum profiles correlate poorly with other guides on the same gene. The analogous question for coherence has a null answer once magnitude is in the model: in Replogle, gene-level coherence does not predict whether two guides agree on direction ($\rho = 0.006$, $[-0.220, 0.225]$). In Papalexi, guides targeting the same gene agree on coherence (ICC$(1) = 0.890$) and on effect magnitude to the same degree ($0.889$; \textbf{SI Appendix, Table~\ref{tab:between_guide}}). Consistency in coherence is consistency in effect size.

\subsection{The residual signal and its scale}

After conditioning on effect size and cell quality, residual coherence is negatively associated with Hallmark Apoptosis and p53 signalling and with DDIT3 (\textbf{Fig.~\ref{fig:pathway_forest}}; \textbf{SI Appendix, Tables~\ref{tab:stress_markers} and \ref{tab:pathway_stress}}). Point estimates are negative in all five scoreable datasets for both pathways. Three associations survive every geometric covariate model: apoptosis in Norman and Replogle, p53 in Replogle. The two pathways are not one signal counted twice: they share $11$ of $113$ union genes, and their scores correlate only weakly in the datasets that enter the gate (\textbf{SI Appendix, Table~\ref{tab:apoptosis_p53_independence}}). DDIT3 is negative in all fifteen dataset--model cells and survives only in Replogle.

The direction matches a simple mechanistic sketch. Perturbations that leave cells in configurations no coordinated program resolves might engage apoptotic and p53-dependent responses~\citep{vousden_2009_p53,tabas_2011_er_stress}, and coherence would then fall as those programs activate. The data are consistent with that sketch. They do not establish it.

Cas9-induced double-strand breaks activate a p53-mediated DNA-damage response~\citep{ihry_2018_p53,haapaniemi_2018_crispr}, so a p53 association could in principle be editing damage rather than perturbation response. Most of the surviving associations are not from nuclease screens. Both apoptosis survivors, and the Replogle p53 result, come from catalytically dead Cas9 (Norman CRISPRa; Replogle CRISPRi). Only Dixit p53 is from a Cas9 nuclease screen, and there the confound cannot be excluded.

Three limits follow. Effect size accounts for $89$ to $96$ percent of coherence variance, so the associations live on a residual of four to eleven percent, with partial $R^2$ of $0.08$ to $0.14$ inside that residual --- under one percent of the variation in coherence, in absolute terms. Five of the six intersection survivors are in Replogle ($n = 1{,}832$), roughly eight times the next-largest screen; apoptosis replicates in Norman ($n = 236$), and nothing survives the intersection in the three smaller datasets. That is a statement about what is detectable at genome-scale $n$, not a claim that the smaller screens contradict it.

The associations also depend on how effect size is defined. They survive centroid displacement, energy distance, and both together, and they do not survive transcriptome-wide impact: in Replogle the partials for apoptosis and p53 are both $+0.005$, with intervals that include zero. That is not because TRADE merely restates the pathway scores. Its correlations with those scores ($0.203$ and $0.260$) are lower than those of centroid displacement ($0.254$ and $0.327$), and it is less correlated with coherence than centroid displacement is ($0.912$ versus $0.977$; \textbf{SI Appendix, Tables~\ref{tab:trade_competitor} and \ref{tab:twi_pathway}}). It is a differently oriented covariate, not a more conservative one. Whether residual coherence carries stress-pathway signal therefore depends on which family of effect-size measure defines the residual.

Most of the survivors also reverse sign. In Replogle the unconditioned correlations are $+0.187$ for apoptosis, $+0.239$ for p53 and $+0.350$ for DDIT3, all negative once effect size is controlled. That is expected when the covariate explains most of the outcome and is itself correlated with the predictor. It also means these are properties of a residual, not relationships one would see by correlating coherence with pathway activity directly.

\subsection{Implications for perturbation heterogeneity metrics}

Single-cell CRISPR screens are increasingly summarized with statistics that go beyond mean effect size. Directional coherence is a natural one: it uses information that pseudobulk discards, it has a clear geometric definition, and it needs no extra experiment. Those properties are not enough. A statistic can be well defined, cheap to compute, and still mostly recover a quantity already in routine use.

Redundancy with effect size should be reported as a matter of course. We did not find a published perturbation-heterogeneity metric accompanied by its correlation with the corresponding effect magnitude; without that number a reader cannot tell whether a reported association belongs to the new statistic or to effect size measured another way. The same issue is familiar in expression variability, where cell-to-cell variation has to be separated from mean expression before it can be interpreted~\citep{eling_2019_challenges}. The correlation is cheap.

One effect-size measure is not enough to claim independence. Centroid displacement and energy distance agree closely here ($\rho = 0.950$ to $1.000$), and either alone would have supported the apoptosis and p53 associations; a differential-expression measure does not. A metric built from displacement geometry will tend to look independent of other geometric size measures. Testing against at least one measure constructed differently is what separates a shared-construction artifact from a finding.

Two of the analyses we ran---split-half agreement, and combinatorial versus single-gene coherence---give strong positive results that vanish under conditioning. Both are natural analyses, and both would have been easy to report without that step. Split-half agreement is circular by construction, a form of double dipping~\citep{kriegeskorte_2009_circular}: the halves come from the same cells and are compared to the same control reference, so a perturbation whose displacement vectors already align will produce agreeing half-samples regardless of reproducibility.

A more portable observation concerns efficiency filtering. Restricting to cells that responded to their assigned guide is common, and it depends on a classifier. On Norman the two standard callers agree on $68.4$ percent of cells ($\kappa = 0.374$), and they move the coherence--magnitude correlation in opposite directions. Estimates of the unperturbed fraction in K562 Perturb-seq span roughly $0$--$25$ percent in simulation-based work and $42$--$64$ percent under the classifier used here. Results reported under a single caller, without that sensitivity, leave a reader unable to tell how much of the result is the classifier.

\subsection{Context dependence}

Several associations change sign between datasets, and where they do is informative.

Two reversals cross experimental context in the direction a context-dependent biology would predict. XBP1 is positively associated with residual coherence in Norman and negatively in Replogle (CRISPRa versus CRISPRi in the same cell line). mTORC1 and ROS reverse between Norman and Dixit, which differ in both modality and cell type. Gene activation and gene interference engage different programs~\citep{Gilbert2013}, and directional dependencies between the two have been shown in paired activation and knockout screens~\citep{boettcher_2018_dual}. We treat these as observations, not as isolated tests of modality: neither comparison holds the rest of the experimental design fixed.

ATF4 is a different case. It survives with a positive coefficient in Adamson UPR ($+0.340$ and $+0.315$ under two of three models) and a negative one in Replogle ($-0.106$ to $-0.111$). Both screens are CRISPRi in K562, so modality and cell type cannot explain the reversal. Adamson UPR is an $87$-perturbation panel enriched for unfolded-protein-response machinery; Replogle is genome-scale. A marker's relationship to residual coherence inside a pathway-focused panel need not match its relationship across a genome-wide set. That is a limit on generalizing from targeted screens, not an instance of biological context dependence, and we do not treat ATF4 as a marker of residual coherence (\textbf{SI Appendix, Table~\ref{tab:stress_markers}}).

Dixit is an outlier in more than one analysis: the weakest coherence--magnitude correlation ($0.843$), the largest residual after centroid displacement ($0.289$, falling to $0.108$ under energy distance), the lowest concordance with diffusion-map and scGPT embeddings (\textbf{SI Appendix, Fig.~\ref{fig:scgpt_concordance} and Table~\ref{tab:embedding_concordance}}), and the only dataset whose perturbation rankings move under responder restriction (\textbf{Fig.~\ref{fig:efficiency_filter}}). It is also the only non-K562 dataset among the well-powered screens, and the earliest Perturb-seq experiment here. Its bone-marrow-derived dendritic cells were profiled under lipopolysaccharide stimulation, where paracrine signalling drives substantial cell-to-cell variation~\citep{shalek_2014_singlecell}. We cannot separate that biology from an early protocol, and we note the recurrence rather than attribute it.

Four of the six datasets are K562. The other two contribute $122$ of $2{,}285$ perturbations. Consistency across modalities is largely consistency within one cell line. Cancer cell lines differ in transcriptional state~\citep{klijn_2015_comprehensive} and diverge over passage and between laboratories~\citep{ben_david_2018_evolution}, so even the K562 result should not be assumed to transfer.

\subsection{Limitations}

The dataset base is narrow in a specific way. Generality across cell types rests on Dixit ($n = 98$, BMDCs) and Papalexi ($n = 24$, THP-1). The surviving pathway associations are concentrated in Replogle; at $n = 1{,}832$ an association explaining under one percent of coherence variance is detectable, and at $n = 98$ it is not. The paper does not show that those associations generalize beyond genome-scale screens.

Conditioning on cell quality is conservative in a direction we cannot quantify. Mitochondrial fraction and detected-gene count are standard quality metrics but are partly downstream of cell state~\citep{luecken_2019_bestpractices}, so regressing them out removes real signal with technical artifact, and the surviving estimates are lower bounds of unknown tightness. We condition on them because residual coherence remains associated with cell quality after magnitude is matched away (\textbf{Fig.~\ref{fig:approach_b_balance}}). The trade is between two biases.

In Papalexi ($n = 24$) the quality-covariate models are rank-deficient in bootstrap resamples, so that dataset contributes point estimates only and is excluded from every survival gate. Coherence rankings are also not numerically reproducible at that scale under GPU inference: three scGPT runs on identical inputs gave concordance $0.575$, $0.633$ and $0.630$, while effect-magnitude concordance was bit-identical (\textbf{SI Appendix, Fig.~\ref{fig:seed_reproducibility}}). Deterministic settings reduced the spread and did not remove it.

Coherence is a single scalar per perturbation. It does not capture subpopulation structure, bifurcating responses, or dose-dependent heterogeneity, and a bimodal mixture of strong responders and near-control cells can look like a unimodal response with the same mean angular spread. It is also defined relative to a representation, and rankings transfer imperfectly, least well under embeddings built to recover continuous trajectories~\citep{saelens_2019_trajectory}.

The associations are correlative. Showing that incoherent responses cause apoptotic or p53 signalling, rather than co-occur with it, would require manipulating coherence independently of magnitude---for example titrated knockdown with mismatched guides~\citep{jost_2020_titrating} or inducible depletion. We have not done that. Any such experiment would also need to specify in advance which effect-size measure it holds fixed.

\subsection{Recommendations}

The following are what we would tell a reader intending to compute directional coherence on their own data, or to evaluate a study that reports it.

\subsubsection{Report the correlation with effect size, and condition on it before reporting anything else} 
In these datasets coherence and effect magnitude correlate at $0.843$ to $0.977$, so an unconditioned association between coherence and any other quantity is likely to be an association with effect size. This applies to comparisons between perturbation classes as much as to correlations with expression: two of the five comparisons we ran produced strong positive results that vanished entirely under conditioning.

\subsubsection{Condition on cell quality as well} 
Residual coherence remains associated with mitochondrial fraction and detected-gene count after effect magnitude is balanced, so associations conditioned on effect size alone will partly reflect library quality. Conditioning on both is conservative, since these covariates are themselves partly downstream of cell state \citep{heumos_2023_bestpractices}, but the alternative is a confound rather than a clean estimate.

\subsubsection{Test against an effect-size measure built differently} 
Centroid displacement and energy distance are both computed from the same displacement vectors as coherence, and they agree with each other closely here. A measure derived from differential expression rather than geometry gave a different answer for the pathway associations. Where a result depends on the effect-size family, that dependence should be reported rather than resolved by choosing the more favorable measure.

\subsubsection{Do not use split-half agreement as a reproducibility endpoint}
Both halves are drawn from the same cells and compared against the same control reference, so tightly aligned displacement vectors produce agreeing half-samples by construction. Independent reagents in independent cells are the minimum for a non-circular test, and even that shares the control reference.

\subsubsection{Treat responder filtering as a sensitivity analysis, not a correction} 
Two standard classifiers agree on $68$ percent of individual cells ($\kappa = 0.374$), and the direction of the effect on the coherence--magnitude correlation differs between them. Report results with all assigned cells as primary, with responder-restricted values alongside, and state which classifier was used.

\subsubsection{Do not compare coherence values across representations without checking concordance} 
Perturbation rankings transfer imperfectly between embeddings, and least well under embeddings built to linearize continuous trajectories, where concordance fell to $0.44$ while effect-magnitude concordance remained at $0.68$. This is a specific instance of a general problem: low-dimensional embeddings distort the high-dimensional pairwise relationships they are derived from \citep{chari_2021_specious}, and a directional statistic is more exposed to that distortion than a distance is. Where coherence is computed in a learned or nonlinear embedding, the corresponding effect-magnitude concordance should be reported alongside it, since without that comparator a moderate value cannot be attributed to the statistic rather than to the representation.

\subsubsection{Interpret coherence rankings cautiously below roughly $n = 50$ perturbations} 
Confidence intervals for conditioned estimates were not estimable at $n = 24$, and coherence rankings at that scale were not numerically reproducible across repeated GPU inference on identical inputs. Effect-magnitude rankings were stable in the same comparisons.

\subsubsection{Implementation} 
Shesha\textsubscript{P} is implemented in the open-source \href{https://pypi.org/project/shesha-geometry/}{\texttt{shesha-geometry}} package~\citep{shesha2026}, which computes coherence and the effect-size measures used here from an AnnData object~\citep{Virshup2024} carrying perturbation labels and a control condition, together with the rank-based partial correlation and bootstrap routines these recommendations call for. The frozen analysis configuration, the per-dataset output tables, and the code reproducing every figure are available at [URL]; running the checks above on a new dataset requires no additional implementation.

%% file: code.tex
All analyses reported in this paper, including figure generation code,  are available at \url{https://github.com/prashantcraju/shesha-crispr} and archived on Zenodo \url{https://doi.org/10.5281/zenodo.22116628} \citep{github}. All five datasets are publicly available through  pertpy~\citep{pertpy} and require no login or registration. The Shesha perturbation coherence metric is implemented in the  open-source Python package \texttt{shesha-geometry}, freely available  on PyPI (\url{https://pypi.org/project/shesha-geometry}; \cite{shesha2026}) under an MIT license with accompanying tutorials.

%% file: acknowledgements-new.tex
The author wishes to thank Padma K. and Annapoorna Raju for generously supporting the computational resources used in this work. The author extends gratitude to the many institutions and individuals whose open-source datasets, frameworks, and models were used in our work.

%% file: si-new.tex

\section{Extended Methods}

\subsection{Datasets and preprocessing}

Six single-cell CRISPR perturbation datasets were accessed via the pertpy Python package (version 1.0.4)~\citep{pertpy}. Table~\ref{tab:datasets} summarizes the datasets. Modalities are CRISPRa (Norman), CRISPRi (Adamson UPR, Adamson pilot, Replogle), and CRISPR-KO (Dixit, Papalexi). The frozen Sp table comprises $n = 2{,}285$ perturbations; pathway and stress partials use five scoreable datasets ($n = 2{,}277$) after excluding the Adamson pilot ($n = 8$). Papalexi was additionally rebuilt from GEO (GSE153056) as an independent check that scores are not pertpy-loader--dependent. Each dataset was preprocessed independently to prevent batch effects. All manuscript-facing tables share one frozen preprocess and Sp-scoring path. The pipeline for each dataset consisted of:

\begin{enumerate}
\item Quality filtering: cells with fewer than 100 detected genes were removed.
\item Library-size normalization and log1p~\citep{wolf2018scanpy} were applied only when \texttt{.X} contained raw counts. Matrix scale was pinned per dataset from inspection of \texttt{.X}: Norman and Dixit are already log-transformed (normalize+log1p skipped); Adamson UPR/pilot, Papalexi, and Replogle are raw counts and were normalized to the median library size (\texttt{target\_sum=None}) before \texttt{log1p}.
\item Frozen linear embedding: Adamson UPR/pilot and Papalexi used \texttt{scanpy.pp.highly\_variable\_genes(n\_top\_genes=\\2000, subset=True)} followed by \texttt{scanpy.pp.pca(n\_comps\\=50)}. Norman, Dixit, and Replogle used variance/mean-ranked genes (all retained cells for Dixit; a seed-320 sample of 20,000 cells for Norman and Replogle) followed by \texttt{TruncatedSVD(n\_components=50, random\_state=320, \\n\_iter=5)}.
\item Cell materialization: perturbations with fewer than 50 cells excluded; large groups downsampled in an order-invariant manner by hashing \texttt{seed|obs\_name} (at most 100 cells per perturbation and 5{,}000 controls; seed 320).
\end{enumerate}

All downstream coherence and magnitude computations were performed on the 50-dimensional frozen linear embedding.

\subsection{Control group identification}

Control group assignment used a multi-stage matching protocol to accommodate heterogeneous labeling conventions across datasets:

\begin{enumerate}
\item \textbf{Exact match} (case-insensitive): labels matching ``control'', ``ctrl'', ``non-targeting'', ``NT'', ``unperturbed''.
\item \textbf{Delimiter-aware regex}: for short tokens (e.g., ``NT'' in ``NTg1''), split on common delimiters and match components.
\item \textbf{Substring matching}: for longer keywords embedded in complex labels.
\end{enumerate}

Dataset-specific handling:
\begin{itemize}
\item \textbf{Replogle 2022:} Labels containing ``non-targeting'' or beginning with ``chr'' were assigned to control. Labels containing ``pos\_control'' were removed. Gene names were extracted by splitting on underscore and taking the first token.
\item \textbf{Papalexi 2021:} The \texttt{gene\_target} column was copied from MuData global metadata. Non-targeting guides (NTg1 through NTg7) were pooled into a single NT control group (2,386 cells). Gene-level Sp from the GEO rebuild matches the frozen gene-grain table (Spearman $\rho \approx 1$ on $n = 24$ shared perturbations).
\item \textbf{Norman 2019:} The \texttt{perturbation\_name} column was used directly. Cells labeled ``control'' served as the control group.
\item \textbf{Adamson 2016 UPR:} Guide labels were aggregated to gene before scoring; Gal4 / mod non-targeting labels served as control.
\end{itemize}

\subsection{Perturbation coherence and effect magnitude}

For a perturbation $p$ applied to $n_p$ cells, let $\mathbf{x}_i \in \mathbb{R}^{50}$ denote the frozen linear-embedding coordinates of perturbed cell $i$ and $\boldsymbol{\mu}_{\text{ctrl}}$ the centroid of all control cells in the same dataset. The shift vector for cell $i$ is:
\begin{equation}
\mathbf{d}_i = \mathbf{x}_i - \boldsymbol{\mu}_{\text{ctrl}}
\end{equation}
The mean perturbation direction is:
\begin{equation}
\bar{\mathbf{d}} = \frac{1}{n_p} \sum_{i=1}^{n_p} \mathbf{d}_i
\end{equation}
Shesha perturbation coherence ($S_p$) is defined as:
\begin{equation}
S_p = \frac{1}{n_p} \sum_{i=1}^{n_p}
\frac{\mathbf{d}_i \cdot \bar{\mathbf{d}}}
{\|\mathbf{d}_i\| \, \|\bar{\mathbf{d}}\|}
\end{equation}
Effect magnitude is:
\begin{equation}
M_p = \|\bar{\mathbf{d}}\|
\end{equation}

The perturbation coherence metric adapts the principle of geometric self-consistency from the general Shesha framework~\citep{raju2026geometric, raju2026canary, shesha2026}. In that framework, Shesha-FS measures split-half RDM consistency across representations. Here, $S_p$ measures perturbation coherence directly within a single representation, specializing the principle to perturbation biology.

\subsection{E-distance competitor}

E-distance~\citep{Peidli2022} is scored on the same hash-selected cells and 50-PC embedding as $S_p$, using the full frozen control set (at most 5{,}000 cells). The implementation is the pertpy Euclidean $U$-statistic with off-diagonal within-group means; for every dataset it is checked against an independent sklearn pairwise-distance reference. The signed estimate is retained and is never labelled magnitude. Residual $S_p$ variance is $1-R^2$ from rank-OLS of rank($S_p$) on rank(size). Method id: \texttt{energy\_distance\_u\_statistic\_euclidean}. Per-dataset correlations are in \textbf{Table~\ref{tab:edistance_competitor}}.

\subsection{TRADE transcriptome-wide impact}

TWI~\citep{Nadig2024} is $\mathrm{Var}(\beta)$ from an ashr half-uniform mixture~\citep{Stephens2016} fit to Welch $\log_2$FC and SE on the frozen log-normalized transcriptome (same cells as $S_p$; no HVG subset, no PCA). scanpy \texttt{log1p} is converted from natural log to $\log_2$ by $/ \ln 2$. Genes with non-finite SE, $|\log_2\mathrm{FC}| > 10$, or matching the perturbed target are dropped. TWI values were computed via a Python implementation of the ashr half-uniform mixture variance (\texttt{mixcompdist='halfuniform'}, \texttt{prior='uniform'}), numerically consistent with R \texttt{ashr}~\citep{Stephens2016}. DE method id: \\\texttt{welch\_log2fc\_on\_frozen\_log\_transcriptome}; fit id: \texttt{trade\_univariate\_ashr\_halfuniform}. TRADE is a robustness column and is not folded into the three-model QC intersection. A diagnostic Spearman of TWI versus each Hallmark score (centroid magnitude as comparator) is required before interpreting TWI-conditioned pathway partials. The Adamson pilot ($n = 8$) is uninformative for pathway claims. Per-dataset correlations are in \textbf{Table~\ref{tab:trade_competitor}}.

\subsection{Minimum cell count filtering}

Perturbations with fewer than 50 cells were excluded under the frozen filter. The Adamson pilot ($n = 8$) is retained in the six-dataset Sp table but excluded from pathway and stress partials.

\subsection{Partial correlations}

All partial correlations reported in the revision use rank-based partial Spearman correlation: variables are ranked before residualization (\texttt{pingouin.partial\_corr}, \texttt{method='spearman'}, or an equivalent numpy rank$\rightarrow$residualize$\rightarrow$Pearson path). Legacy Spearman correlations of raw-scale residuals are not used.

\subsection{Cell-quality covariates}

Per-perturbation QC covariates are the mean of \texttt{percent\_mito}, \texttt{n\_genes}, and \texttt{n\_counts} over the same frozen cells used for $S_p$. Mitochondrial percentage is taken from existing obs columns when present (fraction converted to percent if the maximum is $\le 1.5$) and otherwise computed from \texttt{MT-}genes. When $n < 40$, only mitochondrial percentage is used. Primary pathway and stress survivors are the intersection of $|$centroid$+$QC, $|$E-distance$+$QC, and the joint model.

\subsection{Perturbation-efficiency sensitivity}

To test whether $S_p$--magnitude redundancy is driven by unperturbed cells in assigned guide pools, $S_p$ was recomputed after retaining only putative responder cells. The primary caller is Song et al.'s perturbation-response score (PS; within-perturbation rescaled PS $\geq 0.5$; computed via the Python reimplementation of scMAGeCK described in Extended Methods); Mixscape is secondary (Norman only). Primary endpoints are (i)~$S_p$--magnitude Spearman $\rho$ before versus after filtering and (ii)~rank correlation between $S_p$ on all cells and $S_p$ on responders. Absolute $\Delta S_p$ after filtering is logged but not interpreted as recovered coherence, because both callers select along the mean signature direction. The Adamson pilot was excluded by the same thin-$n$ floor used elsewhere ($n < 30$); Papalexi ($n = 24$) is reported in \textbf{Table~\ref{tab:efficiency_ps}}. On Norman, per-cell PS--Mixscape agreement was modest (Jaccard $= 0.506$, Cohen's $\kappa = 0.374$), so ``responder'' is treated as method-defined and this arm is sensitivity-only (not a replacement for the frozen $S_p$ table).

\subsection{Association survival gate}

Magnitude-conditioned and QC-conditioned associations were adopted when bootstrap CIs excluded 0, $|\rho| > 0.1$, and BH-FDR $< 0.05$. Post hoc conservative rules: (i)~demote rows whose nearer $|\mathrm{CI}|$ bound is strictly $< \varepsilon = 0.02$ to indeterminate; (ii)~mito-only QC covariates when $n < 40$; (iii)~point-estimate-only / exclude from the gate when $n < 30$. These post hoc rules and the adopted $\varepsilon=0.02$ threshold are stated explicitly; no separate $\varepsilon$-sensitivity table is reported.

\subsection{Unbiased discordance GSEA (Approach B)}

The matched pseudobulk design and pre-specified balance criteria are described in Methods. In the frozen run, magnitude balanced after matching (SMD $+0.013$ in Replogle and $-0.024$ in Norman), but the gate failed on cell quality in both datasets (Replogle detected genes SMD $-0.46$; Norman UMIs and detected genes SMD $<-1.2$). Approach B was therefore dropped and Approach A carries the pathway result (Table~\ref{tab:approach_b_balance}). The independent persistence of cell-quality imbalance after magnitude matching supports including QC covariates in the primary partial models. Prior unmatched and cell-level GSEA outputs are retired.

\subsection{Reproducibility}
\label{sec:si-reproducibility}

Frozen real-data tables share configuration version \texttt{2026-07-29.1} (random seed 320) and an Sp digest over (\texttt{dataset}, \texttt{perturbation}, \texttt{stability}, \texttt{magnitude}); the third field is the legacy schema name for Shesha\textsubscript{P} coherence, not a second metric. Downstream joins abort on stamp, row-count ($n = 2{,}285$), dataset-set, or digest mismatch. Frozen manuscript Sp is computed by local \texttt{pipeline\_core.calculate\_sp}. Auxiliary scripts that import \texttt{shesha.bio} use \\\texttt{shesha-geometry==0.2.27}  (commit \texttt{dc0b4b1ea80e7b6ab76b61d4553dcd9757fd6aa1}; \url{https://github.com/prashantcraju/shesha}). Simulation knobs are versioned separately (\texttt{SYNTHETIC\_CONFIG\_VERSION=2026-07-25.9}) and do not change real-data numbers; the synthetic arm is closed for manuscript use (\texttt{SYNTH\_MANUSCRIPT\_USABLE=False}). Responder-filter sensitivity outputs are stamped \texttt{EFFICIENCY\_FILTER\_VERSION=2026-08-09.5} and  \\ \texttt{analysis\_role=sensitivity\_only\_not\_a\_freeze} and do not replace the frozen Sp table. Retired real-data stamps (\texttt{2026-07-25.1}, \texttt{2026-07-17.5}) must not be mixed with the current freeze. Calendar aliases are resolved before digest checks: \texttt{2026-08-04.1} maps to the current \texttt{2026-07-29.1} freeze with the same digest, whereas \texttt{2026-08-03.1} maps to the retired \texttt{2026-07-25.1} run and remains retired. See Table~\ref{tab:pipeline_supersession}. E-distance scores use method \\ \texttt{energy\_distance\_u\_statistic\_euclidean} on the frozen embedding. TRADE TWI uses \\ \texttt{welch\_log2fc\_on\_frozen\_log\_transcriptome} and \texttt{trade\_univariate\_ashr\_halfuniform} (Python ashr implementation; numerically consistent with R \texttt{ashr}); downstream joins do not replace the frozen $S_p$ table.


\section{Robustness Analyses}

\subsection{Distance metric robustness}
All methods produce comparable or stronger correlations relative to the standard Euclidean metric (Table~\ref{tab:distance_robustness}), confirming that the magnitude--coherence relationship is robust to the choice of distance metric. Whitened (Mahalanobis-scaled) and $k$-NN matched methods produce equal or higher correlations across all five primary datasets (Adamson, Dixit, Norman, Papalexi, Replogle), consistent with these methods reducing noise from batch effects and control group heterogeneity. Papalexi 2021 (CRISPR-KO, THP-1, $n = 24$) was analyzed with a minimum of 10 cells per perturbation to preserve sample size; all three metrics yield Spearman $\rho > 0.67$ with 95\% bootstrap CIs excluding zero (Supplementary Figure~S4). The Whitened method substantially improves the Dixit correlation ($\rho: 0.75 \rightarrow 0.97$), suggesting that residual covariance structure in the frozen linear-embedding space attenuates the Euclidean relationship in that dataset.

\subsection{Linear-embedding dimensionality ablation}

Coherence was recomputed at 10, 20, 30, 50, and 100 embedding dimensions, refitting each dataset's frozen backend at every dimension (scanpy PCA for Adamson UPR/pilot and Papalexi; sparse TruncatedSVD for Norman, Dixit, and Replogle). The 50-dimensional rows reproduce the frozen main table rather than importing its values. For Replogle, Spearman $\rho$ increased from $0.963$ at 10 dimensions to $0.980$ at 100 dimensions, with $\rho=0.977$ at the frozen 50-dimensional setting. Results for all datasets are reported in Table~\ref{tab:pca_ablation}.

\subsection{Seed reproducibility}

Cross-seed Spearman correlation of perturbation-level coherence rankings across 15 random seeds (320, 1991, 9, 7258, 7, 2222, 724, 3, 12, 108, 18, 11, 1754, 411, 103) was near-perfect: Norman mean $r = 0.99997$ (range $[0.9999, 1.0000]$, 105 pairwise comparisons); Dixit mean $r = 0.99963$ (range $[0.9993, 0.9999]$). The magnitude--coherence Spearman $\rho$ varied by less than $0.001$ across seeds for both datasets (Norman: $0.959$--$0.960$; Dixit: $0.844$--$0.844$). All reported results use seed 320.

\subsection{Leave-one-out influence}

The maximum $|\Delta\rho|$ upon removing any single perturbation was $0.0019$ for Norman (most influential: HES7; LOO $\rho$ range $[0.959, 0.961]$) and $0.0106$ for Dixit (most influential: INTERGENIC1144056+INTERGENIC1216445; LOO $\rho$ range $[0.840, 0.854]$), indicating that no individual perturbation drives the overall correlation.


\section{Pooled linear companion}

Cross-dataset generalization is the pooled Spearman after within-dataset $z$-scoring (Figure~2g; $\rho = 0.974$). \textbf{Table~\ref{tab:mixed_effects}} reports the linear companion on those same axes: both magnitude and Shesha\textsubscript{P} are $z$-scored within dataset, then $\mathrm{sp}_z = \beta_0 + \beta_1\,\mathrm{mag}_z$ is fit with dataset fixed effects. The slope is $0.956$ (SE $0.006$). Dataset intercepts are $\approx 0$ because the outcome was $z$-scored within dataset; that is not a random-effect variance estimate, and the table does not include spread or cell count. The Methods random-intercept specification (raw $S_p$ on globally $z$-scored magnitude, spread, and $\log n_{\mathrm{cells}}$) is a different model and is not the source of the numbers in \textbf{Table~\ref{tab:mixed_effects}}.


\section{Extended Stress Marker Analysis}

\subsection{Partial correlations}

\textbf{Table~\ref{tab:stress_markers}} reports signed rank-partial Spearman correlations between perturbation coherence and four stress/UPR markers across the five scoreable datasets. The four models condition on magnitude, centroid magnitude plus QC, E-distance plus QC, or both size measures plus QC. Partials use the rank-based procedure described in Extended Methods.

\subsection{Modality analysis}

After conditioning on centroid magnitude and QC, DDIT3 estimates were negative in all five scoreable datasets spanning CRISPRa, CRISPR-KO, and CRISPRi: Norman $\rho=-0.136$ [95\% CI $-0.261$, $0.001$], Adamson UPR $\rho=-0.036$ [$-0.224$, $0.153$], Dixit $\rho=-0.353$ [$-0.539$, $-0.140$], Papalexi $\rho=-0.235$ (descriptive point estimate only at $n=24$), and Replogle $\rho=-0.297$ [$-0.337$, $-0.253$]. Norman's rounded centroid+QC interval extends to $+0.001$ and therefore crosses zero; its E-distance and joint upper bounds ($-0.007$ and $-0.005$) also lie within the pre-specified $\varepsilon=0.02$ knife edge, yielding an indeterminate classification. Only Replogle passed the intersection criterion across all conditioned models. Thus the frozen analysis does not support modality as an explanation for DDIT3 heterogeneity. HSPA5 did not pass the intersection criterion in any dataset. Full results are provided in \textbf{Table~\ref{tab:stress_markers}}.


\section{scGPT Validation Protocol}

\subsection{Model and embedding}

Cell embeddings were generated using the scGPT ``Whole Human'' pretrained checkpoint (\texttt{scGPT\_human}), downloaded from the official repository (\url{https://github.com/bowang-lab/scGPT}). The checkpoint contains three files: \texttt{best\_model.pt}, \texttt{vocab.json}, and \texttt{args.json}.

Input data consisted of raw counts (the \texttt{counts} layer was extracted from each AnnData object; data were not log-normalized prior to embedding). Embeddings were generated using \texttt{embed\_data()} with \texttt{gene\_col="index"}, \texttt{batch\_size=64}, and \texttt{use\_fast\_transformer=False}.

\subsection{Reproducibility settings}

All scGPT computations used deterministic mode:
\begin{itemize}
\item Python random seed: 320
\item NumPy random seed: 320
\item PyTorch manual seed: 320 (CPU and CUDA)
\item \texttt{torch.backends.cudnn.deterministic = True}
\item \texttt{torch.backends.cudnn.benchmark = False}
\end{itemize}

\subsection{Coherence and magnitude computation}

Coherence and magnitude were computed from scGPT embeddings using \texttt{shesha.bio.compute\_stability()} and\\ \texttt{shesha.bio.compute\_magnitude()} with \texttt{perturbation\_key='perturbation\_name'}, \texttt{control\_label='control'}, and \texttt{metric='cosine'} (coherence) or \texttt{metric='euclidean'} (magnitude). Bootstrap confidence intervals: 10,000 resamples, seed 320, percentile method. The first function retains its historical API name; its returned mean-cosine quantity is Shesha\textsubscript{P} coherence throughout this manuscript.

\subsection{Datasets}

scGPT~\citep{Cui2024} analysis was performed on Norman 2019, Adamson 2016 UPR, Dixit 2016, and Replogle 2022. Control-labeled Adamson rows were excluded, leaving $n=87$ scored perturbations. The Adamson pilot was omitted because $n=8$; Papalexi was omitted because scGPT ranks were not numerically reproducible at $n=24$.

\subsection{Results}

Within-scGPT magnitude--coherence correlations were $0.935$ (Norman), $0.912$ (Adamson UPR), $0.854$ (Dixit), and $0.851$ (Replogle). Cross-representation concordance against the frozen linear embedding was, respectively, $0.514$, $0.828$, $0.472$, and $0.750$ for coherence and $0.629$, $0.873$, $0.549$, and $0.784$ for magnitude. Because both score types change across representations, we interpret this as embedding sensitivity and do not assign it a coherence-specific biological mechanism. See \textbf{Tables~\ref{tab:embedding_magnitude} and \ref{tab:embedding_concordance}}.

\subsection{Diffusion maps and PHATE}

Diffusion maps used scanpy neighbors on 100 PCs ($k = 15$) with the trivial DC0 component dropped~\citep{Haghverdi2016}. PHATE used $k = 5$, internal PCA $= 100$, and SGD-MDS~\citep{Moon2019}; SMACOF is the convergence control when SGD does not converge. Manuscript concordance is each embedding's $S_p$ (and separately its magnitude) versus frozen linear-embedding $S_p$ (respectively $M_p$), not versus a re-derived linear-embedding score. Within-embedding $S_p$--magnitude correlations are a separate column. Papalexi is excluded from scGPT concordance because ranks are not numerically reproducible at $n = 24$.


\section{Combinatorial Analysis}

The Norman 2019 dataset contains $n=105$ single-gene and $n=131$ combinatorial perturbations. Combinatorial status was associated with higher raw coherence (mean $S_p=0.490$ vs $0.334$; Mann--Whitney $p=8.4\times10^{-11}$; Spearman $\rho=0.424$), but combinatorial perturbations also had larger effects (magnitude SMD $=0.924$; IQR overlap $=0.139$). After conditioning on magnitude, the association was null ($\rho=-0.048$, 95\% CI $-0.176$, $0.091$); adding QC and cell count gave $\rho=-0.030$ [$-0.162$, $0.113$]. Thus the raw combinatorial association does not provide evidence for greater coherence beyond effect size. Norman selected gene pairs expected to interact, which further limits generalization. Within the combinatorial set, erythroid-containing pairs also had higher raw $S_p$ (median $0.568$ vs $0.465$; one-sided Mann--Whitney $p=0.0035$), but this contrast was null after magnitude residualization ($p=0.518$).


\section{Perturbation-Efficiency Sensitivity}

Primary arm (Song PS) on Norman, Replogle, Dixit, and Adamson UPR. On Norman and Replogle, Sp--magnitude correlations were essentially unchanged after responder filtering ($0.950 \rightarrow 0.902$; $0.978 \rightarrow 0.976$) and Sp rankings were preserved ($\rho = 0.918$ and $0.916$). Adamson UPR was partial ($0.963 \rightarrow 0.951$; rank $\rho = 0.865$); Dixit changed more ($0.841 \rightarrow 0.847$; rank $\rho = 0.642$) and is the coherent outlier across efficiency filtering, residual Sp variance, and embedding concordance. Mixscape on Norman is secondary only: it rescored $149/236$ perturbations (87 with no responders), so its ``all-cell'' Sp--mag baseline is a survivor subset rather than the frozen Norman value, and the Sp--mag $\Delta$ can flip sign relative to PS. Across callers, $\Delta$ in Sp--mag after filtering is within method-to-method noise; the relationship is not attributed to unperturbed cells in assigned guide pools.

%% file: si_figs_tables-new.tex

 
\begin{figure*}[p]
\centering
\includegraphics[width=\textwidth]{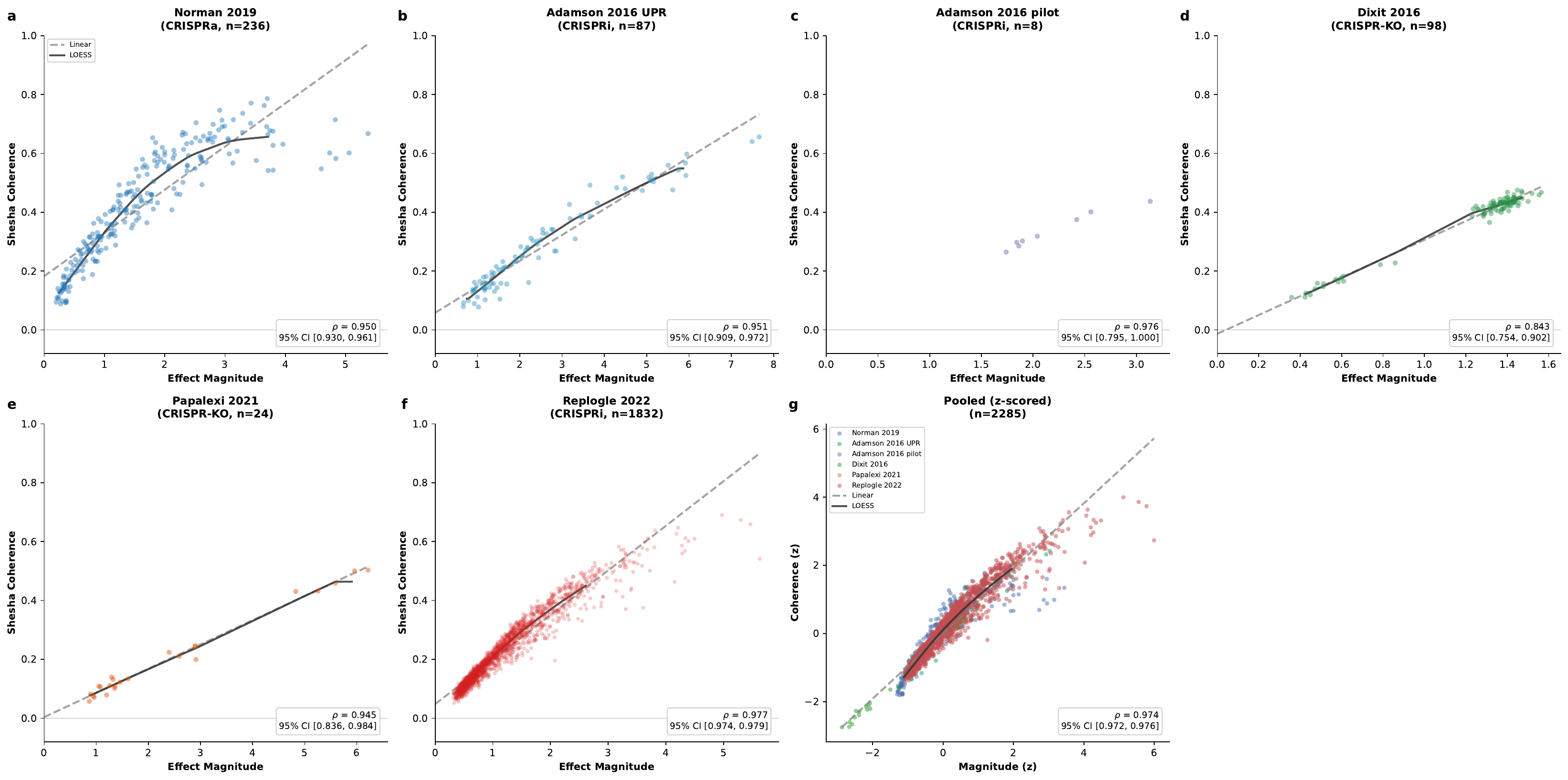}
\caption{\textbf{Coherence increases monotonically with effect magnitude in every dataset, with
deceleration at high magnitude.} Shesha\textsubscript{P} against effect magnitude for (a--f) each dataset and
(g)~pooled after within-dataset $z$-scoring. Dashed grey, ordinary least squares; solid black,
LOESS. Deceleration is clearest in the well-powered datasets (f,~g), consistent with coherence
being a bounded cosine statistic, and the two fits are close in the smaller datasets. Reported
$\rho$ are Spearman rank correlations and are therefore independent of fit form; intervals are
bootstrap 95\% CIs. No fit is shown for the Adamson pilot ($n=8$). LOESS is drawn only over the
range where local density supports estimation.}
\label{fig:si_mag_stab_loess}
\end{figure*}
\clearpage

\begin{figure*}[h]
\centering
\includegraphics[width=\textwidth]{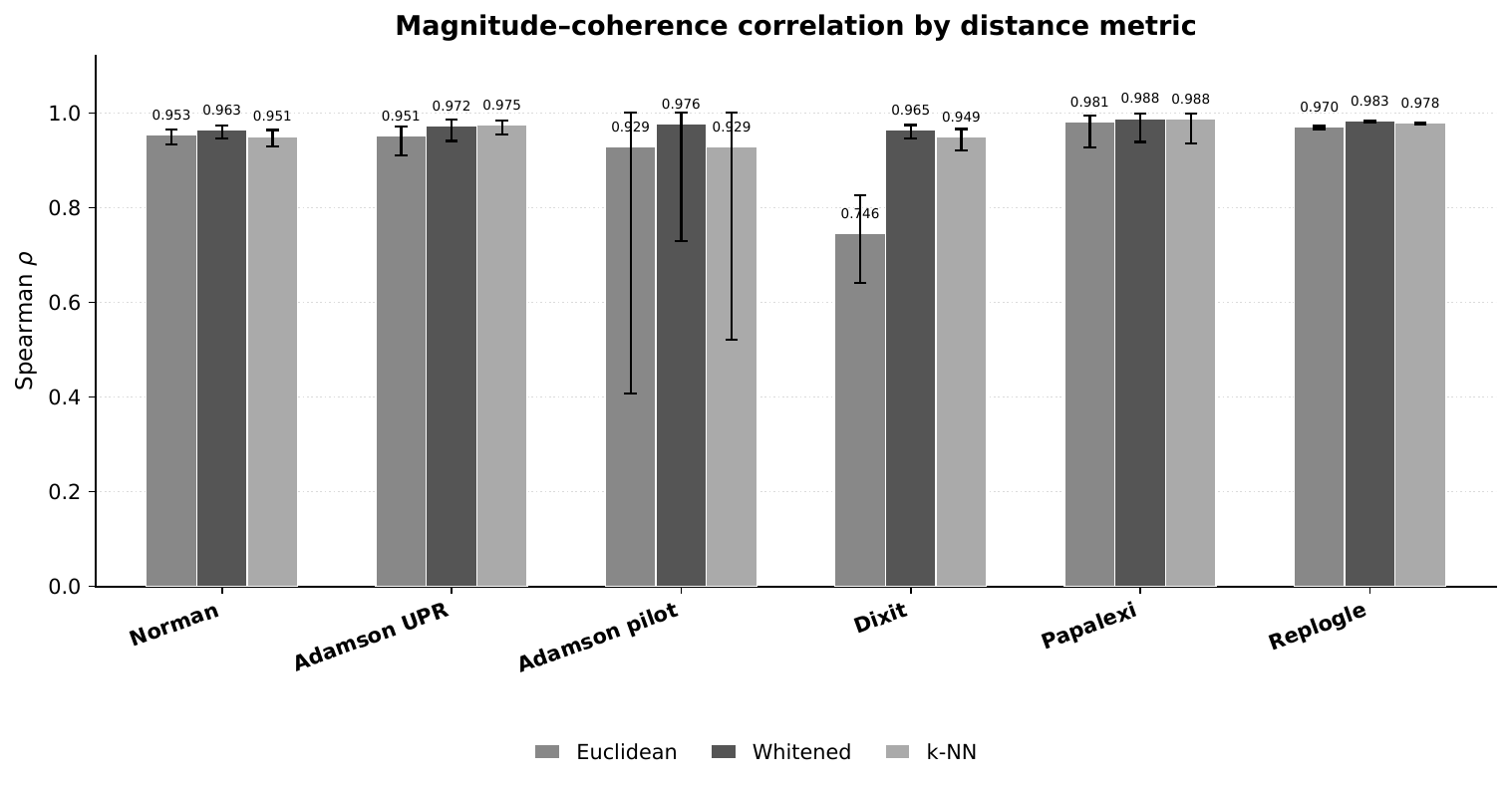}
\caption{\textbf{Magnitude--coherence is not an artifact of Euclidean PCA distance.}
Spearman $\rho$ between Shesha coherence and effect magnitude under three
control-referenced metrics, six datasets: Euclidean centroid, whitened
(Mahalanobis / $\Sigma^{-1/2}$), and $k$-NN matched local controls ($k=50$).
Bars are $95\%$ bootstrap CIs.}
\label{fig:distance_robustness}
\end{figure*}
\clearpage

\begin{figure*}[p]
\centering
\includegraphics[width=.75\textwidth]{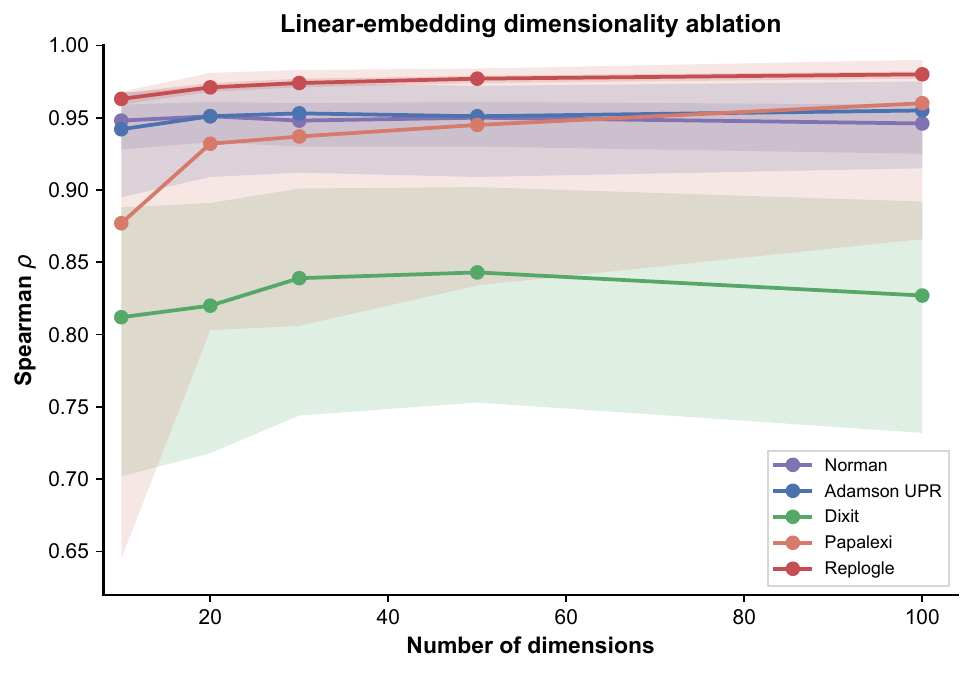}
\caption{\textbf{Linear-embedding dimensionality ablation.}
Magnitude--Shesha Spearman $\rho$ as a function of embedding dimension
(10, 20, 30, 50, 100) under \texttt{2026-07-29.1}. Shaded regions are
95\% bootstrap CIs (10{,}000 iterations, seed 320). Each series refits
the recorded frozen backend (scanpy PCA or sparse TruncatedSVD) on the
freeze cell set; the 50-dimensional point reproduces Table 1. Norman
($n=236$) is stable ($\rho=0.946$--$0.951$) with overlapping CIs.
Adamson UPR ($n=87$) is likewise stable ($\rho=0.942$--$0.955$).
Dixit ($n=98$) is lower throughout ($\rho=0.812$--$0.843$) and does not
increase monotonically. Papalexi ($n=24$) rises from $\rho=0.877$ to
$0.960$ with the widest intervals. Replogle ($n=1{,}832$) remains high
($\rho=0.963$--$0.980$). The choice of 50 dimensions does not drive the
magnitude--Shesha relationship.}
\label{fig:pca_ablation}
\end{figure*}
\clearpage

\begin{figure*}[p]
\centering
\includegraphics[width=\textwidth]{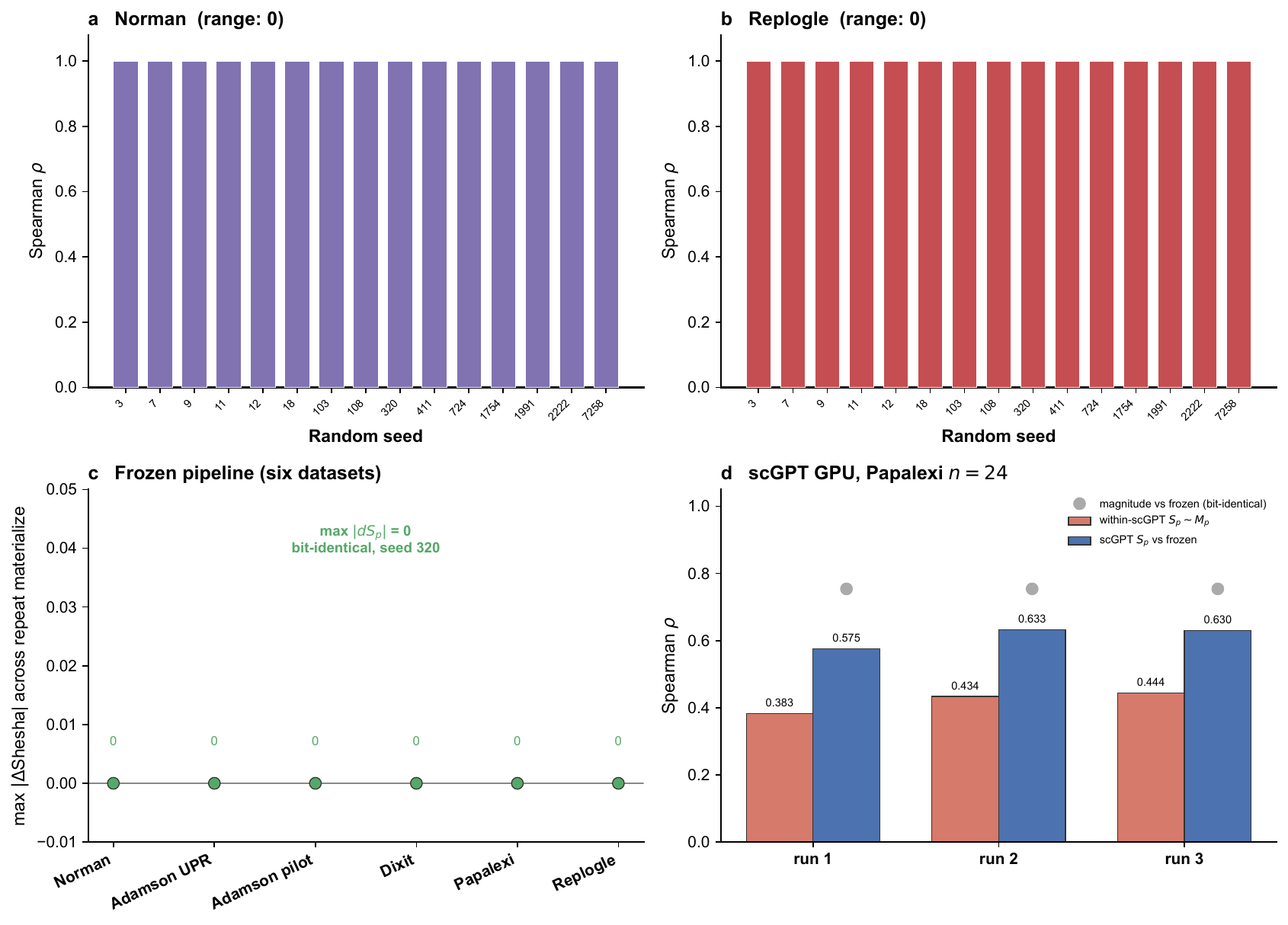}
\caption{\textbf{Seed and freeze reproducibility; scGPT ranks are not stable at small $n$.}
(a,b)~Magnitude--coherence Spearman $\rho=1.000$ at 15 random seeds on Norman and Replogle
($3,7,9,11,12,18,103,108,320,411,724,1754,1991,2222,7258$); range $=0$.
(c)~Repeated materialize + Shesha under the freeze (seed 320) yields $\max|dS_p|=0$
on all six datasets.
(d)~Three locked-h5mu GPU scGPT embeds of Papalexi ($n=24$): within-scGPT $S_p\sim M_p$
spans $0.383$--$0.444$; concordance versus frozen $S_p$ spans $0.575$--$0.633$;
magnitude concordance is bit-identical at $0.7539130434782607$.}
\label{fig:seed_reproducibility}
\end{figure*}
\clearpage

\begin{figure*}[p]
\centering
\includegraphics[width=\textwidth]{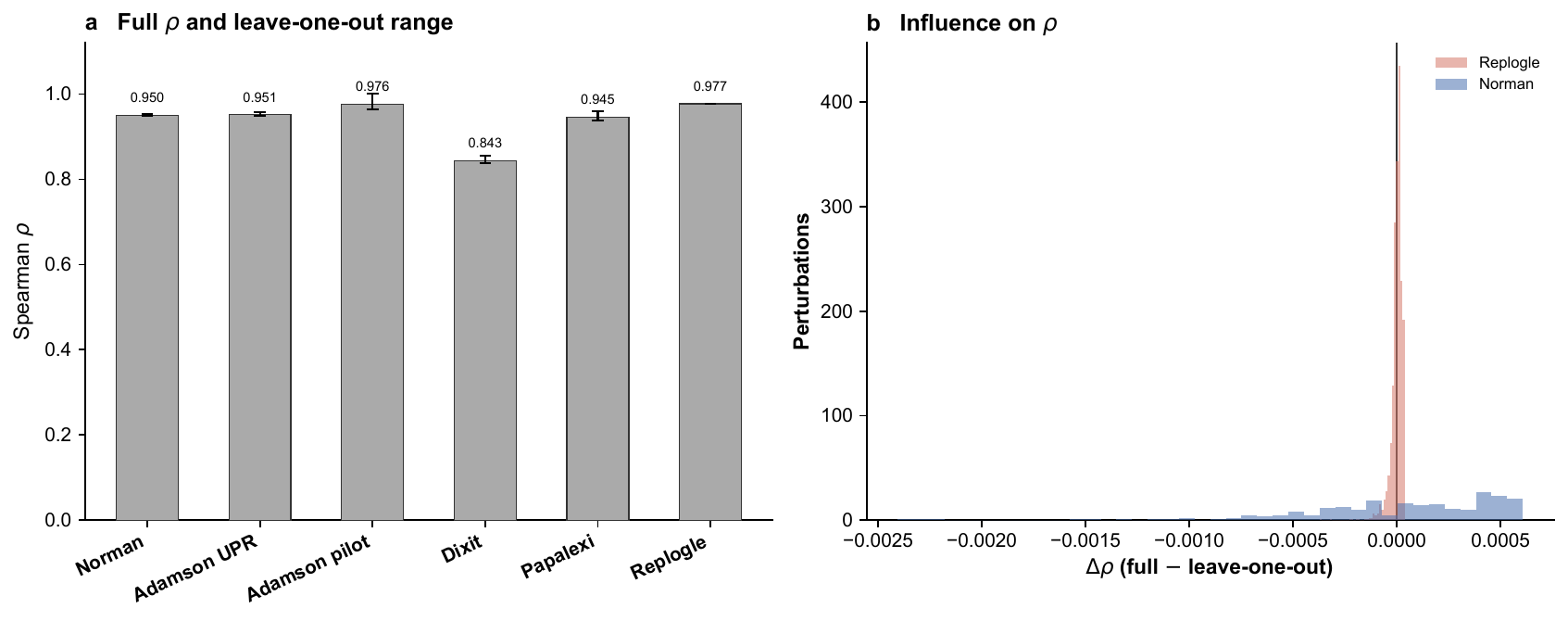}
\caption{\textbf{Leave-one-out influence on the magnitude--coherence Spearman $\rho$.}
(a)~Full $\rho$ (bars) and the range of $\rho$ after dropping each perturbation in turn (error bars), computed from frozen Shesha on all six datasets. Full $\rho$: Norman $0.950$ ($n=236$), Adamson UPR $0.951$ ($n=87$), Adamson pilot $0.976$ ($n=8$), Dixit $0.843$ ($n=98$), Papalexi $0.945$ ($n=24$), Replogle $0.977$ ($n=1832$).
(b)~Per-perturbation $\Delta\rho$ (full $-$ leave-one-out) for Norman and Replogle. Maximum $|\Delta\rho|$ is $0.0024$ on Norman and $0.0004$ on Replogle; the largest swing among the six datasets is $0.024$ on Adamson pilot ($n=8$). No single perturbation drives the correlation.}
\label{fig:loo_influence}
\end{figure*}
\clearpage

\begin{figure*}[p]
\centering
\includegraphics[width=\textwidth]{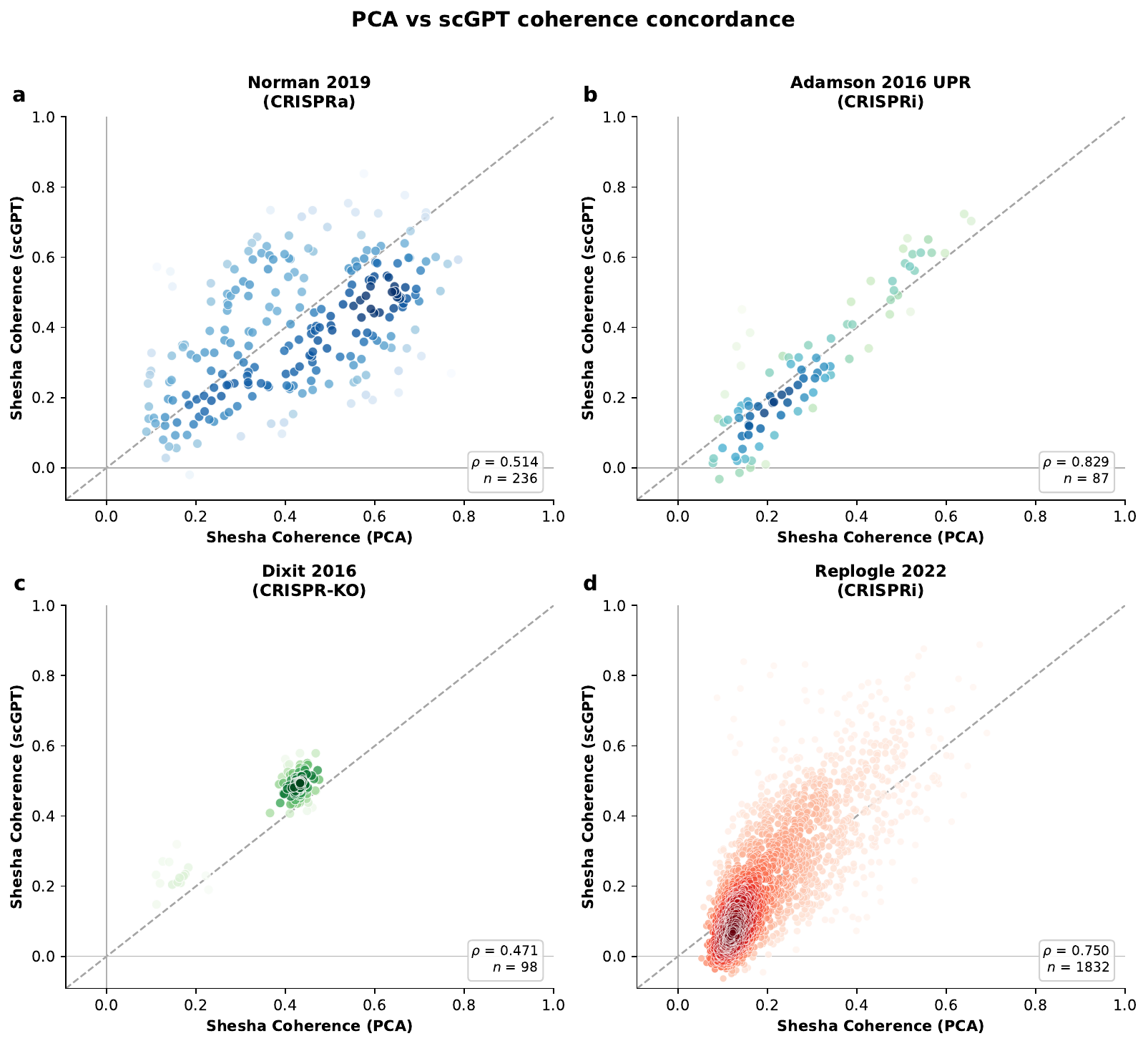}
\caption{\textbf{Cross-representation concordance of Shesha\textsubscript{P} between the frozen PCA
representation and the scGPT embedding.} Each point is one perturbation, plotted as coherence
computed in frozen PCA against coherence recomputed in scGPT. Dashed line is identity. Spearman
$\rho$ and $n$ annotated per panel. Concordance is moderate (0.471--0.829) and points fall
predominantly below identity, so scGPT coherence is systematically lower than PCA coherence. This
is distinct from the within-scGPT redundancy check in Fig.~\ref{fig:magnitude_coherence}, where
both axes are scGPT quantities. Papalexi ($n=24$) and the Adamson pilot ($n=8$) are omitted:
coherence rank statistics are not numerically reproducible under GPU inference at small $n$
(Sec.~\ref{sec:limits-of-measurement}).}
\label{fig:scgpt_concordance}
\end{figure*}
\clearpage

\begin{figure*}[h]
    \centering
        \includegraphics[width=\linewidth]{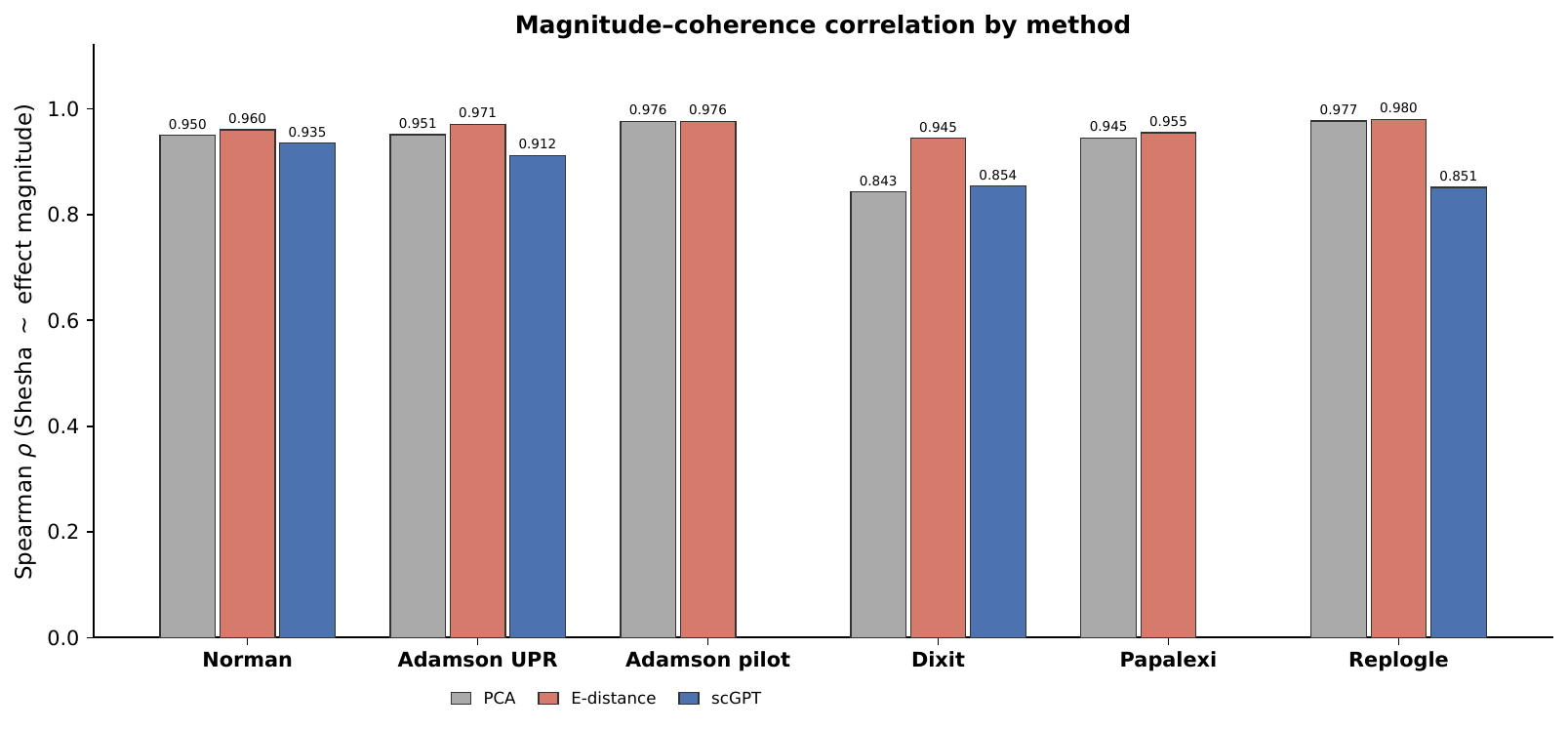}
\caption{\textbf{The coherence--magnitude relationship holds across effect-size definitions and
representations.} Spearman $\rho$ between Shesha\textsubscript{P} and effect magnitude computed three ways:
centroid displacement in frozen PCA, energy distance, and both quantities recomputed inside the
scGPT embedding. Papalexi and the Adamson pilot are omitted from the scGPT bars.}
\label{fig:robustness_methods}
\end{figure*}
\clearpage


\begin{table*}[ht]
\centering
\caption{Dataset overview for the frozen analysis cohort. Frozen cells include
retained perturbed and control cells after filtering for at least 50 cells per
perturbation and hash-stable downsampling to at most 100 cells per perturbation
and 5,000 controls (seed 320). These are analysis-cohort counts, not raw source
dataset totals.}
\label{tab:datasets}
\resizebox{\linewidth}{!}{%

\begin{tabular}{llccrcc}
\hline
Dataset & Modality & Cell type & Scored perturbations &
Frozen cells & Median cells/pert. & Control label \\
\hline
Norman 2019~\citep{norman2019exploring}
 & CRISPRa & K562 & 236 & 28,290 & 100 & control \\
Adamson 2016 UPR~\citep{adamson2016multiplexed}
 & CRISPRi & K562 & 87 & 13,605 & 100 & control \\
Adamson 2016 pilot~\citep{adamson2016multiplexed}
 & CRISPRi & K562 & 8 & 810 & 100 & control \\
Dixit 2016~\citep{dixit2016perturb}
 & CRISPR-KO & BMDC & 98 & 13,664 & 100 & control \\
Papalexi 2021~\citep{papalexi2021characterizing}
 & CRISPR-KO & THP-1 & 24 & 4,786 & 100 & NT \\
Replogle 2022~\citep{Replogle2022}
 & CRISPRi & K562 & 1,832 & 175,013 & 100 & control \\
\hline
\end{tabular}
}
\end{table*}
\clearpage

\begin{table*}[ht]
\centering
\caption{Pipeline supersession. Real-data manuscript tables use a single frozen stamp (\texttt{pipeline\_config.CONFIG\_VERSION}). \texttt{2026-08-04.1} is the same freeze as \texttt{2026-07-29.1} under a calendar label (\texttt{CONFIG\_VERSION\_ALIASES}); do not treat the two names as different digests. Synthetic and efficiency-filter lanes are versioned separately and must not be mixed with freeze numbers.}
\label{tab:pipeline_supersession}
\smallskip
\resizebox{\linewidth}{!}{%
\begin{tabular}{lcc}
\hline
Label & Scope & Status \\
\hline
\texttt{2026-07-29.1} & All real-data tables & Current. Hash-stable downsample; seed 320 \\
\texttt{2026-08-04.1} & Calendar alias of \texttt{2026-07-29.1} & Same freeze and digest. Appears in early Drive/TRADE stamps \\
\texttt{2026-07-25.1} & matrix\_is\_log + positional downsample & Retired. Pathway $|$mag drifted at the same stamp \\
\texttt{2026-08-03.1} & Calendar alias of \texttt{2026-07-25.1} & Retired with that stamp \\
\texttt{2026-07-17.5} & Heuristic Adamson skip-normalize & Retired \\
\texttt{SYNTHETIC 2026-07-25.9} & scDesign3 generator only & Independent of freeze. \texttt{manuscript\_usable=False} \\
\texttt{EFFICIENCY\_FILTER\_VERSION} & PS / Mixscape sensitivity & Does not replace \texttt{frozen\_sp\_scores.csv} \\
\hline
\end{tabular}
}
See \texttt{methods\_config\_supersession.md} and \texttt{pipeline\_config.CONFIG\_VERSION\_ALIASES}.
\end{table*}
\clearpage

\begin{table*}[ht]
\centering
\caption{Per-dataset \texttt{matrix\_is\_log} pin from \texttt{.X} inspection (\texttt{inspect\_matrix\_scale.py}). Integer-like / large-max matrices are raw counts (\texttt{False} $\rightarrow$ normalize + log1p). Non-integer floats with max $\sim$5--10 are already log (\texttt{True} $\rightarrow$ skip). Only Adamson UPR was mis-pinned under the heuristic era.}
\label{tab:matrix_is_log}
\smallskip
\begin{tabular}{lcccc}
\hline
Dataset & max(\texttt{.X}) & frac near-int & pin & verified \\
\hline
Adamson pilot & 47.000 & --- & False & True \\
\hline
\end{tabular}

Pins live in \texttt{DATASETS[*][matrix\_is\_log]}.
\end{table*}
\clearpage

\begin{table*}[ht]
\centering
\caption{Reproducibility verification under \texttt{CONFIG\_VERSION=2026-07-29.1}. The load-bearing claim is bit-identical materialize and Sp across all six datasets. GPU scGPT ranks at $n=24$ are the documented exception and are omitted from the scGPT series.}
\label{tab:repro_verification}
\smallskip
\resizebox{\linewidth}{!}{%
\begin{tabular}{lccc}
\hline
Check & Scope & Result & Detail \\
\hline
Materialize + Sp & six datasets, $n=2{,}285$ & bit-identical & max $|dS_p|=0$ (verified twice) \\
Digest guard & \texttt{assert\_frozen\_sp\_compatible} & aborts & four tested mismatch paths \\
Bootstrap CIs & $n=10{,}000$, seed 320 & bit-identical & \texttt{--verify-bootstrap} \\
Papalexi GEO vs freeze & $n=24$ shared genes & $\rho=1.000$ & independent GSE153056 rebuild \\
Papalexi scGPT (GPU) & $n=24$, three runs & not reproducible & Sp vs freeze $0.575$--$0.633$; mag $\rho=0.753913$ bit-identical \\
Cross-seed frozen-linear Sp & Norman / Replogle, 15 seeds & $r=1$ & analytic path; see \ref{tab:seed_reproducibility} \\
\hline
\end{tabular}
}
\end{table*}
\clearpage

\begin{table*}[ht]
\centering
\caption{Linear slope of coherence on effect magnitude, pooled across datasets.
Ordinary least squares fit of within-dataset $z$-scored
Shesha\textsubscript{P} on within-dataset $z$-scored effect magnitude, with
dataset fixed effects. A unit increase in standardized magnitude corresponds to
a $0.956$ standard-deviation increase in coherence
($\mathrm{SE} = 0.006$), giving a scale-free linear companion to the pooled rank
correlation in Fig.~\ref{fig:magnitude_coherence}g. Because both axes are
$z$-scored within dataset before fitting, the intercept and the dataset fixed
effects are zero by construction and are not interpretable. This fit estimates
the pooled slope only: it does not decompose variance across datasets and does
not include the spread or cell-count terms described in Methods.}
\label{tab:mixed_effects}
\smallskip
\begin{tabular}{lccc}
\hline
Term & coef & SE & Model \\
\hline
Intercept & -0.0000 & 0.0316 & ols\_dataset\_fe\_fallback \\
mag\_z & 0.9557 & 0.0062 & ols\_dataset\_fe\_fallback \\
\hline
\end{tabular}
\end{table*}
\clearpage



\begin{table*}[ht]
\centering
\caption{Coherence--magnitude correlation across effect-size definitions.
Spearman $\rho$ between Shesha\textsubscript{P} and four measures of
perturbation effect size, all computed in the frozen $50$-PC embedding: centroid
Euclidean displacement, which is the measure used throughout the manuscript;
energy distance, which additionally penalizes within-group dispersion; centroid
displacement in a whitened space; and mean distance to $k$-nearest matched
control cells. The relationship is not specific to how effect size is defined,
with correlations of $0.843$ or above under every measure in every dataset. The
largest difference between measures is in Dixit, where energy distance and the
whitened metric give substantially higher values than centroid displacement
($0.945$ and $0.965$ against $0.843$), consistent with a large share of that
dataset's residual under centroid displacement being within-group dispersion
(Fig.~\ref{fig:edistance_competitor}). Values from the pre-freeze pipeline are
not carried over.}
\label{tab:distance_robustness}
\smallskip
\begin{tabular}{lccccc}
\hline
Dataset & $n$ & $\rho(S_p,M_p)$ & $\rho(S_p,\mathrm{Ed})$ & $\rho_{\mathrm{Whitened}}$ & $\rho_{\mathrm{k-NN}}$ \\
\hline
Norman & 236 & 0.950 & 0.960 & 0.963 & 0.951 \\
Adamson UPR & 87 & 0.951 & 0.971 & 0.976 & 0.929 \\
Adamson pilot & 8 & 0.976 & 0.976 & 0.976 & 0.929 \\
Dixit & 98 & 0.843 & 0.945 & 0.965 & 0.949 \\
Papalexi & 24 & 0.945 & 0.955 & 0.988 & 0.988 \\
Replogle & 1,832 & 0.977 & 0.980 & 0.983 & 0.978 \\
\hline
\end{tabular}

Euclidean / centroid magnitude and energy distance from the frozen 50-PC embedding. Pre-freeze Dixit $0.746$ / Papalexi $0.981$ are not carried over.
\end{table*}
\clearpage

\begin{table*}[ht]
\centering
\caption{Linear-embedding dimensionality ablation. Magnitude--Shesha Spearman $\rho$ with 95\% bootstrap CIs (10{,}000 iterations, seed 320) across embedding dimensions under \texttt{2026-07-29.1}. The 50-dimensional row is a live reproduction of Table 1, not an imported value. Every dimension refits the recorded dataset-specific backend (scanpy PCA or sparse TruncatedSVD) on the freeze cell set. Adamson UPR ($n=87$) and Papalexi ($n=24$) belong in this table when the freeze sweep includes them.}
\label{tab:pca_ablation}
\smallskip
\begin{tabular}{lcccc}
\hline
Dataset & Dimensions & $\rho$ & 95\% CI & $p$ \\
\hline
Adamson UPR & 10 & 0.942 & [+0.895, +0.967] & $3.6 \times 10^{-42}$ \\
Adamson UPR & 20 & 0.951 & [+0.909, +0.972] & $4.2 \times 10^{-45}$ \\
Adamson UPR & 30 & 0.953 & [+0.912, +0.974] & $6.5 \times 10^{-46}$ \\
Adamson UPR & 50 & 0.951 & [+0.909, +0.972] & $3.2 \times 10^{-45}$ \\
Adamson UPR & 100 & 0.955 & [+0.915, +0.975] & $7.8 \times 10^{-47}$ \\
\hline
Dixit & 10 & 0.812 & [+0.702, +0.888] & $3.5 \times 10^{-24}$ \\
Dixit & 20 & 0.820 & [+0.718, +0.891] & $5.2 \times 10^{-25}$ \\
Dixit & 30 & 0.839 & [+0.744, +0.901] & $4.3 \times 10^{-27}$ \\
Dixit & 50 & 0.843 & [+0.753, +0.902] & $1.4 \times 10^{-27}$ \\
Dixit & 100 & 0.827 & [+0.732, +0.892] & $1.1 \times 10^{-25}$ \\
\hline
Norman & 10 & 0.948 & [+0.928, +0.959] & $4.0 \times 10^{-118}$ \\
Norman & 20 & 0.951 & [+0.933, +0.961] & $3.2 \times 10^{-121}$ \\
Norman & 30 & 0.948 & [+0.930, +0.960] & $1.3 \times 10^{-118}$ \\
Norman & 50 & 0.950 & [+0.930, +0.961] & $8.0 \times 10^{-120}$ \\
Norman & 100 & 0.946 & [+0.925, +0.959] & $4.3 \times 10^{-116}$ \\
\hline
Papalexi & 10 & 0.877 & [+0.645, +0.968] & $2.0 \times 10^{-8}$ \\
Papalexi & 20 & 0.932 & [+0.803, +0.981] & $3.5 \times 10^{-11}$ \\
Papalexi & 30 & 0.937 & [+0.806, +0.983] & $1.5 \times 10^{-11}$ \\
Papalexi & 50 & 0.945 & [+0.834, +0.984] & $3.6 \times 10^{-12}$ \\
Papalexi & 100 & 0.960 & [+0.866, +0.990] & $1.2 \times 10^{-13}$ \\
\hline
Replogle & 10 & 0.963 & [+0.959, +0.967] & $< 10^{-300}$ \\
Replogle & 20 & 0.971 & [+0.968, +0.974] & $< 10^{-300}$ \\
Replogle & 30 & 0.974 & [+0.971, +0.977] & $< 10^{-300}$ \\
Replogle & 50 & 0.977 & [+0.974, +0.979] & $< 10^{-300}$ \\
Replogle & 100 & 0.980 & [+0.977, +0.982] & $< 10^{-300}$ \\
\hline
\end{tabular}
\end{table*}
\clearpage

\begin{table*}[ht]
\centering
\caption{Seed and runtime reproducibility. The frozen linear-embedding path is deterministic: 15 analytic seeds give identical Spearman $\rho$ (cross-seed $r=1$, not a rounded $0.99997$). The interesting exception is Papalexi scGPT under GPU inference at $n=24$: Shesha ranks move across runs while magnitude concordance is bit-identical. That dataset is omitted from every scGPT series (same thin-$n$ rule as the Adamson pilot, $n=8$).}
\label{tab:seed_reproducibility}
\smallskip
\resizebox{\linewidth}{!}{%
\begin{tabular}{lccccc}
\hline
Dataset & Quantity & Runs & Mean / span & Other & Verdict \\
\hline
Norman / Replogle & Frozen linear Sp & 15 & $1.000000 \pm 0.000000$ & $1.000000$ & analytic; not the $0.99997$ GPU-era figure \\
Papalexi scGPT (GPU) & vs frozen Sp & 3 & 0.613 (span 0.575--0.633) & 0.753913043 & exclude from scGPT series \\
\hline
\end{tabular}
}
\end{table*}
\clearpage


\begin{table*}[ht]
\centering
\caption{Leave-one-out influence on the magnitude--Shesha Spearman correlation. $\Delta\rho$ is $\rho$ after dropping one perturbation minus the full-sample $\rho$. Dixit is $n=98$ under the freeze (the prior version $n=153$ used \texttt{MIN\_CELLS}$=10$). No single perturbation drives the relationship in any scoreable dataset.}
\label{tab:loo_influence}
\smallskip
\resizebox{\linewidth}{!}{%
\begin{tabular}{lcccccc}
\hline
Dataset & $n$ & Full $\rho$ & LOO range & max $|\Delta\rho|$ & Largest $+\Delta\rho$ & Largest $-\Delta\rho$ \\
\hline
Adamson UPR & 87 & 0.951 & [0.950, 0.957] & 0.0060 & \texttt{GBF1} (+0.0060) & \texttt{CARS} (-0.0017) \\
Adamson pilot & 8 & 0.976 & [0.964, 1.000] & 0.0238 & \texttt{CREB1\_pDS269} (+0.0238) & \texttt{62(mod)\_pBA581} (-0.0119) \\
Dixit & 98 & 0.843 & [0.838, 0.856] & 0.0127 & \texttt{INTERGENIC216151+EGR1} (+0.0127) & \texttt{CREB1} (-0.0049) \\
Norman & 236 & 0.950 & [0.949, 0.952] & 0.0024 & \texttt{CEBPA+KLF1} (+0.0024) & \texttt{TGFBR2} (-0.0006) \\
Papalexi & 24 & 0.945 & [0.938, 0.960] & 0.0153 & \texttt{CD86} (+0.0153) & \texttt{CAV1} (-0.0075) \\
Replogle & 1,832 & 0.977 & [0.977, 0.977] & 0.0004 & \texttt{CHMP2A} (+0.0004) & \texttt{EIF2S1} (-0.0000) \\
\hline
\end{tabular}
}
\end{table*}
\clearpage

\begin{table*}[ht]
\centering
\caption{Partial-correlation arithmetic. Several scripts in an earlier version of
this analysis residualized on the raw scale and then took Spearman correlations
of the residuals. That is not equivalent to a rank-based partial correlation when
the relationship between the conditioning variable and the outcome is nonlinear,
which it is here: coherence saturates with effect magnitude
(Fig.~\ref{fig:si_mag_stab_loess}). The manuscript default is therefore the
rank-based partial Spearman, computed as three steps below
(\texttt{pingouin.partial\_corr(..., method='spearman')}, with a numpy fallback in
\texttt{stats\_utils.\_partial\_spearman\_rank\_numpy}). Bootstrap confidence
intervals resample perturbations and repeat all three steps
($10{,}000$ resamples, seed $320$). All pathway and stress-marker survival
verdicts use this estimator.}
\label{tab:partial_corr_arithmetic}
\smallskip
\resizebox{\linewidth}{!}{%
\begin{tabular}{lc}
\hline
Step & Operation \\
\hline
1 & Rank $x$, $y$, and each covariate in $Z$ \\
2 & Residualize $\mathrm{rank}(x)$ and $\mathrm{rank}(y)$ on $[1,\ \mathrm{rank}(Z)]$ by OLS \\
3 & Pearson correlation of those residuals $=$ partial Spearman $\rho$ \\
Legacy (retired) & Residualize raw $x,y$ on raw $Z$, then Spearman. Not equivalent when the magnitude--Shesha map is nonlinear. \\
\hline
\end{tabular}
}

See \texttt{stats\_utils.py}. Survival of pathway / stress rows uses this estimator, not the legacy residual Spearman.
\end{table*}
\clearpage

\begin{table*}[ht]
\centering
\caption{Single-marker stress associations with Shesha\textsubscript{P}. Five
scoreable datasets (Adamson UPR $n = 87$; the $n = 8$ pilot is excluded
throughout) and four covariate models: magnitude; centroid + QC; E-distance + QC;
joint (centroid + E-distance + QC). Survival follows \texttt{ci\_and\_fdr.v1}
and applies to the three QC-conditioned models within each dataset; it does not
imply a common direction across datasets. Magnitude-only rows are reported for
comparison and are not passed through the gate (Status ---). Papalexi
magnitude-only intervals are shown ($10{,}000$ valid bootstrap resamples),
whereas its three QC-conditioned intervals are omitted because the $n = 24$ QC
design is rank-deficient and yielded zero valid resamples. $\star$, shown on the
joint row, marks marker--dataset pairs surviving all three QC-conditioned models
(Replogle ATF4, DDIT3 and XBP1). ATF4 illustrates why single markers are
unreliable here: it survives positively in Adamson UPR (centroid+QC $\rho =
+0.340$; E-distance+QC $+0.315$) and negatively in Replogle ($-0.106$, $-0.106$,
joint $-0.111$), with both screens CRISPRi in K562. HSPA5 survives nowhere under
the QC-conditioned models.}
\label{tab:stress_markers}
\smallskip
\begin{tabular}{lcccccc}
\hline
Marker & Dataset & $n$ & Model & $\rho$ & 95\% CI & Status \\
\hline
DDIT3 & Norman & 236 & magnitude & -0.237 & [-0.351, -0.107] & survives \\
 & Norman & 236 & centroid+QC & -0.136 & [-0.261, +0.001] & does\_not\_survive \\
 & Norman & 236 & edistance+QC & -0.145 & [-0.259, -0.007] & indeterminate \\
 & Norman & 236 & centroid+edistance+QC & -0.143 & [-0.265, -0.005] & indeterminate \\
 & Adamson UPR & 87 & magnitude & -0.067 & [-0.237, +0.120] & does\_not\_survive \\
 & Adamson UPR & 87 & centroid+QC & -0.036 & [-0.224, +0.153] & does\_not\_survive \\
 & Adamson UPR & 87 & edistance+QC & -0.070 & [-0.253, +0.129] & does\_not\_survive \\
 & Adamson UPR & 87 & centroid+edistance+QC & -0.131 & [-0.303, +0.113] & does\_not\_survive \\
 & Dixit & 98 & magnitude & -0.314 & [-0.498, -0.112] & survives \\
 & Dixit & 98 & centroid+QC & -0.353 & [-0.539, -0.140] & survives \\
 & Dixit & 98 & edistance+QC & -0.194 & [-0.436, +0.072] & does\_not\_survive \\
 & Dixit & 98 & centroid+edistance+QC & -0.132 & [-0.410, +0.151] & does\_not\_survive \\
 & Papalexi & 24 & magnitude & -0.248 & [-0.584, +0.142] & does\_not\_survive \\
 & Papalexi & 24 & centroid+QC & -0.235 & --- & descriptive\_small\_n \\
 & Papalexi & 24 & edistance+QC & -0.349 & --- & descriptive\_small\_n \\
 & Papalexi & 24 & centroid+edistance+QC & -0.394 & --- & descriptive\_small\_n \\
 & Replogle & 1,832 & magnitude & -0.250 & [-0.288, -0.208] & survives \\
 & Replogle & 1,832 & centroid+QC & -0.297 & [-0.337, -0.253] & survives \\
 & Replogle & 1,832 & edistance+QC & -0.287 & [-0.328, -0.239] & survives \\
 & Replogle & 1,832 & centroid+edistance+QC & -0.303 & [-0.343, -0.260] & survives$^\star$ \\
 \hline
ATF4 & Norman & 236 & magnitude & +0.395 & [+0.275, +0.487] & survives \\
 & Norman & 236 & centroid+QC & +0.140 & [+0.008, +0.258] & indeterminate \\
 & Norman & 236 & edistance+QC & +0.146 & [+0.016, +0.262] & indeterminate \\
 & Norman & 236 & centroid+edistance+QC & +0.123 & [-0.007, +0.251] & does\_not\_survive \\
 & Adamson UPR & 87 & magnitude & +0.250 & [-0.018, +0.473] & does\_not\_survive \\
 & Adamson UPR & 87 & centroid+QC & +0.340 & [+0.052, +0.534] & survives \\
 & Adamson UPR & 87 & edistance+QC & +0.315 & [+0.040, +0.512] & survives \\
 & Adamson UPR & 87 & centroid+edistance+QC & +0.150 & [-0.088, +0.436] & does\_not\_survive \\
 & Dixit & 98 & magnitude & -0.176 & [-0.356, +0.009] & does\_not\_survive \\
 & Dixit & 98 & centroid+QC & -0.244 & [-0.438, -0.032] & survives \\
 & Dixit & 98 & edistance+QC & -0.218 & [-0.424, +0.007] & does\_not\_survive \\
 & Dixit & 98 & centroid+edistance+QC & -0.305 & [-0.483, -0.065] & survives \\
 & Papalexi & 24 & magnitude & +0.009 & [-0.431, +0.497] & does\_not\_survive \\
 & Papalexi & 24 & centroid+QC & -0.152 & --- & descriptive\_small\_n \\
 & Papalexi & 24 & edistance+QC & -0.259 & --- & descriptive\_small\_n \\
 & Papalexi & 24 & centroid+edistance+QC & -0.279 & --- & descriptive\_small\_n \\
 & Replogle & 1,832 & magnitude & -0.136 & [-0.175, -0.096] & survives \\
 & Replogle & 1,832 & centroid+QC & -0.106 & [-0.148, -0.062] & survives \\
 & Replogle & 1,832 & edistance+QC & -0.106 & [-0.148, -0.063] & survives \\
 & Replogle & 1,832 & centroid+edistance+QC & -0.111 & [-0.152, -0.068] & survives$^\star$ \\
\hline
\end{tabular}
\end{table*}
\clearpage

\begin{table*}[ht]
\centering
\caption*{Table \ref{tab:stress_markers} Continued}
\label{tab:stress_markers2}
\smallskip
\begin{tabular}{lcccccc}
\hline
Marker & Dataset & $n$ & Model & $\rho$ & 95\% CI & Status \\
\hline
XBP1 & Norman & 236 & magnitude & +0.287 & [+0.159, +0.396] & survives \\
 & Norman & 236 & centroid+QC & +0.089 & [-0.044, +0.212] & does\_not\_survive \\
 & Norman & 236 & edistance+QC & +0.062 & [-0.071, +0.187] & does\_not\_survive \\
 & Norman & 236 & centroid+edistance+QC & +0.009 & [-0.113, +0.156] & does\_not\_survive \\
 & Adamson UPR & 87 & magnitude & +0.123 & [-0.087, +0.327] & does\_not\_survive \\
 & Adamson UPR & 87 & centroid+QC & +0.251 & [-0.009, +0.454] & does\_not\_survive \\
 & Adamson UPR & 87 & edistance+QC & +0.209 & [-0.034, +0.398] & does\_not\_survive \\
 & Adamson UPR & 87 & centroid+edistance+QC & +0.112 & [-0.097, +0.368] & does\_not\_survive \\
 & Dixit & 98 & magnitude & -0.139 & [-0.328, +0.064] & does\_not\_survive \\
 & Dixit & 98 & centroid+QC & -0.143 & [-0.341, +0.075] & does\_not\_survive \\
 & Dixit & 98 & edistance+QC & -0.110 & [-0.323, +0.125] & does\_not\_survive \\
 & Dixit & 98 & centroid+edistance+QC & -0.067 & [-0.291, +0.167] & does\_not\_survive \\
 & Papalexi & 24 & magnitude & -0.163 & [-0.527, +0.370] & does\_not\_survive \\
 & Papalexi & 24 & centroid+QC & -0.256 & --- & descriptive\_small\_n \\
 & Papalexi & 24 & edistance+QC & -0.373 & --- & descriptive\_small\_n \\
 & Papalexi & 24 & centroid+edistance+QC & -0.394 & --- & descriptive\_small\_n \\
 & Replogle & 1,832 & magnitude & -0.239 & [-0.279, -0.198] & survives \\
 & Replogle & 1,832 & centroid+QC & -0.196 & [-0.239, -0.153] & survives \\
 & Replogle & 1,832 & edistance+QC & -0.199 & [-0.240, -0.155] & survives \\
 & Replogle & 1,832 & centroid+edistance+QC & -0.201 & [-0.242, -0.158] & survives$^\star$ \\
 \hline
HSPA5 & Norman & 236 & magnitude & -0.036 & [-0.170, +0.104] & does\_not\_survive \\
 & Norman & 236 & centroid+QC & -0.125 & [-0.255, +0.020] & does\_not\_survive \\
 & Norman & 236 & edistance+QC & -0.151 & [-0.276, -0.007] & indeterminate \\
 & Norman & 236 & centroid+edistance+QC & -0.166 & [-0.285, -0.010] & indeterminate \\
 & Adamson UPR & 87 & magnitude & +0.048 & [-0.140, +0.244] & does\_not\_survive \\
 & Adamson UPR & 87 & centroid+QC & +0.116 & [-0.098, +0.315] & does\_not\_survive \\
 & Adamson UPR & 87 & edistance+QC & +0.056 & [-0.153, +0.260] & does\_not\_survive \\
 & Adamson UPR & 87 & centroid+edistance+QC & -0.055 & [-0.223, +0.217] & does\_not\_survive \\
 & Dixit & 98 & magnitude & -0.141 & [-0.345, +0.067] & does\_not\_survive \\
 & Dixit & 98 & centroid+QC & -0.083 & [-0.292, +0.120] & does\_not\_survive \\
 & Dixit & 98 & edistance+QC & +0.011 & [-0.230, +0.221] & does\_not\_survive \\
 & Dixit & 98 & centroid+edistance+QC & +0.001 & [-0.246, +0.210] & does\_not\_survive \\
 & Papalexi & 24 & magnitude & -0.152 & [-0.656, +0.394] & does\_not\_survive \\
 & Papalexi & 24 & centroid+QC & -0.205 & --- & descriptive\_small\_n \\
 & Papalexi & 24 & edistance+QC & -0.252 & --- & descriptive\_small\_n \\
 & Papalexi & 24 & centroid+edistance+QC & -0.264 & --- & descriptive\_small\_n \\
 & Replogle & 1,832 & magnitude & -0.128 & [-0.172, -0.083] & survives \\
 & Replogle & 1,832 & centroid+QC & -0.094 & [-0.139, -0.049] & does\_not\_survive \\
 & Replogle & 1,832 & edistance+QC & -0.074 & [-0.120, -0.028] & does\_not\_survive \\
 & Replogle & 1,832 & centroid+edistance+QC & -0.078 & [-0.123, -0.033] & does\_not\_survive \\
\hline
\end{tabular}
\end{table*}
\clearpage

\begin{table*}[ht]
\centering
\caption{Pathway-level partial correlations with Shesha\textsubscript{P}.
Rank-partial Spearman between coherence and Hallmark Apoptosis and p53 scores,
for every dataset--pathway pair under four covariate models ($5 \times 2 \times
4 = 40$ rows). The three quality-controlled models condition on effect size
together with per-perturbation mitochondrial fraction, detected genes, and total
counts; the magnitude-only rows condition on effect size alone and are reported
for comparison rather than passed through the survival gate (Status ---).
Adamson UPR is included at $n = 87$; the $n = 8$ pilot is excluded throughout.
Status follows the gate defined in Methods: an association is marked as
surviving when the interval excludes zero, $|\rho| > 0.1$, and BH-FDR $< 0.05$,
with Benjamini--Hochberg applied across the nine scored features within each
dataset and model. Norman p53 is indeterminate because its nearer interval bound
falls inside the knife-edge margin ($\varepsilon = 0.02$). Papalexi ($n = 24$)
is point-estimate only, as its quality-covariate models are rank-deficient in
bootstrap resamples. Hallmark UPR is not among the manuscript pathways and is
not shown.}
\label{tab:pathway_stress}
\smallskip
\resizebox{\linewidth}{!}{%
\begin{tabular}{lccccccc}
\hline
Pathway & Dataset & $n$ & Model & $\rho$ & 95\% CI & $p$ & Status \\
\hline
Apoptosis & Norman & 236 & magnitude & -0.349 & --- & $3.8 \times 10^{-8}$ & --- \\
Apoptosis & Norman & 236 & centroid+QC & -0.378 & [-0.503, -0.211] & $2.7 \times 10^{-9}$ & survives \\
Apoptosis & Norman & 236 & edistance+QC & -0.376 & [-0.492, -0.211] & $2.4 \times 10^{-9}$ & survives \\
Apoptosis & Norman & 236 & centroid+edistance+QC & -0.329 & [-0.459, -0.176] & $2.2 \times 10^{-7}$ & survives \\
Apoptosis & Adamson UPR & 87 & magnitude & -0.299 & --- & $0.005$ & --- \\
Apoptosis & Adamson UPR & 87 & centroid+QC & -0.356 & [-0.534, -0.134] & $9.7 \times 10^{-4}$ & survives \\
Apoptosis & Adamson UPR & 87 & edistance+QC & -0.308 & [-0.495, -0.073] & $0.004$ & survives \\
Apoptosis & Adamson UPR & 87 & centroid+edistance+QC & -0.141 & [-0.416, +0.053] & $0.192$ & does\_not\_survive \\
Apoptosis & Dixit & 98 & magnitude & -0.057 & --- & $0.577$ & --- \\
Apoptosis & Dixit & 98 & centroid+QC & -0.133 & [-0.325, +0.088] & $0.202$ & does\_not\_survive \\
Apoptosis & Dixit & 98 & edistance+QC & -0.146 & [-0.343, +0.088] & $0.151$ & does\_not\_survive \\
Apoptosis & Dixit & 98 & centroid+edistance+QC & -0.154 & [-0.382, +0.099] & $0.131$ & does\_not\_survive \\
Apoptosis & Papalexi & 24 & magnitude & -0.447 & --- & $0.032$ & --- \\
Apoptosis & Papalexi & 24 & centroid+QC & -0.363 & --- & $0.097$ & descriptive\_small\_n \\
Apoptosis & Papalexi & 24 & edistance+QC & -0.232 & --- & $0.275$ & descriptive\_small\_n \\
Apoptosis & Papalexi & 24 & centroid+edistance+QC & -0.221 & --- & $0.299$ & descriptive\_small\_n \\
Apoptosis & Replogle & 1,832 & magnitude & -0.294 & --- & $6.1 \times 10^{-38}$ & --- \\
Apoptosis & Replogle & 1,832 & centroid+QC & -0.291 & [-0.329, -0.248] & $5.6 \times 10^{-37}$ & survives \\
Apoptosis & Replogle & 1,832 & edistance+QC & -0.274 & [-0.313, -0.231] & $7.3 \times 10^{-33}$ & survives \\
Apoptosis & Replogle & 1,832 & centroid+edistance+QC & -0.289 & [-0.328, -0.248] & $1.5 \times 10^{-36}$ & survives \\
\hline
p53 & Norman & 236 & magnitude & -0.333 & --- & $1.7 \times 10^{-7}$ & --- \\
p53 & Norman & 236 & centroid+QC & -0.152 & [-0.274, -0.015] & $0.020$ & indeterminate \\
p53 & Norman & 236 & edistance+QC & -0.147 & [-0.262, -0.010] & $0.024$ & indeterminate \\
p53 & Norman & 236 & centroid+edistance+QC & -0.112 & [-0.237, +0.014] & $0.086$ & does\_not\_survive \\
p53 & Adamson UPR & 87 & magnitude & -0.063 & --- & $0.565$ & --- \\
p53 & Adamson UPR & 87 & centroid+QC & -0.071 & [-0.272, +0.147] & $0.524$ & does\_not\_survive \\
p53 & Adamson UPR & 87 & edistance+QC & -0.102 & [-0.296, +0.123] & $0.345$ & does\_not\_survive \\
p53 & Adamson UPR & 87 & centroid+edistance+QC & -0.124 & [-0.315, +0.121] & $0.254$ & does\_not\_survive \\
p53 & Dixit & 98 & magnitude & -0.247 & --- & $0.015$ & --- \\
p53 & Dixit & 98 & centroid+QC & -0.331 & [-0.490, -0.140] & $0.001$ & survives \\
p53 & Dixit & 98 & edistance+QC & -0.269 & [-0.451, -0.047] & $0.007$ & survives \\
p53 & Dixit & 98 & centroid+edistance+QC & -0.241 & [-0.452, -0.004] & $0.017$ & indeterminate \\
p53 & Papalexi & 24 & magnitude & -0.398 & --- & $0.060$ & --- \\
p53 & Papalexi & 24 & centroid+QC & -0.276 & --- & $0.214$ & descriptive\_small\_n \\
p53 & Papalexi & 24 & edistance+QC & -0.206 & --- & $0.335$ & descriptive\_small\_n \\
p53 & Papalexi & 24 & centroid+edistance+QC & -0.196 & --- & $0.360$ & descriptive\_small\_n \\
p53 & Replogle & 1,832 & magnitude & -0.400 & --- & $2.1 \times 10^{-71}$ & --- \\
p53 & Replogle & 1,832 & centroid+QC & -0.382 & [-0.415, -0.343] & $1.7 \times 10^{-64}$ & survives \\
p53 & Replogle & 1,832 & edistance+QC & -0.356 & [-0.396, -0.311] & $5.1 \times 10^{-56}$ & survives \\
p53 & Replogle & 1,832 & centroid+edistance+QC & -0.379 & [-0.413, -0.341] & $1.1 \times 10^{-63}$ & survives \\
\hline
\end{tabular}
}
\end{table*}
\clearpage

\begin{table*}[ht]
\centering
\caption{The apoptosis and p53 associations are not one result reported twice.
Hallmark Apoptosis and Hallmark p53 Pathway share $11$ of $113$ union genes
(Jaccard $0.097$), and their composite scores correlate weakly to moderately in
every dataset contributing to the survival gate ($\rho = 0.145$ to $0.423$).
Papalexi ($\rho = 0.723$) is excluded from the gate because its
quality-covariate models are rank-deficient at $n = 24$, and it contributes to
neither pathway result. Dixit is the clearest case for independence: the two
scores correlate at $0.423$, yet the partial correlations diverge, with p53
surviving conditioning and apoptosis null. The gene-set Jaccard is a property of
the Hallmark sets and is identical across datasets; the score correlation is
dataset-specific and is the informative column.}
\label{tab:apoptosis_p53_independence}
\smallskip
\begin{tabular}{lccccc}
\hline
Dataset & $n$ & score $\rho$ & Jaccard (full) & shared/union & gate-relevant \\
\hline
Norman & 236 & 0.153 & 0.097 & 11/113 & true \\
Adamson UPR & 87 & 0.145 & 0.097 & 11/113 & true \\
Dixit & 98 & 0.423 & 0.097 & 11/113 & true \\
Papalexi & 24 & 0.723 & 0.097 & 11/113 & false \\
Replogle & 1,832 & 0.295 & 0.097 & 11/113 & true \\
\hline
\end{tabular}
\end{table*}
\clearpage

\begin{table*}[ht]
\centering
\caption{E-distance competitor on the frozen 50-PC embedding. $\rho$ is Spearman; residual fractions are $1-R^{2}$ from rank-OLS of $\mathrm{rank}(S_p)$ on $\mathrm{rank}(\mathrm{size})$. CIs use 2{,}000 bootstrap replicates, seed 320. Dixit's residual drop $0.289\rightarrow 0.108$ is the dispersion correction.}
\label{tab:edistance_competitor}
\smallskip
\begin{tabular}{lcccccc}
\hline
Dataset & $n$ & $\rho(M_p,\mathrm{Ed})$ & $\rho(S_p,M_p)$ & $\rho(S_p,\mathrm{Ed})$ & frac$|$cent. & frac$|$Ed \\
\hline
Norman & 236 & 0.998 & 0.950 & 0.960 & 0.098 & 0.078 \\
Adamson UPR & 87 & 0.995 & 0.951 & 0.971 & 0.095 & 0.058 \\
Adamson pilot & 8 & 1.000 & 0.976 & 0.976 & 0.047 & 0.047 \\
Dixit & 98 & 0.950 & 0.843 & 0.945 & 0.289 & 0.108 \\
Papalexi & 24 & 0.991 & 0.945 & 0.955 & 0.107 & 0.088 \\
Replogle & 1,832 & 0.994 & 0.977 & 0.980 & 0.046 & 0.039 \\
\hline
\end{tabular}
\end{table*}
\clearpage

\begin{table*}[ht]
\centering
\caption{TRADE transcriptome-wide impact (TWI) competitor (R \texttt{ashr} half-uniform mixture). The informative contrast is $\rho(S_p,\mathrm{TWI})$ vs $\rho(M_p,\mathrm{TWI})$, not $\rho(S_p,\mathrm{TWI})$ vs $\rho(S_p,M_p)$. Pilot $n=8$ is uninformative.}
\label{tab:trade_competitor}
\smallskip
\begin{tabular}{lccccc}
\hline
Dataset & $n$ & $\rho(S_p,M_p)$ & $\rho(S_p,\mathrm{TWI})$ & $\rho(M_p,\mathrm{TWI})$ & $\rho(S_p,\mathrm{Ed})$ \\
\hline
Norman & 236 & 0.950 & 0.688 & 0.812 & 0.960 \\
Adamson UPR & 87 & 0.951 & 0.954 & 0.968 & 0.971 \\
Adamson pilot & 8 & 0.976 & 0.857 & 0.833 & 0.976 \\
Dixit & 98 & 0.843 & 0.452 & 0.513 & 0.945 \\
Papalexi & 24 & 0.945 & 0.858 & 0.832 & 0.955 \\
Replogle & 1,832 & 0.977 & 0.912 & 0.919 & 0.980 \\
\hline
\end{tabular}
\end{table*}
\clearpage

\begin{table*}[ht]
\centering
\caption{TWI versus Hallmark pathway-score diagnostic. Checks whether TRADE's transcriptome-wide impact merely restates the apoptosis / p53 scores used in Approach A.}
\label{tab:twi_pathway}
\smallskip
\begin{tabular}{lcccc}
\hline
Dataset & Pathway & $n$ & $\rho(\mathrm{TWI},\mathrm{score})$ & $p$ \\
\hline
Adamson UPR & Apoptosis & 87 & -0.061 & $0.576$ \\
Adamson UPR & ROS & 87 & -0.230 & $0.032$ \\
Adamson UPR & UPR & 87 & +0.335 & $0.002$ \\
Adamson UPR & mTORC1 & 87 & +0.113 & $0.298$ \\
Adamson UPR & p53 & 87 & +0.114 & $0.294$ \\
\hline
Adamson pilot & Apoptosis & 8 & +0.167 & $0.693$ \\
Adamson pilot & ROS & 8 & -0.643 & $0.086$ \\
Adamson pilot & UPR & 8 & -0.024 & $0.955$ \\
Adamson pilot & mTORC1 & 8 & -0.881 & $0.004$ \\
Adamson pilot & p53 & 8 & +0.381 & $0.352$ \\
\hline
Dixit & Apoptosis & 98 & -0.218 & $0.031$ \\
Dixit & ROS & 98 & +0.096 & $0.345$ \\
Dixit & UPR & 98 & -0.157 & $0.124$ \\
Dixit & mTORC1 & 98 & -0.351 & $3.9 \times 10^{-4}$ \\
Dixit & p53 & 98 & -0.403 & $4.0 \times 10^{-5}$ \\
\hline
Norman & Apoptosis & 236 & -0.002 & $0.970$ \\
Norman & ROS & 236 & -0.391 & $4.6 \times 10^{-10}$ \\
Norman & UPR & 236 & -0.085 & $0.192$ \\
Norman & mTORC1 & 236 & -0.780 & $1.4 \times 10^{-49}$ \\
Norman & p53 & 236 & +0.542 & $1.9 \times 10^{-19}$ \\
\hline
Papalexi & Apoptosis & 24 & -0.458 & $0.024$ \\
Papalexi & ROS & 24 & +0.009 & $0.968$ \\
Papalexi & UPR & 24 & -0.254 & $0.231$ \\
Papalexi & mTORC1 & 24 & +0.047 & $0.828$ \\
Papalexi & p53 & 24 & -0.271 & $0.200$ \\
\hline
Replogle & Apoptosis & 1,832 & +0.203 & $1.8 \times 10^{-18}$ \\
Replogle & ROS & 1,832 & -0.132 & $1.3 \times 10^{-8}$ \\
Replogle & UPR & 1,832 & -0.150 & $1.2 \times 10^{-10}$ \\
Replogle & mTORC1 & 1,832 & -0.574 & $5.9 \times 10^{-161}$ \\
Replogle & p53 & 1,832 & +0.260 & $1.2 \times 10^{-29}$ \\
\hline
\end{tabular}
\end{table*}
\clearpage

\begin{table*}[ht]
\centering
\caption{Restricting to responder cells leaves the coherence--magnitude
relationship unchanged. Spearman $\rho$ between Shesha\textsubscript{P} and
effect magnitude computed on all assigned cells and on cells classified as
responders by perturbation scoring, together with the rank correlation between
the two sets of perturbation-level coherence values. Changes to the
coherence--magnitude correlation span $-0.048$ to $+0.026$ with no consistent
direction, so unperturbed cells in assigned guide pools do not account for the
coupling between the two quantities. Rank preservation is classified against
pre-specified thresholds (preserved, $\rho \ge 0.9$; change, $\rho < 0.7$);
Dixit is the only dataset whose perturbation rankings change materially.
Perturbations are dropped when too few responder cells remain, which accounts
for the difference between the two sample-size columns and is largest in
Replogle ($171$ of $1{,}832$). Because responder classification selects cells
whose displacement aligns with an estimated perturbation signature, coherence
itself necessarily rises under this restriction; that increase is not reported
as a result. These values are a sensitivity analysis and do not replace the
frozen estimates, which use all assigned cells throughout.}
\label{tab:efficiency_ps}
\smallskip
\begin{tabular}{lccccccc}
\hline
Dataset & $n$ all & $n$ resp. & $\rho$ all & $\rho$ resp. & rank $\rho$ & 95\% CI & Verdict \\
\hline
Norman & 236 & 234 & 0.950 & 0.902 & 0.918 & [+0.890, +0.937] & preserved \\
Adamson UPR & 87 & 80 & 0.951 & 0.951 & 0.865 & [+0.789, +0.909] & partial \\
Dixit & 98 & 96 & 0.843 & 0.847 & 0.642 & [+0.502, +0.745] & change \\
Papalexi & 24 & 24 & 0.945 & 0.971 & 0.785 & [+0.516, +0.907] & partial \\
Replogle & 1,832 & 1,661 & 0.977 & 0.976 & 0.916 & [+0.908, +0.923] & preserved \\
\hline
\end{tabular}
\end{table*}
\clearpage

\begin{table*}[ht]
\centering
\caption{Mixscape efficiency-filter arm, Norman only. Repeating the
responder-restriction analysis with Mixscape as an independent caller. Mixscape
returns no responders for $87$ of the $236$ perturbations, so the comparison is
made on the remaining $149$ and both columns are computed on that subset. The
all-cell correlation of $0.853$ is therefore a property of this subset and is not
comparable to the frozen Norman value of $0.950$ computed on all $236$
perturbations. Within the subset, restricting to responders changes the
coherence--magnitude correlation by $+0.030$, opposite in sign to the $-0.048$
obtained with perturbation scoring on the same dataset; the two callers agree on
only $68.4$ percent of individual cells
(Table~\ref{tab:responder_agreement}). Perturbation rankings are preserved
between the full and responder-only analyses within the subset
($\rho = 0.891$).}
\label{tab:efficiency_mixscape}
\smallskip
\begin{tabular}{lcccccc}
\hline
Dataset & $n$ scored & $n$ resp. & $\rho$ all & $\rho$ resp. & rank $\rho$ & Note \\
\hline
Norman & 149 & 149 & 0.853 & 0.883 & 0.891 & 149/236 denominator; not a frozen baseline \\
\hline
\end{tabular}
\end{table*}
\clearpage

\begin{table*}[ht]
\centering
\caption{The two standard responder classifiers disagree on roughly half the
cells they call. Per-cell responder assignments from perturbation scoring and
from Mixscape, compared on the $23{,}290$ Norman cells scored by both. Mixscape
is the more conservative caller, and its responder set is close to but not
nested within that of perturbation scoring: $2{,}512$ cells are called
responders by Mixscape alone. Because which cells count as responders is
substantially method-defined, responder-only reanalysis is reported as a
sensitivity analysis conditional on the classifier rather than as a correction
applied to the frozen estimates, and the direction of the change in the
coherence--magnitude correlation differs between the two callers in this dataset
(Fig.~\ref{fig:efficiency_filter}).}
\label{tab:responder_agreement}
\smallskip
\begin{tabular}{lc}
\hline
Quantity & Value \\
\hline
$n$ shared cells & 23,290 \\
PS responder fraction & 0.531 \\
Mixscape responder fraction & 0.431 \\
Both & 7,521 \\
PS only & 4,836 \\
Mixscape only & 2,512 \\
Jaccard & 0.506 \\
Cohen's $\kappa$ & 0.374 \\
Raw agreement & 0.684 \\
\hline
\end{tabular}
\end{table*}
\clearpage

\begin{table*}[ht]
\centering
\caption{Coherence does not predict between-guide agreement beyond effect size.
Split-half agreement is circular as a measure of reproducibility, because both
halves are drawn from the same cells and compared against the same control
reference. We therefore used independent guides targeting the same gene, which
are separate reagents in separate cells. For each gene with at least two guides
passing the cell-count filter, we correlated gene-level Shesha\textsubscript{P}
with the cosine between the two guides' mean response directions, conditioning
on effect magnitude. The partial correlation is null in both datasets. In
Replogle this endpoint is available for $104$ of the $1{,}832$ scored
perturbations, so it covers a small and not necessarily representative fraction
of the screen. The third row asks a different question: whether guides targeting
the same gene give similar coherence values at all. They do
(ICC$(1) = 0.890$), but effect magnitude agrees between guides to the same
degree (ICC$(1) = 0.889$, difference $0.001$), so between-guide consistency in
coherence carries no information beyond consistency in effect size. Pairwise
Spearman correlations over unordered guide pairs are not reported, because the
value depends on which guide of each pair is assigned to $x$; ICC$(1)$ is the
well-defined concordance statistic for this design. Both datasets share a
control reference across guides, so this is independent-reagent rather than
fully independent replication.}
\label{tab:between_guide}
\smallskip
\resizebox{\linewidth}{!}{%
\begin{tabular}{lccccc}
\hline
Dataset & Endpoint & $n$ & Stat & 95\% CI & Note \\
\hline
Replogle & gene Sp $\sim$ between-guide cosine $|$ magnitude & 104 genes & $0.006$ & $[-0.220,\ 0.225]$ & null; $104/1{,}832$ of the screen \\
Papalexi & gene Sp $\sim$ between-guide cosine $|$ magnitude & 23 genes & $0.183$ & $[-0.317,\ 0.644]$ & null (GEO rebuild) \\
Papalexi & ICC(1) of guide-level Sp & 23 genes & $0.890$ & $[0.752,\ 0.951]$ & ICC(mag)$=0.889$; $\Delta=0.001$ \\
\hline
\end{tabular}
}
\end{table*}
\clearpage

\begin{table*}[ht]
\centering
\caption{Papalexi GEO reporting ladder (GSE153056, no pertpy). The prior version $n=25$ is the gene-level inventory at \texttt{MIN\_CELLS}$=10$; the freeze uses $50$ throughout and yields $n=24$ --- the same cutoff change that took Dixit from $153$ to $98$. The independent rebuild reproduces frozen gene Sp at $\rho=1.000$ ($n=24$).}
\label{tab:papalexi_geo_ladder}
\smallskip
\begin{tabular}{lc}
\hline
Rung & Count \\
\hline
Guides deposited (GSE153056) & 112 \\
Guides / genes at \texttt{MIN\_CELLS}=10 (sensitivity) & 96 / 25 \\
Guides / genes at frozen \texttt{MIN\_CELLS}=50 & 83 / 24 \\
GEO vs frozen gene Sp ($n$ shared) & $\rho=1.000$ ($n=24$) \\
\hline
\end{tabular}
\end{table*}
\clearpage

\begin{table*}[ht]
\centering
\caption{Within-representation magnitude--Shesha correlations (one of the two quantities formerly mixed in \texttt{tab:scgpt\_comparison}). Frozen linear is the manuscript Euclidean column (scanpy PCA or sparse TruncatedSVD according to the dataset-specific backend); E-distance and scGPT are the same relationship scored in those representations. scGPT omitted for Papalexi ($n=24$, GPU ranks not reproducible) and the Adamson pilot ($n=8$). Diffusion-map / PHATE within-arm rows appear when those CSVs exist; concordance against the frozen linear embedding is Table~\ref{tab:embedding_concordance}.}
\label{tab:embedding_magnitude}
\smallskip
\begin{tabular}{lcccc}
\hline
Dataset & $n$ & Frozen linear $\rho(S_p,M_p)$ & E-distance $\rho$ & scGPT $\rho$ \\
\hline
Norman & 236 & 0.950 & 0.960 & 0.935 \\
Adamson UPR & 87 & 0.951 & 0.971 & 0.912 \\
Adamson pilot & 8 & 0.976 & 0.976 & omit \\
Dixit & 98 & 0.843 & 0.945 & 0.854 \\
Papalexi & 24 & 0.945 & 0.955 & omit \\
Replogle & 1,832 & 0.977 & 0.980 & 0.851 \\
\hline
\end{tabular}
\end{table*}
\clearpage

\begin{table*}[ht]
\centering
\caption{Three-arm embedding concordance against the frozen linear embedding (scanpy PCA or
sparse TruncatedSVD, according to the frozen dataset-specific backend). Left block: Shesha\textsubscript{P}
computed in each alternative embedding against frozen Shesha\textsubscript{P}. Right block: effect magnitude
computed in each embedding against frozen effect magnitude. Diffusion maps are the only
representation in which coherence transfers as well as or better than magnitude in every dataset;
under PHATE and scGPT magnitude transfers better throughout, and the PHATE gap is widest where
coherence is weakest (Replogle $0.44$ vs $0.68$). MDS convergence does not account for the PHATE
pattern (Table~\ref{tab:phate_convergence}). scGPT is omitted for Papalexi ($n=24$) and the
Adamson pilot ($n=8$): coherence rank statistics are not numerically reproducible under GPU
inference at those sample sizes.}
\label{tab:embedding_concordance}
\smallskip
\begin{tabular}{lccccccc}
\hline
Dataset & $n$ & DM $S_p$ & PHATE $S_p$ & scGPT $S_p$ & DM $M_p$ & PHATE $M_p$ & scGPT $M_p$ \\
\hline
Norman & 236 & 0.858 & 0.882 & 0.514 & 0.886 & 0.891 & 0.629 \\
Adamson UPR & 87 & 0.948 & 0.705 & 0.829 & 0.949 & 0.775 & 0.873 \\
Adamson pilot & 8 & 0.690 & 0.667 & omit & 0.476 & 0.762 & omit \\
Dixit & 98 & 0.641 & 0.684 & 0.471 & 0.538 & 0.794 & 0.549 \\
Papalexi & 24 & 0.852 & 0.506 & omit & 0.808 & 0.844 & omit \\
Replogle & 1,832 & 0.920 & 0.439 & 0.750 & 0.902 & 0.679 & 0.784 \\
\hline
\end{tabular}
\end{table*}
\clearpage

\begin{table*}[ht]
\centering
\caption{PHATE convergence control. MDS solver convergence does not predict concordance with the
frozen embedding: the three datasets whose embeddings converged span $\rho = 0.51$ to $0.68$,
while the three that did not span $0.44$ to $0.88$. Norman failed to converge and has the highest
PHATE concordance in the set; Papalexi converged and has one of the lowest. The PHATE degradation
in Figure~4 is therefore a property of the embedding rather than optimizer failure. No rows are
excluded.}
\label{tab:phate_convergence}
\smallskip
\begin{tabular}{lcccc}
\hline
Dataset & $n$ & Solver & Converged & $\rho$ vs frozen $S_p$ \\
\hline
Adamson pilot & 8 & SGD-MDS & yes & 0.667 \\
Dixit & 98 & SGD-MDS & yes & 0.684 \\
Papalexi & 24 & SGD-MDS & yes & 0.506 \\
Norman & 236 & SGD-MDS & no ($-3.5\%$) & 0.882 \\
Adamson UPR & 87 & SGD-MDS & no ($-4.0\%$) & 0.705 \\
Replogle & 1{,}832 & SGD-MDS & no ($-5.1\%$) & 0.439 \\
\hline
\end{tabular}
\end{table*}
\clearpage

\begin{table*}[ht]
\centering
\caption{Approach B covariate balance before and after caliper matching on residual Shesha quartiles. Gate: $|\mathrm{SMD}|\le 0.25$ (inclusive) and $\ge 15$ pairs. Magnitude balances in both datasets; cell-quality covariates do not (Replogle $n$ genes $-0.46$; Norman UMIs / genes $<-1.2$). Replogle $n$ cells after matching is $+0.249$ and therefore passes. Because the gate fails, Approach B is dropped and Approach A carries the pathway result.}
\label{tab:approach_b_balance}
\smallskip
\begin{tabular}{lccccc}
\hline
Dataset & Stage & Covariate & $n$ pairs & SMD & Gate \\
\hline
Replogle & before & magnitude & 458 & +0.084 & pass \\
Replogle & before & n\_cells & 458 & +0.266 & fail \\
Replogle & before & n\_counts & 458 & -0.178 & pass \\
Replogle & before & n\_genes & 458 & -0.489 & fail \\
Replogle & before & percent\_mito & 458 & +0.100 & pass \\
\hline
Replogle & after & magnitude & 422 & +0.013 & pass \\
Replogle & after & n\_cells & 422 & +0.249 & pass \\
Replogle & after & n\_counts & 422 & -0.153 & pass \\
Replogle & after & n\_genes & 422 & -0.460 & fail \\
Replogle & after & percent\_mito & 422 & +0.106 & pass \\
\hline\hline
Norman & before & magnitude & 59 & +0.046 & pass \\
Norman & before & n\_cells & 59 & -0.566 & fail \\
Norman & before & n\_counts & 59 & -1.172 & fail \\
Norman & before & n\_genes & 59 & -1.250 & fail \\
Norman & before & percent\_mito & 59 & +0.141 & pass \\
\hline
Norman & after & magnitude & 43 & -0.024 & pass \\
Norman & after & n\_cells & 43 & -0.533 & fail \\
Norman & after & n\_counts & 43 & -1.248 & fail \\
Norman & after & n\_genes & 43 & -1.342 & fail \\
Norman & after & percent\_mito & 43 & +0.134 & pass \\
\hline
\end{tabular}
\end{table*}
\clearpage


\begin{table*}[ht]
\centering
\caption{Calibration of the attempted simulation arm. A perturbation simulator
was implemented to test whether coherence distinguishes heterogeneity arising
from variable perturbation efficiency from heterogeneity arising from cells
engaging different transcriptional programs. It was not usable, for two reasons
recorded here. The effect scale needed to reproduce observed coherence depends
on which reference perturbation the pipeline selects: at a fixed
\texttt{effect\_scale} of $1.5$, simulated coherence was $0.593$ against the
SLU7 reference and $0.144$ against GRPEL1, a fourfold change with no
modification to the generator. No portable calibration constant therefore
exists, and re-tuning the scale to whichever reference the current freeze
selects would be fitting a constant to a known target. Separately, the generator
could not hold multi-program response magnitude fixed while varying the number
of programs, so the identifiability question the arm was built to answer was not
testable as constructed. The comparison of simulated against observed coherence
is reported for transparency and is not a pre-specified acceptance criterion:
the tolerance was set after the arm had already produced a passing value. Arm
tables are retained in the analysis outputs but are not cited in the manuscript.}
\label{tab:synthetic_calibration}
\smallskip
\begin{tabular}{ll}
\hline
Item & Value \\
\hline
Simulator configuration & \texttt{2026-07-25.9} \\
\texttt{effect\_scale} & $1.5$ \\
Simulated coherence, SLU7 reference & $0.593$ \\
Simulated coherence, GRPEL1 reference & $0.144$ \\
Observed coherence, GRPEL1 & $0.397$ \\
Efficiency arm, coherence across 16 cells & $0.110$--$0.144$ \\
Manuscript status & attempted, not used \\
\hline
\end{tabular}

\smallskip
\raggedright
The simulation arm is isolated from the real-data pipeline: scoring calls
\texttt{preprocess(..., matrix\_is\_log=False)} and \texttt{calculate\_sp} only,
and the real-data freeze remains \texttt{2026-07-29.1}.
\end{table*}
\clearpage

\begin{table*}[ht]
\centering
\caption{Centroid-based distances do not approximate the scMAGeCK
perturbation-response score. Spearman correlations in Replogle 2022 between
scMAGeCK PS and two per-cell distance measures sometimes used as substitutes for
it: mean Euclidean distance from the control centroid, and the same distance
whitened by the inverse control covariance. Both correlate weakly with the
algorithm-derived score, so a centroid distance should not be treated as a
lower-fidelity approximation to PS. Responder classification in this paper uses
scMAGeCK PS directly (\textbf{Fig.~\ref{fig:efficiency_filter}}).}
\label{tab:ps_proxy_caution}
\smallskip
\begin{tabular}{lc}
\hline
Comparison & $\rho$ \\
\hline
Euclidean centroid vs scMAGeCK PS & $0.097$ \\
Mahalanobis centroid vs scMAGeCK PS & $0.149$ \\
\hline
\end{tabular}
\end{table*}
\clearpage

%% file: biblio.bib
@software{shesha2026,
  title = {{Shesha: Self-Consistency Metrics for Representational Stability}},
  author = {Raju, Prashant C.},
  year = {2026},
  howpublished = {Zenodo},
  doi = {10.5281/zenodo.18227453},
  url = {https://doi.org/10.5281/zenodo.18227453},
  copyright = {MIT License},
  notes = {10.5281/zenodo.18227453}
}

@article{raju2026geometric,
  title = {{Geometric Stability: The Missing Axis of Representations}},
  author = {Raju, Prashant C.},
  journal = {arXiv preprint arXiv:2601.09173},
  year = {2026}
}

@inproceedings{raju2026canary,
  title = {{The Geometric Canary: Predicting Steerability and Detecting Drift via Representational Stability}},
  author = {Raju, Prashant C.},
  booktitle= {ICML Mechanistic Interpretability Workshop},
  year= {2026}
}

@article{Nadig2024,
  title = {Transcriptome-wide analysis of differential expression in perturbation atlases},
  volume = {57},
  ISSN = {1546-1718},
  url = {http://dx.doi.org/10.1038/s41588-025-02169-3},
  DOI = {10.1038/s41588-025-02169-3},
  number = {5},
  journal = {Nature Genetics},
  publisher = {Springer Science and Business Media LLC},
  author = {Nadig,  Ajay and Replogle,  Joseph M. and Pogson,  Angela N. and Murthy,  Mukundh and McCarroll,  Steven A. and Weissman,  Jonathan S. and Robinson,  Elise B. and O’Connor,  Luke J.},
  year = {2025},
  month = apr,
  pages = {1228–1237}
}

@article{Haghverdi2016,
  title = {Diffusion pseudotime robustly reconstructs lineage branching},
  volume = {13},
  ISSN = {1548-7105},
  url = {http://dx.doi.org/10.1038/nmeth.3971},
  DOI = {10.1038/nmeth.3971},
  number = {10},
  journal = {Nature Methods},
  publisher = {Springer Science and Business Media LLC},
  author = {Haghverdi,  Laleh and B\"{u}ttner,  Maren and Wolf,  F Alexander and Buettner,  Florian and Theis,  Fabian J},
  year = {2016},
  month = Aug,
  pages = {845–848}
}

@article{Haghverdi2015,
  title = {Diffusion maps for high-dimensional single-cell analysis of differentiation data},
  volume = {31},
  ISSN = {1367-4803},
  url = {http://dx.doi.org/10.1093/bioinformatics/btv325},
  DOI = {10.1093/bioinformatics/btv325},
  number = {18},
  journal = {Bioinformatics},
  publisher = {Oxford University Press (OUP)},
  author = {Haghverdi,  Laleh and Buettner,  Florian and Theis,  Fabian J.},
  year = {2015},
  month = May,
  pages = {2989–2998}
}

@article{Peidli2022,
  title = {{scPerturb: harmonized single-cell perturbation data}},
  volume = {21},
  ISSN = {1548-7105},
  url = {http://dx.doi.org/10.1038/s41592-023-02144-y},
  DOI = {10.1038/s41592-023-02144-y},
  number = {3},
  journal = {Nature Methods},
  publisher = {Springer Science and Business Media LLC},
  author = {Peidli,  Stefan and Green,  Tessa D. and Shen,  Ciyue and Gross,  Torsten and Min,  Joseph and Garda,  Samuele and Yuan,  Bo and Schumacher,  Linus J. and Taylor-King,  Jake P. and Marks,  Debora S. and Luna,  Augustin and Bl\"{u}thgen,  Nils and Sander,  Chris},
  year = {2024},
  month = jan,
  pages = {531–540}
}

@article{Stephens2016,
  title = {False discovery rates: a new deal},
  ISSN = {1468-4357},
  url = {http://dx.doi.org/10.1093/biostatistics/kxw041},
  DOI = {10.1093/biostatistics/kxw041},
  journal = {Biostatistics},
  publisher = {Oxford University Press (OUP)},
  author = {Stephens,  Matthew},
  year = {2016},
  month = Oct,
  pages = {kxw041}
}

@article{Seale2026,
  title = {MUSICiAn: genome-wide identification of genes involved in DNA repair via control-free mutational spectra analysis},
  volume = {8},
  ISSN = {2631-9268},
  url = {http://dx.doi.org/10.1093/nargab/lqaf202},
  DOI = {10.1093/nargab/lqaf202},
  number = {1},
  journal = {NAR Genomics and Bioinformatics},
  publisher = {Oxford University Press (OUP)},
  author = {Seale,  Colm and Barazas,  Marco and van Schendel,  Robin and Tijsterman,  Marcel and Gon\c{c}alves,  Joana P},
  year = {2026},
  month = Jan 
}

@article{szekely_2013_energy,
  title = {Energy statistics: A class of statistics based on distances},
  volume = {143},
  ISSN = {0378-3758},
  url = {http://dx.doi.org/10.1016/j.jspi.2013.03.018},
  DOI = {10.1016/j.jspi.2013.03.018},
  number = {8},
  journal = {Journal of Statistical Planning and Inference},
  publisher = {Elsevier BV},
  author={Sz{\'e}kely, G{\'a}bor J and Rizzo, Maria L},
  year = {2013},
  month = Aug,
  pages = {1249–1272}
}

@article{doench_2016_optimized,
  title = {Optimized sgRNA design to maximize activity and minimize off-target effects of CRISPR-Cas9},
  volume = {34},
  ISSN = {1546-1696},
  url = {http://dx.doi.org/10.1038/nbt.3437},
  DOI = {10.1038/nbt.3437},
  number = {2},
  journal = {Nature Biotechnology},
  publisher = {Springer Science and Business Media LLC},
  author = {Doench,  John G and Fusi,  Nicolo and Sullender,  Meagan and Hegde,  Mudra and Vaimberg,  Emma W and Donovan,  Katherine F and Smith,  Ian and Tothova,  Zuzana and Wilen,  Craig and Orchard,  Robert and Virgin,  Herbert W and Listgarten,  Jennifer and Root,  David E},
  year = {2016},
  month = Jan,
  pages = {184–191}
}

@article{luecken_2019_bestpractices,
  title = {Current best practices in single‐cell RNA‐seq analysis: a tutorial},
  volume = {15},
  ISSN = {1744-4292},
  url = {http://dx.doi.org/10.15252/msb.20188746},
  DOI = {10.15252/msb.20188746},
  number = {6},
  journal = {Molecular Systems Biology},
  publisher = {Springer Science and Business Media LLC},
  author = {Luecken,  Malte D and Theis,  Fabian J},
  year = {2019},
  month = Jun
}

@article{saelens_2019_trajectory,
  title = {A comparison of single-cell trajectory inference methods},
  volume = {37},
  ISSN = {1546-1696},
  url = {http://dx.doi.org/10.1038/s41587-019-0071-9},
  DOI = {10.1038/s41587-019-0071-9},
  number = {5},
  journal = {Nature Biotechnology},
  publisher = {Springer Science and Business Media LLC},
  author = {Saelens,  Wouter and Cannoodt,  Robrecht and Todorov,  Helena and Saeys,  Yvan},
  year = {2019},
  month = Apr,
  pages = {547–554}
}

@article{heumos_2023_bestpractices,
  title = {Best practices for single-cell analysis across modalities},
  volume = {24},
  ISSN = {1471-0064},
  url = {http://dx.doi.org/10.1038/s41576-023-00586-w},
  DOI = {10.1038/s41576-023-00586-w},
  number = {8},
  journal = {Nature Reviews Genetics},
  publisher = {Springer Science and Business Media LLC},
  author = {Heumos,  Lukas and Schaar,  Anna C. and Lance,  Christopher and Litinetskaya,  Anastasia and Drost,  Felix and Zappia,  Luke and L\"{u}cken,  Malte D. and Strobl,  Daniel C. and Henao,  Juan and Curion,  Fabiola and Aliee,  Hananeh and Ansari,  Meshal and Badia-i-Mompel,  Pau and B\"{u}ttner,  Maren and Dann,  Emma and Dimitrov,  Daniel and Dony,  Leander and Frishberg,  Amit and He,  Dongze and Hediyeh-zadeh,  Soroor and Hetzel,  Leon and Ibarra,  Ignacio L. and Jones,  Matthew G. and Lotfollahi,  Mohammad and Martens,  Laura D. and M\"{u}ller,  Christian L. and Nitzan,  Mor and Ostner,  Johannes and Palla,  Giovanni and Patro,  Rob and Piran,  Zoe and Ramírez-Suástegui,  Ciro and Saez-Rodriguez,  Julio and Sarkar,  Hirak and Schubert,  Benjamin and Sikkema,  Lisa and Srivastava,  Avi and Tanevski,  Jovan and Virshup,  Isaac and Weiler,  Philipp and Schiller,  Herbert B. and Theis,  Fabian J.},
  year = {2023},
  month = Mar,
  pages = {550–572}
}

@article{chari_2021_specious,
  title = {The specious art of single-cell genomics},
  volume = {19},
  ISSN = {1553-7358},
  url = {http://dx.doi.org/10.1371/journal.pcbi.1011288},
  DOI = {10.1371/journal.pcbi.1011288},
  number = {8},
  journal = {PLOS Computational Biology},
  publisher = {Public Library of Science (PLoS)},
  author = {Chari,  Tara and Pachter,  Lior},
  editor = {Papin,  Jason A.},
  year = {2023},
  month = Aug,
  pages = {e1011288}
}

@article{jost_2020_titrating,
  title = {Titrating gene expression using libraries of systematically attenuated CRISPR guide RNAs},
  volume = {38},
  ISSN = {1546-1696},
  url = {http://dx.doi.org/10.1038/s41587-019-0387-5},
  DOI = {10.1038/s41587-019-0387-5},
  number = {3},
  journal = {Nature Biotechnology},
  publisher = {Springer Science and Business Media LLC},
  author = {Jost,  Marco and Santos,  Daniel A. and Saunders,  Reuben A. and Horlbeck,  Max A. and Hawkins,  John S. and Scaria,  Sonia M. and Norman,  Thomas M. and Hussmann,  Jeffrey A. and Liem,  Christina R. and Gross,  Carol A. and Weissman,  Jonathan S.},
  year = {2020},
  month = Jan,
  pages = {355–364}
}

@article{peidli_2024_scperturb,
  title = {scPerturb: harmonized single-cell perturbation data},
  volume = {21},
  ISSN = {1548-7105},
  url = {http://dx.doi.org/10.1038/s41592-023-02144-y},
  DOI = {10.1038/s41592-023-02144-y},
  number = {3},
  journal = {Nature Methods},
  publisher = {Springer Science and Business Media LLC},
  author = {Peidli,  Stefan and Green,  Tessa D. and Shen,  Ciyue and Gross,  Torsten and Min,  Joseph and Garda,  Samuele and Yuan,  Bo and Schumacher,  Linus J. and Taylor-King,  Jake P. and Marks,  Debora S. and Luna,  Augustin and Bl\"{u}thgen,  Nils and Sander,  Chris},
  year = {2024},
  month = Jan,
  pages = {531–540}
}

@article{shalek_2014_singlecell,
  title = {Single-cell RNA-seq reveals dynamic paracrine control of cellular variation},
  volume = {510},
  ISSN = {1476-4687},
  url = {http://dx.doi.org/10.1038/nature13437},
  DOI = {10.1038/nature13437},
  number = {7505},
  journal = {Nature},
  publisher = {Springer Science and Business Media LLC},
  author = {Shalek,  Alex K. and Satija,  Rahul and Shuga,  Joe and Trombetta,  John J. and Gennert,  Dave and Lu,  Diana and Chen,  Peilin and Gertner,  Rona S. and Gaublomme,  Jellert T. and Yosef,  Nir and Schwartz,  Schraga and Fowler,  Brian and Weaver,  Suzanne and Wang,  Jing and Wang,  Xiaohui and Ding,  Ruihua and Raychowdhury,  Raktima and Friedman,  Nir and Hacohen,  Nir and Park,  Hongkun and May,  Andrew P. and Regev,  Aviv},
  year = {2014},
  month = Jun,
  pages = {363–369}
}

@article{klijn_2015_comprehensive,
  title = {A comprehensive transcriptional portrait of human cancer cell lines},
  volume = {33},
  ISSN = {1546-1696},
  url = {http://dx.doi.org/10.1038/nbt.3080},
  DOI = {10.1038/nbt.3080},
  number = {3},
  journal = {Nature Biotechnology},
  publisher = {Springer Science and Business Media LLC},
  author = {Klijn,  Christiaan and Durinck,  Steffen and Stawiski,  Eric W and Haverty,  Peter M and Jiang,  Zhaoshi and Liu,  Hanbin and Degenhardt,  Jeremiah and Mayba,  Oleg and Gnad,  Florian and Liu,  Jinfeng and Pau,  Gregoire and Reeder,  Jens and Cao,  Yi and Mukhyala,  Kiran and Selvaraj,  Suresh K and Yu,  Mamie and Zynda,  Gregory J and Brauer,  Matthew J and Wu,  Thomas D and Gentleman,  Robert C and Manning,  Gerard and Yauch,  Robert L and Bourgon,  Richard and Stokoe,  David and Modrusan,  Zora and Neve,  Richard M and de Sauvage,  Frederic J and Settleman,  Jeffrey and Seshagiri,  Somasekar and Zhang,  Zemin},
  year = {2014},
  month = Dec,
  pages = {306–312}
}

@article{ben_david_2018_evolution,
  title = {Genetic and transcriptional evolution alters cancer cell line drug response},
  volume = {560},
  ISSN = {1476-4687},
  url = {http://dx.doi.org/10.1038/s41586-018-0409-3},
  DOI = {10.1038/s41586-018-0409-3},
  number = {7718},
  journal = {Nature},
  publisher = {Springer Science and Business Media LLC},
  author = {Ben-David,  Uri and Siranosian,  Benjamin and Ha,  Gavin and Tang,  Helen and Oren,  Yaara and Hinohara,  Kunihiko and Strathdee,  Craig A. and Dempster,  Joshua and Lyons,  Nicholas J. and Burns,  Robert and Nag,  Anwesha and Kugener,  Guillaume and Cimini,  Beth and Tsvetkov,  Peter and Maruvka,  Yosef E. and O’Rourke,  Ryan and Garrity,  Anthony and Tubelli,  Andrew A. and Bandopadhayay,  Pratiti and Tsherniak,  Aviad and Vazquez,  Francisca and Wong,  Bang and Birger,  Chet and Ghandi,  Mahmoud and Thorner,  Aaron R. and Bittker,  Joshua A. and Meyerson,  Matthew and Getz,  Gad and Beroukhim,  Rameen and Golub,  Todd R.},
  year = {2018},
  month = Aug,
  pages = {325–330}
}

@article{Gilbert2013,
  title = {CRISPR-Mediated Modular RNA-Guided Regulation of Transcription in Eukaryotes},
  volume = {154},
  ISSN = {0092-8674},
  url = {http://dx.doi.org/10.1016/j.cell.2013.06.044},
  DOI = {10.1016/j.cell.2013.06.044},
  number = {2},
  journal = {Cell},
  publisher = {Elsevier BV},
  author = {Gilbert,  Luke A. and Larson,  Matthew H. and Morsut,  Leonardo and Liu,  Zairan and Brar,  Gloria A. and Torres,  Sandra E. and Stern-Ginossar,  Noam and Brandman,  Onn and Whitehead,  Evan H. and Doudna,  Jennifer A. and Lim,  Wendell A. and Weissman,  Jonathan S. and Qi,  Lei S.},
  year = {2013},
  month = Jul,
  pages = {442–451}
}

@article{boettcher_2018_dual,
  title = {Dual gene activation and knockout screen reveals directional dependencies in genetic networks},
  volume = {36},
  ISSN = {1546-1696},
  url = {http://dx.doi.org/10.1038/nbt.4062},
  DOI = {10.1038/nbt.4062},
  number = {2},
  journal = {Nature Biotechnology},
  publisher = {Springer Science and Business Media LLC},
  author = {Boettcher,  Michael and Tian,  Ruilin and Blau,  James A and Markegard,  Evan and Wagner,  Ryan T and Wu,  David and Mo,  Xiulei and Biton,  Anne and Zaitlen,  Noah and Fu,  Haian and McCormick,  Frank and Kampmann,  Martin and McManus,  Michael T},
  year = {2018},
  month = Jan,
  pages = {170–178}
}

@article{eling_2019_challenges,
  title = {Challenges in measuring and understanding biological noise},
  volume = {20},
  ISSN = {1471-0064},
  url = {http://dx.doi.org/10.1038/s41576-019-0130-6},
  DOI = {10.1038/s41576-019-0130-6},
  number = {9},
  journal = {Nature Reviews Genetics},
  publisher = {Springer Science and Business Media LLC},
  author = {Eling,  Nils and Morgan,  Michael D. and Marioni,  John C.},
  year = {2019},
  month = May,
  pages = {536–548}
}

@article{kriegeskorte_2009_circular,
  title = {Circular analysis in systems neuroscience: the dangers of double dipping},
  volume = {12},
  ISSN = {1546-1726},
  url = {http://dx.doi.org/10.1038/nn.2303},
  DOI = {10.1038/nn.2303},
  number = {5},
  journal = {Nature Neuroscience},
  publisher = {Springer Science and Business Media LLC},
  author = {Kriegeskorte,  Nikolaus and Simmons,  W Kyle and Bellgowan,  Patrick S F and Baker,  Chris I},
  year = {2009},
  month = Apr,
  pages = {535–540}
}

@article{ihry_2018_p53,
  title = {p53 inhibits CRISPR–Cas9 engineering in human pluripotent stem cells},
  volume = {24},
  ISSN = {1546-170X},
  url = {http://dx.doi.org/10.1038/s41591-018-0050-6},
  DOI = {10.1038/s41591-018-0050-6},
  number = {7},
  journal = {Nature Medicine},
  publisher = {Springer Science and Business Media LLC},
  author = {Ihry,  Robert J. and Worringer,  Kathleen A. and Salick,  Max R. and Frias,  Elizabeth and Ho,  Daniel and Theriault,  Kraig and Kommineni,  Sravya and Chen,  Julie and Sondey,  Marie and Ye,  Chaoyang and Randhawa,  Ranjit and Kulkarni,  Tripti and Yang,  Zinger and McAllister,  Gregory and Russ,  Carsten and Reece-Hoyes,  John and Forrester,  William and Hoffman,  Gregory R. and Dolmetsch,  Ricardo and Kaykas,  Ajamete},
  year = {2018},
  month = Jun,
  pages = {939–946}
}

@article{haapaniemi_2018_crispr,
  title = {CRISPR–Cas9 genome editing induces a p53-mediated DNA damage response},
  volume = {24},
  ISSN = {1546-170X},
  url = {http://dx.doi.org/10.1038/s41591-018-0049-z},
  DOI = {10.1038/s41591-018-0049-z},
  number = {7},
  journal = {Nature Medicine},
  publisher = {Springer Science and Business Media LLC},
  author = {Haapaniemi,  Emma and Botla,  Sandeep and Persson,  Jenna and Schmierer,  Bernhard and Taipale,  Jussi},
  year = {2018},
  month = Jun,
  pages = {927–930}
}

@article{costamattioli_2020_isr,
  title = {The integrated stress response: From mechanism to disease},
  volume = {368},
  ISSN = {1095-9203},
  url = {http://dx.doi.org/10.1126/science.aat5314},
  DOI = {10.1126/science.aat5314},
  number = {6489},
  journal = {Science},
  publisher = {American Association for the Advancement of Science (AAAS)},
  author = {Costa-Mattioli,  Mauro and Walter,  Peter},
  year = {2020},
  month = Apr 
}

@article{symmons_2016_luck,
  title = {What’s Luck Got to Do with It: Single Cells,  Multiple Fates,  and Biological Nondeterminism},
  volume = {62},
  ISSN = {1097-2765},
  url = {http://dx.doi.org/10.1016/j.molcel.2016.05.023},
  DOI = {10.1016/j.molcel.2016.05.023},
  number = {5},
  journal = {Molecular Cell},
  publisher = {Elsevier BV},
  author = {Symmons,  Orsolya and Raj,  Arjun},
  year = {2016},
  month = Jun,
  pages = {788–802}
}

@book{mardia_2000_directional,
  title = {Directional Statistics},
  ISBN = {9780470316979},
  ISSN = {1940-6347},
  url = {http://dx.doi.org/10.1002/9780470316979},
  DOI = {10.1002/9780470316979},
  journal = {Wiley Series in Probability and Statistics},
  publisher = {Wiley},
  author = {Mardia,  Kanti V. and Jupp,  Peter E.},
  year = {1999},
  month = Jan 
}

@article{Garbulowski2024,
  title = {GeneSPIDER2: large scale GRN simulation and benchmarking with perturbed single-cell data},
  volume = {6},
  ISSN = {2631-9268},
  url = {http://dx.doi.org/10.1093/nargab/lqae121},
  DOI = {10.1093/nargab/lqae121},
  number = {3},
  journal = {NAR Genomics and Bioinformatics},
  publisher = {Oxford University Press (OUP)},
  author = {Garbulowski,  Mateusz and Hillerton,  Thomas and Morgan,  Daniel and Se\c{c}ilmiş,  Deniz and Sonnhammer,  Lisbet and Tj\"{a}rnberg,  Andreas and Nordling,  Torbj\"{o}rn E M and Sonnhammer,  Erik L L},
  year = {2024},
  month = Jul
}

@article{vousden_2009_p53,
  title = {Blinded by the Light: The Growing Complexity of p53},
  volume = {137},
  ISSN = {0092-8674},
  url = {http://dx.doi.org/10.1016/j.cell.2009.04.037},
  DOI = {10.1016/j.cell.2009.04.037},
  number = {3},
  journal = {Cell},
  publisher = {Elsevier BV},
  author = {Vousden,  Karen H. and Prives,  Carol},
  year = {2009},
  month = May,
  pages = {413–431}
}

@article{tabas_2011_er_stress,
  title = {Integrating the mechanisms of apoptosis induced by endoplasmic reticulum stress},
  volume = {13},
  ISSN = {1476-4679},
  url = {http://dx.doi.org/10.1038/ncb0311-184},
  DOI = {10.1038/ncb0311-184},
  number = {3},
  journal = {Nature Cell Biology},
  publisher = {Springer Science and Business Media LLC},
  author = {Tabas,  Ira and Ron,  David},
  year = {2011},
  month = Mar,
  pages = {184–190}
}

@article{norman2019exploring,
  title={Exploring genetic interaction manifolds constructed from rich single-cell phenotypes},
  author={Norman, Thomas M and Horlbeck, Max A and Replogle, Joseph M and Ge, Y Alexander and Xu, Alex and Jost, Marco and Gilbert, Luke A and Weissman, Jonathan S},
  journal={Science},
  volume = {365},
  ISSN = {1095-9203},
  DOI = {10.1126/science.aax4438},
  number = {6455},
  pages = {786–793},
  year={2019}
}

@article{adamson2016multiplexed,
  title = {A Multiplexed Single-Cell {CRISPR} Screening Platform Enables Systematic Dissection of the Unfolded Protein Response},
  author = {Adamson,  Britt and Norman,  Thomas M. and Jost,  Marco and Cho,  Min Y. and Nuñez,  James K. and Chen,  Yuwen and Villalta,  Jacqueline E. and Gilbert,  Luke A. and Horlbeck,  Max A. and Hein,  Marco Y. and Pak,  Ryan A. and Gray,  Andrew N. and Gross,  Carol A. and Dixit,  Atray and Parnas,  Oren and Regev,  Aviv and Weissman,  Jonathan S.},
  journal={Cell},
  volume = {167},
  ISSN = {0092-8674},
  DOI = {10.1016/j.cell.2016.11.048},
  number = {7},
  pages = {1867--1882.e21},
  year={2016}
}

@article{dixit2016perturb,
  title={Perturb-Seq: dissecting molecular circuits with scalable single-cell {RNA} profiling of pooled genetic screens},
  author = {Dixit,  Atray and Parnas,  Oren and Li,  Biyu and Chen,  Jenny and Fulco,  Charles P. and Jerby-Arnon,  Livnat and Marjanovic,  Nemanja D. and Dionne,  Danielle and Burks,  Tyler and Raychowdhury,  Raktima and Adamson,  Britt and Norman,  Thomas M. and Lander,  Eric S. and Weissman,  Jonathan S. and Friedman,  Nir and Regev,  Aviv},
  journal={Cell},
  volume = {167},
  ISSN = {0092-8674},
  DOI = {10.1016/j.cell.2016.11.038},
  number = {7},
  pages = {1853--1866.e17},
  year={2016}
}

@article{papalexi2021characterizing,
  title={Characterizing the molecular regulation of inhibitory immune checkpoints with multimodal single-cell screens},
  author = {Papalexi,  Efthymia and Mimitou,  Eleni P. and Butler,  Andrew W. and Foster,  Samantha and Bracken,  Bernadette and Mauck,  William M. and Wessels,  Hans-Hermann and Hao,  Yuhan and Yeung,  Bertrand Z. and Smibert,  Peter and Satija,  Rahul},
  journal={Nature Genetics},
  volume = {53},
  ISSN = {1546-1718},
  DOI = {10.1038/s41588-021-00778-2},
  number = {3},
  pages = {322–331},
  year={2021}
}

@article{pertpy,
  title = {Pertpy: an end-to-end framework for perturbation analysis},
  volume = {23},
  ISSN = {1548-7105},
  url = {http://dx.doi.org/10.1038/s41592-025-02909-7},
  DOI = {10.1038/s41592-025-02909-7},
  number = {2},
  journal = {Nature Methods},
  publisher = {Springer Science and Business Media LLC},
  author = {Heumos,  Lukas and Ji,  Yuge and May,  Lilly and Green,  Tessa D. and Peidli,  Stefan and Zhang,  Xinyue and Wu,  Xichen and Ostner,  Johannes and Schumacher,  Antonia and Hrovatin,  Karin and M\"{u}ller,  Michaela and Chong,  Faye and Sturm,  Gregor and Tejada,  Alejandro and Dann,  Emma and Dong,  Mingze and Pinto,  Gon\c{c}alo and Bahrami,  Mojtaba and Gold,  Ilan and Rybakov,  Sergei and Namsaraeva,  Altana and Moinfar,  Amir Ali and Zheng,  Zihe and Roellin,  Eljas and Mekki,  Isra and Sander,  Chris and Lotfollahi,  Mohammad and Schiller,  Herbert B. and Theis,  Fabian J.},
  year = {2025},
  month = dec,
  pages = {350–359}
}

@article{wolf2018scanpy,
  title = {{SCANPY}: large-scale single-cell gene expression data analysis},
  volume = {19},
  ISSN = {1474-760X},
  DOI = {10.1186/s13059-017-1382-0},
  number = {1},
  journal = {Genome Biology},
  author = {Wolf,  F. Alexander and Angerer,  Philipp and Theis,  Fabian J.},
  year = {2018}
}

@article{Moon2019,
  title = {Visualizing structure and transitions in high-dimensional biological data},
  volume = {37},
  ISSN = {1546-1696},
  url = {http://dx.doi.org/10.1038/s41587-019-0336-3},
  DOI = {10.1038/s41587-019-0336-3},
  number = {12},
  journal = {Nature Biotechnology},
  publisher = {Springer Science and Business Media LLC},
  author = {Moon,  Kevin R. and van Dijk,  David and Wang,  Zheng and Gigante,  Scott and Burkhardt,  Daniel B. and Chen,  William S. and Yim,  Kristina and Elzen,  Antonia van den and Hirn,  Matthew J. and Coifman,  Ronald R. and Ivanova,  Natalia B. and Wolf,  Guy and Krishnaswamy,  Smita},
  year = {2019},
  month = dec,
  pages = {1482–1492}
}

@article{Cui2024,
  title = {{scGPT}: toward building a foundation model for single-cell multi-omics using generative {AI}},
  volume = {21},
  ISSN = {1548-7105},
  DOI = {10.1038/s41592-024-02201-0},
  number = {8},
  journal = {Nature Methods},
  publisher = {Springer Science and Business Media LLC},
  author = {Cui,  Haotian and Wang,  Chloe and Maan,  Hassaan and Pang,  Kuan and Luo,  Fengning and Duan,  Nan and Wang,  Bo},
  year = {2024},
  month = feb,
  pages = {1470–1480}
}

@article{Replogle2022,
  title = {Mapping information-rich genotype-phenotype landscapes with genome-scale Perturb-seq},
  volume = {185},
  ISSN = {0092-8674},
  DOI = {10.1016/j.cell.2022.05.013},
  number = {14},
  journal = {Cell},
  publisher = {Elsevier BV},
  author = {Replogle,  Joseph M. and Saunders,  Reuben A. and Pogson,  Angela N. and Hussmann,  Jeffrey A. and Lenail,  Alexander and Guna,  Alina and Mascibroda,  Lauren and Wagner,  Eric J. and Adelman,  Karen and Lithwick-Yanai,  Gila and Iremadze,  Nika and Oberstrass,  Florian and Lipson,  Doron and Bonnar,  Jessica L. and Jost,  Marco and Norman,  Thomas M. and Weissman,  Jonathan S.},
  year = {2022},
  month = jul,
  pages = {2559--2575.e28}
}

@software{github,
  title = {prashantcraju/shesha-crispr: v.0.2.0},
  author = {Raju, Prashant C.},
  year = {2026},
  howpublished = {Zenodo},
  doi = {10.5281/ZENODO.22116628},
  url = {https://zenodo.org/doi/10.5281/zenodo.22116628},
  copyright = {MIT License},
  notes = {10.5281/zenodo.22116628}
}

@article{Liberzon2015,
  title = {The Molecular Signatures Database Hallmark Gene Set Collection},
  volume = {1},
  ISSN = {2405-4712},
  url = {http://dx.doi.org/10.1016/j.cels.2015.12.004},
  DOI = {10.1016/j.cels.2015.12.004},
  number = {6},
  journal = {Cell Systems},
  publisher = {Elsevier BV},
  author = {Liberzon,  Arthur and Birger,  Chet and Thorvaldsdóttir,  Helga and Ghandi,  Mahmoud and Mesirov,  Jill P. and Tamayo,  Pablo},
  year = {2015},
  month = Dec,
  pages = {417–425}
}

@article{Virshup2024,
  title = {anndata: Access and store annotated data
matrices},
  volume = {9},
  ISSN = {2475-9066},
  url = {http://dx.doi.org/10.21105/joss.04371},
  DOI = {10.21105/joss.04371},
  number = {101},
  journal = {Journal of Open Source Software},
  publisher = {The Open Journal},
  author = {Virshup,  Isaac and Rybakov,  Sergei and Theis,  Fabian J. and Angerer,  Philipp and Wolf,  F. Alexander},
  year = {2024},
  month = sep,
  pages = {4371}
}

@article{Song2025,
  title = {Decoding heterogeneous single-cell perturbation responses},
  volume = {27},
  ISSN = {1476-4679},
  url = {http://dx.doi.org/10.1038/s41556-025-01626-9},
  DOI = {10.1038/s41556-025-01626-9},
  number = {3},
  journal = {Nature Cell Biology},
  publisher = {Springer Science and Business Media LLC},
  author = {Song,  Bicna and Liu,  Dingyu and Dai,  Weiwei and McMyn,  Natalie F. and Wang,  Qingyang and Yang,  Dapeng and Krejci,  Adam and Vasilyev,  Anatoly and Untermoser,  Nicole and Loregger,  Anke and Song,  Dongyuan and Williams,  Breanna and Rosen,  Bess and Cheng,  Xiaolong and Chao,  Lumen and Kale,  Hanuman T. and Zhang,  Hao and Diao,  Yarui and B\"{u}rckst\"{u}mmer,  Tilmann and Siliciano,  Janet D. and Li,  Jingyi Jessica and Siliciano,  Robert F. and Huangfu,  Danwei and Li,  Wei},
  year = {2025},
  month = feb,
  pages = {493–504}
}

@article{Weinreb2020,
  title = {Lineage tracing on transcriptional landscapes links state to fate during differentiation},
  volume = {367},
  ISSN = {1095-9203},
  url = {http://dx.doi.org/10.1126/science.aaw3381},
  DOI = {10.1126/science.aaw3381},
  number = {6479},
  journal = {Science},
  publisher = {American Association for the Advancement of Science (AAAS)},
  author = {Weinreb,  Caleb and Rodriguez-Fraticelli,  Alejo and Camargo,  Fernando D. and Klein,  Allon M.},
  year = {2020},
  month = feb 
}
